\documentclass[12pt]{article}
\usepackage{graphicx} 
\usepackage{hyperref}
\usepackage{url}
\usepackage{listings,jvlisting}
\usepackage{amssymb}
\usepackage{tabularx}
\usepackage{amsmath}
\usepackage{mdframed}
\usepackage{graphicx}
\usepackage{setspace}
\usepackage{here}
\usepackage{authblk}
\usepackage{mathtools}
\usepackage{amsthm}
\usepackage{algorithmic}
\usepackage{algorithm}
\usepackage{cite}
\usepackage{lscape}
\usepackage{xr}

\makeatletter

\newcommand*{\addFileDependency}[1]{
\typeout{(#1)}
\@addtofilelist{#1}
\IfFileExists{#1}{}{\typeout{No file #1.}}
}\makeatother

\newcommand*{\myexternaldocument}[1]{%
\externaldocument{#1}%
\addFileDependency{#1.tex}%
\addFileDependency{#1.aux}%
}

\def\be{{\beta}}

\def\de{{\delta}}

\def\la{{\lambda}}

\def\si{{\sigma}}

\def\bbe{{\text{\boldmath $\beta$}}}

\def\c{{\text{\boldmath $c$}}}

\def\t{{\text{\boldmath $t$}}}
\def\u{{\text{\boldmath $u$}}}
\def\v{{\text{\boldmath $v$}}}

\def\x{{\text{\boldmath $x$}}}

\def\R{{\text{\boldmath $R$}}}

\def\X{{\text{\boldmath $X$}}}

\def\Nc{{\cal N}}

\myexternaldocument{supp}

\begin{document}
\title{Bayesian Parametric Methods for Deriving Distribution of Restricted Mean Survival Time}
\author[1]{Keisuke Hanada}
\author[2]{Masahiro Kojima\footnote{Address:10-3 Midori-cho, Tachikawa, Tokyo 190-8562, Japan. Tel: +81-(0)50-5533-8500 \quad
E-Mail: kojima.masahiro@ism.ac.jp}}
\affil[1]{Osaka University}
\affil[2]{The Institute of Statistical Mathematics}

\maketitle
\abstract{\noindent
We propose a Bayesian method for deriving the distribution of restricted mean survival time (RMST) using posterior samples, which accounts for covariates and heterogeneity among clusters based on a parametric model for survival time. We derive an explicit RMST equation by devising an integral of the survival function, allowing for the calculation of not only the mean and credible interval but also the mode, median, and probability of exceeding a certain value. Additionally, We propose two methods: one using random effects to account for heterogeneity among clusters and another utilizing frailty. We developed custom Stan code for the exponential, Weibull, log-normal frailty, and log-logistic models, as they cannot be processed using the brm functions in R. We evaluate our proposed methods through computer simulations and analyze real data from the eight Empowered Action Group states in India to confirm consistent results across states after adjusting for cluster differences.

In conclusion, we derived explicit RMST formulas for parametric models and their distributions, enabling the calculation of the mean, median, mode, and credible interval. Our simulations confirmed the robustness of the proposed methods, and using the shrinkage effect allowed for more accurate results for each cluster.
}
\par\vspace{4mm}

\section{Introduction}

The restricted mean survival time (RMST) up to a pre-specified evaluation time point allows for the assessment of survival time using an intuitive measure of the mean~\cite{uno2014moving}. RMST can serve as a valuable endpoint in Phase II and III randomized controlled studies. Most RMST methodological research is based on frequentist theory~\cite{andersen2003generalised,tian2014predicting,hanada2024random}, allowing for the adjustment of baseline covariates and heterogeneity among clusters. However, these methods do not support deriving the distribution of RMST, estimating the probability of RMST exceeding a specific threshold, or utilizing prior distributions. While frequentist theory offers differences in means and confidence intervals through normal approximation, we aim to derive more exact distributions using Bayesian analysis. Although a non-parametric Bayesian method for RMST estimation has been proposed~\cite{zhang2023bayesian}, it cannot handle covariate adjustment or cluster heterogeneity.

In this paper, we propose a Bayesian method for deriving the distribution of RMST using posterior samples. Because it is difficult to fit RMST directly to a statistical model, we present a Bayesian representation that can account for covariates and heterogeneity among clusters based on a parametric model for survival time. The RMST requires integrating the survival function up to the evaluation point. Although integrating the survival function is challenging for most parametric distributions, we use a formula that facilitates calculating the integral for all survival functions by transforming the existing equation. No explicit equation for RMST has ever been proposed, and this equation allows us to calculate the distribution of RMST using posterior samples. By obtaining the RMST distribution, we can compute not only the mean and confidence interval but also the mode, median, and probability of exceeding a certain value. It is often necessary to account for heterogeneity among clusters or to use shrinkage effects to increase estimation accuracy for clusters with small sample sizes. Hence, we introduce two methods for RMST: one using a random effect to account for heterogeneity among clusters and another utilizing frailty. We evaluate the performance of the RMST distribution using posterior samples in simulations and present the results of an analysis based on real data from the eight Empowered Action Group (EAG) states in India. We develop custom Stan source code to analyze the exponential, Weibull, log-normal frailty models and all log-logistic models, as they cannot be processed using the brm functions.

This paper is organized as follows. Section 2 introduces the explicit RMST formula calculation, Bayesian sampling, and RMST distribution. Section 3 details the settings and results of computer simulations. Section 4 presents an analysis plan and the results of analyzing data from the eight EAG states in India. Finally, Section 5 discusses all the results.

\section{Method}
\subsection{Restricted mean survival time}
The RMST is defined as the integral of any survival function $S(t)$ from $0$ to $\tau$, where $\tau>0$ is the pre-specified time point. We assume a parametric model for the survival time. The RMST requires integrating the survival function, but this integral of the parametric survival function can generally be complex. Therefore, we use the following formula to calculate the RMST.
\begin{align}
\label{eq:formula}
RMST(\tau)=\int^\tau_0 S(t)dt=\int^\tau_0 tf(t)dt+\tau S(\tau),
\end{align}
where $f(t)$ is a density function for a random survival time $T$. Equation (\ref{eq:formula}) implies that the RMST can be calculated for any distribution for which the process for deriving the expectation is known. The equation (\ref{eq:formula}) is obtained from the following integral.
\begin{align}
\int^\tau_0 tf(t)dt=[-tS(t)]^\tau_0+\int^\tau_0 S(t)dt=-\tau S(\tau)+\int^\tau_0 S(t)dt.\nonumber
\end{align}
\medskip
This paper deals with four parametric models: exponential, Weibull, log-logistic, and log-normal. We demonstrate the RMSTs of these models. The RMSTs for models other than these four may be calculated using the formula (\ref{eq:formula}).
\medskip
\\
{\bf [Exponential distribution]}
The RMST for an exponential distribution $T \sim Exp(\lambda)$ can be easily obtained, where a rate parameter $\lambda>0$.
\begin{align}
RMST_E(\tau)=\int^\tau_0 S(t)dt=\frac{1-e^{-\lambda \tau}}{\lambda}.\nonumber
\end{align}
When $\tau$ approaches infinitely, it becomes apparent that the expected value of the exponential distribution is obtained. The derivation is shown in the Appendix \ref{App:exp_RMST} in the Supplemental material. Note that it may not be necessary to apply the equation \ref{eq:formula} because the survival function of the exponential distribution can be easily integrated. 
\medskip
\\
{\bf [Weibull distribution]}
Next, we consider the RMST of the Weibull distribution $T \sim W(\lambda,k)$, where $\la$ is the scale parameter ($\la>0$) and $k$ is the shape parameter ($k>0$). The RMST is challenging to calculate without using formula (\ref{eq:formula}).
\begin{align}
RMST_W(\tau)=\int^\tau_0 S(t)dt=\lambda^{-\frac{1}{k}}\gamma\left(\lambda\tau^k;\frac{1}{k}+1\right)+\tau \exp(-\lambda \tau^k), \nonumber
\end{align}
where $\gamma(z;a)=\int^z_0 t^{a-1}e^{-t}dt$ is the incomplete gamma function. The incomplete gamma function can be computed in software such as R or Python. The derivation of the RMST is found in the Appendix \ref{App:Wei_RMST} in the Supplemental material. By making $\tau$ sufficiently large, the RMST coincides with the expectation of the Weibull distribution. The RMST can be obtained by substituting the estimated values into the parameters. Therefore, the ability to calculate the RMST using an explicit formula, even for distributions where directly integrating the survival function is challenging, is an advantage. We can also see that in subsequent distributions, we can calculate the RMST without using approximations.
\medskip
\\
{\bf [Log-logistic distribution]}
We show the RMST for the log-logistic distribution $T \sim LL(\mu,k)$, where $\mu$ is the scale parameter ($\mu\in\R$) and $k$ is the shape parameter ($k>0$).
\begin{align}
RMST_{LL}(\tau)=\int^\tau_0 S(t)dt=e^{-\frac{\mu}{k}}B\left(\frac{e^\mu\tau^k}{1+e^\mu\tau^k};1+\frac{1}{k},1-\frac{1}{k}\right)+\tau\frac{1}{1+e^\mu \tau^k}, \nonumber
\end{align}
where $B(z;a,b)=\int^{z}_0 t^{a-1}(1-t)^{b-1}dt$ is the incomplete beta function. The incomplete gamma function can also be computed in software such as R or Python. The derivation can be found in the Appendix \ref{App:LL_RMST} in the Supplemental material. Based on the first moment condition, we restrict to $k>1$. Taking the limit as $\tau\rightarrow \infty$, we obtain the expectation of the log-logistic distribution. 
\medskip
\\
{\bf [Log-normal distribution]}
Finally, we show the RMST in a log-normal distribution $T \sim LN(\mu,\sigma^2)$, where the parameters are $\mu\in\R$ and $\sigma^2>0$.
\begin{align}
\label{eq:LL_RMST}
RMST_{LN}(\tau)=\int^\tau_0 S(t)dt=\exp\left\{\mu+\frac{\si^2}{2}\right\}\Phi\left(\frac{\log(\tau)-\mu-\si^2}{\si}\right)+\tau\left(1-\Phi\left(\frac{\log(\tau)-\mu}{\sigma}\right)\right).
\end{align}
The derivation is shown in the Appendix \ref{App:LN_RMST} in the Supplemental material. As $\tau\rightarrow\infty$ on the right-hand side of the equation, equation (\ref{eq:LL_RMST}) becomes the expectation of the log-normal distribution.
\medskip
\\
We also consider the RMST for each cluster with shrinkage effects via the hierarchical Bayesian method. The RMST, including random effects, is shown in Appendix \ref{App:RMST_rand} in the Supplemental material. Additionally, the RMST under the hazard with frailty effects is presented in Appendix \ref{App:RMST_frail} in the Supplemental material.

The Weibull and log-logistic distributions have different representations, and the RMSTs in the different representations are shown in Appendix \ref{App:Ano_W} and \ref{App:Ano_LL} in the Supplemental material.

The Bayesian distribution of RMSTs is given by substituting the posterior samples into the parameters. In the next section, we will introduce the method for generating posterior samples.

\subsection{Posterior probability}\label{sec:post}
We show a posterior probability for each model. We assume that there are $M$ clusters and sample size of $n_i$ for the $i$-th cluster. Additionally, to account for the heterogeneity, we apply the random effect or frailty. We assume that the random effect $u_i$ follows a normal distribution $\Nc(0,\phi^2)$ and the frailty $v_i$ follows a gamma distribution $Gamma\left(\frac{1}{\phi},\frac{1}{\phi}\right)$.
\medskip
\\
{\bf [Exponential model]} We transform the parameter $\lambda$ to account for the $q$ covariates vector $\x$.
\begin{align}
\label{eq:para_transform}
\la_{ij}=\exp\{\x_{ij}^T\bbe\}
\end{align}
where $i$ represents the $i$-th cluster, $j$ represents the $j$-th subject in $i$-th cluster, $\x_{ij}$ includes the coefficient corresponding to the intercept term and the treatment group, and baseline covariates, and $\bbe$ is a $q$-dimensional coefficient parameter vector. The random effect $u_i$ for $i$-th cluster is added to $\x_{ij}^T\bbe$. 

The posterior probability is 
\begin{align}
\pi_r(\bbe,\u,\phi|\t,\X)\propto L_r(\t|\X,\bbe,\u,\phi)g_r(\u|\phi)\pi(\bbe|\c)\pi(\phi|\xi),\nonumber
\end{align}
where
\begin{align}
L_r(\t|\X,\bbe,\u,\phi)&=\prod^M_{i=1}\prod^{n_i}_{j=1}\{f_r(t_{ij}|v_i)\}^{\de_{ij}}\{S_r(t_{ij}|v_i)\}^{1-\de_{ij}}\nonumber\\
&=\prod^M_{i=1}\prod^{n_i}_{j=1}\left[\exp\{\x_{ij}^T\bbe+u_i\}\exp\left\{-\exp(\x_{ij}^T\bbe+u_i)t_{ij}\right\}\right]^{\de_{ij}}\nonumber\\
&\hspace{2cm}\times\left[\exp\left\{-\exp(\x_{ij}^T\bbe+u_i)t_{ij}\right\}\right]^{1-\de_{ij}},\nonumber
\end{align}
the variable $\de_{ij}$ is a censoring indicator that takes the value 1 if an event occurs and 0 if censored. The details of the other functions are shown in the Appendix \ref{App:exp_rand} in the Supplemental material.

The frailty model is defined as the product of the hazard function and the frailty term. The hazard function with the frailty term $v_i$ is
\begin{align}
h(t_{ij}|v_i)=v_ih(t_{ij})=v_i\exp\{\x_{ij}^T\bbe\},\nonumber
\end{align}
where $h(t_{ij})$ is the hazard function of the exponential distribution. 
The posterior probability is
\begin{align}
\pi_f(\bbe,\v,\phi|\t,\X)\propto L_f(\t|\X,\bbe,\v,\phi)g_f(\v|\phi)\pi(\bbe|\c)\pi(\phi|\xi),\nonumber
\end{align}
where
\begin{align}
L_f(\t|\X,\bbe,\v,\phi)&=\prod^M_{i=1}\prod^{n_i}_{j=1}\{f_f(t_{ij}|v_i)\}^{\de_{ij}}\{S_f(t_{ij}|v_i)\}^{1-\de_{ij}}\nonumber\\
&=\prod^M_{i=1}\prod^{n_i}_{j=1}\left[v_j\exp\{\x_{ij}^T\bbe\}\exp\left\{-v_j\exp\{\x_{ij}^T\bbe\}t\right\}\right]^{\de_{ij}}\left[\exp\left\{-v_j\exp\{\x_{ij}^T\bbe\}t\right\}\right]^{1-\de_{ij}}.\nonumber
\end{align}
The details of the other functions are shown in the Appendix \ref{App:exp_frail} in the Supplemental material. 
\medskip
\\
{\bf [Weibull model]}
Next, we consider the Weibull model. We transform the parameter $\lambda$ to the equation (\ref{eq:para_transform}) to account for covariates similarly to an exponential distribution. For the mixed effects model, the posterior probability is 
\begin{align}
\pi_r(\bbe,\u,\phi,k|\t,\X)\propto L_r(\t|\X,\bbe,\u,\phi,k)g_r(\u|\phi)\pi(\bbe|\c)\pi(\phi|\xi)\pi(k|a,b),\nonumber
\end{align}
where
\begin{align}
L_r(\t|\X,\bbe,\u,\phi,k)&=\prod^M_{i=1}\prod^{n_i}_{j=1}\{f_r(t_{ij}|u_i)\}^{\de_{ij}}\{S_r(t_{ij}|u_i)\}^{1-\de_{ij}}\nonumber\\
&=\prod^M_{i=1}\prod^{n_i}_{j=1}\left[\frac{\exp\{\x_{ij}^T\bbe+u_i\}k t_{ij}^{k-1}}{\left(1+\exp(\x_{ij}^T\bbe+u_i) t^k\right)^{2}}\right]^{\de_{ij}}\left[\frac{1}{1+\exp(\x_{ij}^T\bbe+u_i) t^k}\right]^{1-\de_{ij}}.\nonumber
\end{align}
The details of the other functions are shown in the Appendix \ref{App:wei_rand} in the Supplemental material. The hazard function with frailty term $v_i$ is
\begin{align}
h(t_{ij}|v_i)=v_i\exp\{\x_{ij}^T\bbe\} k t^{k-1}.\nonumber
\end{align}
The posterior probability is 
\begin{align}
\pi_f(\bbe,\v,\phi,k|\t,\X)\propto L_f(\t|\X,\bbe,\v,\phi)g_f(\v|\phi)\pi(\bbe|\c)\pi(\phi|\xi)\pi(k|a,b)\nonumber
\end{align}
\begin{align}
L_f(\t|\X,\bbe,\v,\phi,k)&=\prod^M_{i=1}\prod^{n_i}_{j=1}\{f_f(t_{ij}|v_i)\}^{\de_{ij}}\{S_f(t_{ij}|v_i)\}^{1-\de_{ij}}\nonumber\\
&=\prod^M_{i=1}\prod^{n_i}_{j=1}\left[v_j\exp\{\x_{ij}^T\bbe\}kt^{k-1}\exp\left\{-v_j\exp\{\x_{ij}^T\bbe\}t^k\right\}\right]^{\de_{ij}}\nonumber\\
&\hspace{2cm}\times\left[\exp\left\{-v_j\exp\{\x_{ij}^T\bbe\}t^k\right\}\right]^{1-\de_{ij}}.\nonumber
\end{align}
The details of the other functions are shown in the Appendix \ref{App:wei_frail} in the Supplemental material.
\medskip
\\
{\bf [Log-logistic distribution]}
For the log-logistic distribution, we transform the parameter $\mu$ to account for covariates. 
\begin{align}
\mu_{ij}=\x_{ij}^T\bbe.\nonumber
\end{align}
For the random effects model, the posterior probability is 
\begin{align}
\pi_r(\bbe,\u,\phi,k|\t,\X)\propto L_r(\t|\X,\bbe,\u,\phi,k)g_r(\u|\phi)\pi(\bbe|\c)\pi(\phi|\xi)\pi(k|a,b),\nonumber
\end{align}
\begin{align}
L_r(\t|\X,\bbe,\u,\phi,k)&=\prod^M_{i=1}\prod^{n_i}_{j=1}\{f_r(t_{ij}|u_i)\}^{\de_{ij}}\{S_r(t_{ij}|u_i)\}^{1-\de_{ij}}\nonumber\\
&=\prod^M_{i=1}\prod^{n_i}_{j=1}\left[\frac{\exp\{\x_{ij}^T\bbe+u_i\}k t_{ij}^{k-1}}{\left(1+\exp(\x_{ij}^T\bbe+u_i) t^k\right)^{2}}\right]^{\de_{ij}}\left[\frac{1}{1+\exp(\x_{ij}^T\bbe+u_i) t^k}\right]^{1-\de_{ij}}.\nonumber
\end{align}
The details of the other functions are shown in the Appendix \ref{App:LL_rand} in the Supplemental material. For the frailty model, the hazard function with frailty term $v_i$ is
\begin{align}
h(t_{ij}|v_i)=v_i\frac{e^{\x_{ij}^T\bbe}k t^{k-1}}{1+e^{\x_{ij}^T\bbe}t^k}. \nonumber
\end{align}
The posterior probability is
\begin{align}
\pi_f(\bbe,\v,\phi,k|\t,\X)\propto L_f(\t|\X,\bbe,\v,\phi)g_f(\v|\phi)\pi(\bbe|\c)\pi(\phi|\xi)\pi(k|a,b),\nonumber
\end{align}
where
\begin{align}
L_f(\t|\X,\bbe,\v,\phi,k)&=\prod^M_{i=1}\prod^{n_i}_{j=1}\{f(t_{ij}|v_i)\}^{\de_{ij}}\{S(t_{ij}|v_i)\}^{1-\de_{ij}}\nonumber\\
&=\prod^M_{i=1}\prod^{n_i}_{j=1}\left[v_i\exp\{\x_{ij}^T\bbe\}k t_{ij}^{k-1}\left(\frac{1}{1+e^{\x_{ij}^T\bbe} t^k}\right)^{v_i+1}\right]^{\de_{ij}}\left[\left(\frac{1}{1+e^{\x_{ij}^T\bbe} t^k}\right)^{v_i}\right]^{1-\de_{ij}}.\nonumber
\end{align}
The details of the other functions are shown in the Appendix \ref{App:LL_frail} in the Supplemental material.
\medskip
\\
{\bf [Log-normal distribution]}
We transform the parameter $\mu$ to account for covariates.
\begin{align}
\mu_{ij}=\x_{ij}^T\bbe. \nonumber
\end{align}
For the random effect model, the posterior probability is
\begin{align}
\pi_r(\bbe,\u,\phi,\si^2|\t,\X)\propto L_r(\t|\X,\bbe,\u,\phi,\si^2)g_r(\u|\phi)\pi(\bbe|\c)\pi(\phi|\xi)\pi(\si^2|a,b),\nonumber
\end{align}
\begin{align}
L_r(\t|\X,\bbe,\u,\phi,\si^2)&=\prod^M_{i=1}\prod^{n_i}_{j=1}\{f_r(t_{ij}|u_i)\}^{\de_{ij}}\{S_r(t_{ij}|u_i)\}^{1-\de_{ij}}\nonumber\\
&=\prod^M_{i=1}\prod^{n_i}_{j=1}\left[\frac{1}{t\sqrt{2\pi\sigma^2}}\exp\left[-\frac{\{\log(t)-(\x_{ij}^T\bbe+u_i)\}^2}{2\sigma^2}\right]\right]^{\de_{ij}}\nonumber\\
&\hspace{2cm}\times\left[1-\Phi\left(\frac{\log(t)-(\x_{ij}^T\bbe+u_i)}{\sigma}\right)\right]^{1-\de_{ij}}. \nonumber
\end{align}
The details of the other functions are shown in the Appendix \ref{App:LN_rand} in the Supplemental material. For the frailty model, the hazard function with frailty term $v_i$ is
\begin{align}
h(t_{ij}|v_i)=v_i\frac{\frac{1}{t\sqrt{2\pi\sigma^2}}\exp\left\{-\frac{(\log(t)-\mu)^2}{2\sigma^2}\right\}}{1-\Phi\left(\frac{\log(t)-\mu}{\sigma}\right)}.
\end{align}
The posterior probability is
\begin{align}
\pi_f(\bbe,\v,\phi,\si^2|\t,\X)\propto L_f(\t|\X,\bbe,\v,\phi)g_f(\v|\phi)\pi(\bbe|\c)\pi(\phi|\xi)\pi(\si^2|a,b),\nonumber
\end{align}
where
\begin{align}
L_f(\t|\X,\bbe,\v,\phi,k)&=\prod^M_{i=1}\prod^{n_i}_{j=1}\{f(t_{ij}|v_i)\}^{\de_{ij}}\{S(t_{ij}|v_i)\}^{1-\de_{ij}}\nonumber\\
&=\prod^M_{i=1}\prod^{n_i}_{j=1}\left[v_i\frac{1}{t\sqrt{2\pi\sigma^2}}\exp\left\{-\frac{(\log(t)-\mu)^2}{2\sigma^2}\right\}\left(1-\Phi\left(\frac{\log(t)-\mu}{\sigma}\right)\right)^{v_i-1}\right]^{\de_{ij}}\nonumber\\
&\hspace{2cm}\times\left[\left(1-\Phi\left(\frac{\log(t)-\mu}{\sigma}\right)\right)^{v_i}\right]^{1-\de_{ij}}.\nonumber
\end{align}
The details of the other functions are shown in the Appendix \ref{App:LN_frail} in the Supplemental material.

If random or frailty effects are not included, a function with $u_i$ set to 0 or $v_i$ set to 1 can be used, and the posterior probabilities can be calculated by excluding the distribution of the effects.

\subsection{Distribution of RMST}
The distribution of RMST is obtained using the posterior sample. We are interested in the RMST for each group and the difference between groups. For $\bbe$, when the intercept is $\be_0$ and the coefficient parameter for the dose group is $\be_1$, the coefficient parameters reflected in the RMST are $\be_0$ and $\be_1$.
\medskip
\\
{\bf [Exponential model]}
The distribution of RMST for each group is obtained below
\begin{align}
\label{eq:RMST-E}
RMST_E(\tau,x_1,\be_0^\ast,\be_1^\ast)=\frac{1-e^{-\exp(\be_0^\ast+x_1\be_1^\ast) \tau}}{\exp(\be_0^\ast+x_1\be_1^\ast)},
\end{align}
where $\ast$ denotes the posterior sample, and $x_1$ is 0 if the group is the control and 1 if the group is the treatment. The difference between the RMSTs is
\begin{align}
RMST_E(\tau,1,\be_0^\ast,\be_1^\ast)-RMST_E(\tau,0,\be_0^\ast,\be_1^\ast).\nonumber
\end{align}
\medskip
\\
{\bf [Weibull model]}
The distribution of RMST for each group is obtained below
\begin{align}
\label{eq:RMST-W}
RMST_W(\tau,x_1,\be_0^\ast,\be_1^\ast)=&\exp\left(-\frac{\be_0^\ast+x_1\be_1^\ast}{k^\ast}\right)\gamma\left(\exp(\be_0^\ast+x_1\be_1^\ast)\tau^{k^\ast};\frac{1}{k^\ast}+1\right)\nonumber\\
&\hspace{2cm}+\tau \exp\left(-\exp(\be_0^\ast+x_1\be_1^\ast)\tau^{k^\ast}\right).
\end{align}
The difference between the RMSTs is
\begin{align}
RMST_W(\tau,1,\be_0^\ast,\be_1^\ast)-RMST_W(\tau,0,\be_0^\ast,\be_1^\ast).\nonumber
\end{align}
\medskip
\\
{\bf [Log-logistic model]}
The distribution of RMST for each group is obtained below
\begin{align}
\label{eq:RMST-LL}
RMST_{LL}(\tau,x_1,\be_0^\ast,\be_1^\ast)=e^{-\frac{\be_0^\ast+x_1\be_1^\ast}{k^\ast}}B\left(\frac{e^{\be_0^\ast+x_1\be_1^\ast}\tau^{k^\ast}}{1+e^{\be_0^\ast+x_1\be_1^\ast}\tau^{k^\ast}};1+\frac{1}{k^\ast},1-\frac{1}{k^\ast}\right)+\tau\frac{1}{1+e^{\be_0^\ast+x_1\be_1^\ast}\tau^{k^\ast}}.
\end{align}
The difference between the RMSTs is
\begin{align}
RMST_{LL}(\tau,1,\be_0^\ast,\be_1^\ast)-RMST_{LL}(\tau,0,\be_0^\ast,\be_1^\ast). \nonumber
\end{align}
\medskip
\\
{\bf [Log-normal model]}
The distribution of RMST for each group is obtained below
\begin{align}
\label{eq:RMST-LN}
RMST_{LN}(\tau,x_1,\be_0^\ast,\be_1^\ast)&=\exp\left\{\be_0^\ast+x_1\be_1^\ast+\frac{\si^{2\ast}}{2}\right\}\Phi\left(\frac{\log(\tau)-(\be_0^\ast+x_1\be_1^\ast-\si^{2\ast})}{\si^\ast}\right) \nonumber\\
&\hspace{1.5cm}+\tau\left(1-\Phi\left(\frac{\log(\tau)-(\be_0^\ast+x_1\be_1^\ast)}{\si^\ast}\right)\right).
\end{align}
The difference between the RMSTs is
\begin{align}
RMST_{LN}(\tau,1,\be_0^\ast,\be_1^\ast)-RMST_{LN}(\tau,0,\be_0^\ast,\be_1^\ast). \nonumber
\end{align}

The RMST for each cluster, using shrinkage effects via random or frailty effects, is shown in the Appendices \ref{App:RMST_rand} and \ref{App:RMST_frail} in the Supplemental material, as it follows a similar RMST formula to the ones introduced so far.

\subsection{Theoretical aspect}
We show the consistency of the RMST for any models when the estimated parameters are consistent. The consistency of the RMST implies that the RMST derived from the posterior distribution approaches the true value as the number of subjects increases.
Furthermore, it is known that the posterior probability has consistency if the true parameter is in support of the prior and the consistent Bayes estimator exists with rich enough information set~\cite{doob1949application, gelman2013philosophy, schervish2012theory}.

Let $\hat{\bbe}_n = (\hat{\beta}_{n0}, \hat{\beta}_{n1})$ be the estimators by the posterior distribution with $n$ subjects and $\bbe = (\beta_0, \beta_1)$ be the coefficient parameters.
If $\hat{\bbe}_n \to_p \bbe$, then $\int_0^{\tau} S(t|\hat{\bbe}_n) dt \to_p \int_0^{\tau} S(t|\bbe) dt$.

Here, we show the proof. We can calculate the below inequality.
\begin{align*}
\left| \int_0^{\tau} S(t|\hat{\bbe}_n) dt - \int_0^{\tau} S(t|\bbe) dt \right| 
&= \left| \int_0^{\tau} \left\{ S(t|\hat{\bbe}_n) - S(t|\bbe) \right\} dt \right| \\
&\le \int_0^{\tau} \left| S(t|\hat{\bbe}_n) - S(t|\bbe) \right| dt \\
&= \int_0^{\tau} \left| F(t|\hat{\bbe}_n) - F(t|\bbe) \right| dt
\end{align*}
Because the maximum difference for distribution functions is $1$, there exists a constant $t_c \in [0, \tau]$ that satisfies 
\begin{align*}
    \int_0^{\tau} \left| F(t|\hat{\bbe}_n) - F(t|\bbe) \right| dt &\le \tau \left| F(t_c|\hat{\bbe}_n) - F(t_c|\bbe) \right|.
\end{align*}
By the first degree Taylor approximating polynomial of $F(t|\hat{\bbe}_n)$ at $\hat{\bbe}_n = \bbe$, the inequality can be transformed as
\begin{align*}
    \tau \left| F(t_c|\hat{\bbe}_n) - F(t_c|\bbe) \right| 
    &= \tau \left| \left\{ \left. \frac{\partial F(t_c|\bbe')}{\partial \beta_0} \right|_{\bbe=\bbe'} \right\} (\hat{\beta}_{n0} - \beta_0) + \left\{ \left. \frac{\partial F(t_c|\bbe')}{\partial \beta_1} \right|_{\bbe=\bbe'} \right\} (\hat{\beta}_{n1} - \beta_1) \right|  \\
    &\le \tau \left| \left\{ \left.\frac{\partial F(t_c|\bbe)}{\partial \beta_0}\right|_{\bbe=\bbe'} \right\} \right| \left|\hat{\beta}_{n0} - \beta_0\right| + \tau \left| \left\{ \left. \frac{\partial F(t_c|\bbe)}{\partial \beta_1}\right|_{\bbe=\bbe'} \right\} \right| \left|\hat{\beta}_{n1} - \beta_1 \right| \\
    &\to 0 \,\,\, (n \to \infty),
\end{align*}
where $\bbe'$ is a constant that satisfied $|\beta'_i-\beta_i|<|\beta'_i-\hat{\beta}_i| \,\,\, (i=0,1)$.
Thus, $\left| \int_0^{\tau} S(t|\hat{\bbe}_n) dt - \int_0^{\tau} S(t|\bbe) dt \right| \to 0 \,\,\, (n\to\infty)$, and then the RMST has consistent.

\subsection{Model selection}
We utilize model selection methods to determine the distribution, covariates, and random and frailty effects that best describe the time-to-event data. Model selection is conducted using the posterior samples. Established information criteria such as widely applicable information criterion (WAIC)~\cite{watanabe2010asymptotic}, widely applicable Bayesian information criterion (WBIC)~\cite{watanabe2013widely}, deviance information criterion (DIC)~\cite{spiegelhalter2002bayesian}, and leave-one-out cross-validation (LOO)~\cite{vehtari2017practical} are available. In particular, WAIC and LOO can be calculated using the \texttt{brms} package in R.

\section{Simulation}
\subsection{Simulation configuration}
We investigated the performance of our proposed method through computer simulations based on two data scenarios. For comparison purposes, we computed the bias, mean squared error (MSE), the difference between mode and true RMST (Mode), and the difference between median and true RMST (Median). Specifically, our primary interest was the bias of the difference in RMST between groups across exponential, Weibull, log-logistic, and log-normal distribution. We assumed that a censoring probability is $0.1$, and all survival times larger than $100$ months were censored. We considered different sample sizes ranging from small to large $(N=64, 512, 2048)$. The Markov chain Monte Carlo (MCMC) specification included two chains, each with 2000 iterations and 1000 burn-ins. The number of simulations was set to 100 due to the extensive computation time involved in MCMC.

The simulation datasets for Scenario A were generated from a log-logistic model. The scale parameter was as follows:
\begin{align}
\label{eq:simmodel-ll}
    \mu_{ij} &= -(\beta_0 + \beta_1 x_{1ij} + \beta_2 x_{2ij} + u_{i}),
\end{align}
where $\beta_0=5$, $\beta_1=-0.2$, $\beta_2=1$, $u_i$ represented the $i$-th cluster random effect drawn from a normal distribution with mean $0$ and variance $0.1$, $x_{1ij}$ denoted the treatment arm ($0$ for the control group and $1$ for the treatment group), and $x_{2ij}$ was a continuous covariate distributed as standard normal. The number of clusters was four, and the shape parameter $k$ was set to $2$.
The true model-based RMST value for each group in Scenario A was $82.69$ and $87.99$ by equation \eqref{eq:RMST-LL}, respectively. Thus, the true difference between the two RMSTs was $-5.30$.

The simulation datasets of Scenario B were generated from a log-normal model. The parameters were given as like equation \eqref{eq:simmodel-ll}, where $\beta_0=3$, $\beta_1=-0.5$ and $\beta_2=1$. The variance parameter $\sigma^2$ was set to $1$. The true model-based RMST values for each group in Scenario B were $19.14$ and $29.51$ by equation \eqref{eq:RMST-LN}, respectively. Thus, the true difference between the RMSTs was $-10.37$.

A supplementary simulation, Scenario C, was also conducted. Refer to Appendix \ref{t:sim-C} in the Supplemental material for details on Scenario C. All survival curves were shown in Supplemental Figure \ref{Sim:data} in the Supplemental material.

\subsection{Simulation results}
The simulation results were in Table \ref{t:sim} for Scenarios A and B, and Supplemental Table \ref{t:sim-C} for scenario C in the Supplemental material.
In Scenario A, where the mixed-effects log-logistic model was the true model, the bias and MSE of the log-logistic model were small regardless of the number of subjects. However, when the number of subjects was large, the Weibull and log-normal models had smaller bias and MSE than the log-logistic model. The exponential model, which is incorrect in Scenario A, exhibited some bias, but the MSE tended to be similar to the log-logistic model. Regarding the Mode and Median, the log-logistic model with frailty effects had the smallest values among the models.

In Scenario B, where the log-normal model with random effects is the true model, the log-normal model had a smaller bias and MSE than the Weibull and exponential models. The log-logistic model produced results similar to those in Scenario A, but the Weibull and exponential models had a systematic bias. Despite the log-normal being the true model, the log-logistic model showed a smaller Mode and Median.

\begin{table}[H]
  \begin{center}
\caption{Results of simulation in each scenario \label{t:sim}}
\begin{tabular}{|cccccccccccccc|}
\hline
\multicolumn{1}{|c|}{}       & \multicolumn{1}{c|}{}     & \multicolumn{3}{c|}{Exponential}           & \multicolumn{3}{c|}{Log-Logistic}          & \multicolumn{3}{c|}{Log-Normal}             & \multicolumn{3}{c|}{Weibull} \\
\multicolumn{1}{|c|}{}       & \multicolumn{1}{c|}{n}    & T     & F     & \multicolumn{1}{c|}{M}     & T     & F     & \multicolumn{1}{c|}{M}     & T     & F      & \multicolumn{1}{c|}{M}     & T        & F       & M       \\ \hline
\multicolumn{14}{|l|}{Scenario A: Log-Logistic model with random effect}                                                                                                                                                        \\ \hline
\multicolumn{1}{|c|}{Bias}   & \multicolumn{1}{c|}{64}   & 1.57  & 2.04  & \multicolumn{1}{c|}{1.46}  & 1.06  & 1.16  & \multicolumn{1}{c|}{1.24}  & 1.17  & 0.96   & \multicolumn{1}{c|}{1.35}  & 1.28     & 0.42    & 1.08    \\
\multicolumn{1}{|c|}{}       & \multicolumn{1}{c|}{512}  & 0.94  & 1.28  & \multicolumn{1}{c|}{0.85}  & 0.30  & -0.14 & \multicolumn{1}{c|}{0.35}  & 0.18  & 1.09   & \multicolumn{1}{c|}{0.23}  & 0.61     & 0.11    & 0.57    \\
\multicolumn{1}{|c|}{}       & \multicolumn{1}{c|}{2048} & 1.10  & 1.38  & \multicolumn{1}{c|}{0.98}  & 0.39  & -0.25 & \multicolumn{1}{c|}{0.44}  & 0.39  & 2.83   & \multicolumn{1}{c|}{0.40}  & 0.77     & 0.22    & 0.64    \\ \hline
\multicolumn{1}{|c|}{MSE}    & \multicolumn{1}{c|}{64}   & 30.81 & 24.66 & \multicolumn{1}{c|}{31.18} & 31.44 & 36.05 & \multicolumn{1}{c|}{31.13} & 32.13 & 38.69  & \multicolumn{1}{c|}{32.44} & 34.42    & 49.65   & 35.02   \\
\multicolumn{1}{|c|}{}       & \multicolumn{1}{c|}{512}  & 6.45  & 6.10  & \multicolumn{1}{c|}{6.21}  & 6.65  & 7.97  & \multicolumn{1}{c|}{5.95}  & 6.33  & 5.84   & \multicolumn{1}{c|}{5.66}  & 7.18     & 8.01    & 7.07    \\
\multicolumn{1}{|c|}{}       & \multicolumn{1}{c|}{2048} & 2.68  & 3.19  & \multicolumn{1}{c|}{2.35}  & 2.46  & 2.79  & \multicolumn{1}{c|}{2.40}  & 2.22  & 9.61   & \multicolumn{1}{c|}{2.21}  & 2.47     & 1.93    & 2.20    \\ \hline
\multicolumn{1}{|c|}{Mode}   & \multicolumn{1}{c|}{64}   & 1.77  & 3.07  & \multicolumn{1}{c|}{2.20}  & 1.44  & 2.02  & \multicolumn{1}{c|}{2.26}  & 1.10  & 1.84   & \multicolumn{1}{c|}{1.95}  & 1.47     & 1.84    & 2.10    \\
\multicolumn{1}{|c|}{}       & \multicolumn{1}{c|}{512}  & 0.93  & 1.89  & \multicolumn{1}{c|}{1.31}  & 0.27  & 0.04  & \multicolumn{1}{c|}{0.88}  & 0.15  & 1.68   & \multicolumn{1}{c|}{0.61}  & 0.69     & 0.84    & 1.03    \\
\multicolumn{1}{|c|}{}       & \multicolumn{1}{c|}{2048} & 1.14  & 1.62  & \multicolumn{1}{c|}{1.30}  & 0.35  & -0.17 & \multicolumn{1}{c|}{0.69}  & 0.39  & 3.26   & \multicolumn{1}{c|}{0.61}  & 0.79     & 0.70    & 0.97    \\ \hline
\multicolumn{1}{|c|}{Median} & \multicolumn{1}{c|}{64}   & 1.64  & 2.37  & \multicolumn{1}{c|}{1.73}  & 1.14  & 1.41  & \multicolumn{1}{c|}{1.54}  & 1.21  & 1.30   & \multicolumn{1}{c|}{1.57}  & 1.32     & 0.85    & 1.44    \\
\multicolumn{1}{|c|}{}       & \multicolumn{1}{c|}{512}  & 0.95  & 1.50  & \multicolumn{1}{c|}{1.03}  & 0.30  & -0.10 & \multicolumn{1}{c|}{0.54}  & 0.18  & 1.31   & \multicolumn{1}{c|}{0.38}  & 0.61     & 0.38    & 0.78    \\
\multicolumn{1}{|c|}{}       & \multicolumn{1}{c|}{2048} & 1.11  & 1.49  & \multicolumn{1}{c|}{1.12}  & 0.39  & -0.24 & \multicolumn{1}{c|}{0.53}  & 0.39  & 2.97   & \multicolumn{1}{c|}{0.48}  & 0.77     & 0.42    & 0.78    \\ \hline
\multicolumn{14}{|l|}{Scenario B: Log-Normal model with random effect}                                                                                                                                                          \\ \hline
\multicolumn{1}{|c|}{Bias}   & \multicolumn{1}{c|}{64}   & -1.38 & -1.42 & \multicolumn{1}{c|}{-0.91} & -0.29 & -0.61 & \multicolumn{1}{c|}{-0.45} & -0.49 & 0.02   & \multicolumn{1}{c|}{-0.42} & -1.28    & -0.19   & -0.73   \\
\multicolumn{1}{|c|}{}       & \multicolumn{1}{c|}{512}  & -1.48 & -1.74 & \multicolumn{1}{c|}{-1.30} & -0.52 & -0.37 & \multicolumn{1}{c|}{-0.77} & -0.88 & -0.84  & \multicolumn{1}{c|}{-0.86} & -1.44    & -0.87   & -1.23   \\
\multicolumn{1}{|c|}{}       & \multicolumn{1}{c|}{2048} & -1.19 & -1.33 & \multicolumn{1}{c|}{-0.91} & -0.13 & -0.13 & \multicolumn{1}{c|}{-0.30} & -0.52 & -0.61  & \multicolumn{1}{c|}{-0.47} & -1.12    & -0.53   & -0.82   \\ \hline
\multicolumn{1}{|c|}{MSE}    & \multicolumn{1}{c|}{64}   & 59.42 & 53.76 & \multicolumn{1}{c|}{51.03} & 38.26 & 53.88 & \multicolumn{1}{c|}{38.17} & 39.61 & 35.20  & \multicolumn{1}{c|}{37.46} & 58.34    & 47.91   & 51.58   \\
\multicolumn{1}{|c|}{}       & \multicolumn{1}{c|}{512}  & 10.56 & 11.41 & \multicolumn{1}{c|}{9.63}  & 6.27  & 6.71  & \multicolumn{1}{c|}{6.48}  & 6.46  & 6.47   & \multicolumn{1}{c|}{6.33}  & 10.50    & 8.70    & 9.95    \\
\multicolumn{1}{|c|}{}       & \multicolumn{1}{c|}{2048} & 4.75  & 4.40  & \multicolumn{1}{c|}{3.69}  & 2.33  & 2.78  & \multicolumn{1}{c|}{2.31}  & 2.61  & 3.10   & \multicolumn{1}{c|}{2.49}  & 4.66     & 3.24    & 3.64    \\ \hline
\multicolumn{1}{|c|}{Mode}   & \multicolumn{1}{c|}{64}   & -1.11 & -0.38 & \multicolumn{1}{c|}{0.00}  & 0.11  & 0.16  & \multicolumn{1}{c|}{0.53}  & -0.64 & 0.82   & \multicolumn{1}{c|}{0.40}  & -1.27    & 0.98    & -0.05   \\
\multicolumn{1}{|c|}{}       & \multicolumn{1}{c|}{512}  & -1.52 & -1.82 & \multicolumn{1}{c|}{-1.14} & -0.57 & -0.20 & \multicolumn{1}{c|}{-0.41} & -0.80 & -0.86  & \multicolumn{1}{c|}{-0.66} & -1.36    & -0.68   & -1.14   \\
\multicolumn{1}{|c|}{}       & \multicolumn{1}{c|}{2048} & -1.18 & -1.54 & \multicolumn{1}{c|}{-1.05} & -0.13 & -0.12 & \multicolumn{1}{c|}{-0.13} & -0.53 & -0.58  & \multicolumn{1}{c|}{-0.50} & -1.12    & -0.72   & -0.97   \\ \hline
\multicolumn{1}{|c|}{Median} & \multicolumn{1}{c|}{64}   & -1.28 & -1.17 & \multicolumn{1}{c|}{-0.68} & -0.22 & -0.45 & \multicolumn{1}{c|}{-0.18} & -0.48 & 0.26   & \multicolumn{1}{c|}{-0.20} & -1.18    & 0.23    & -0.45   \\
\multicolumn{1}{|c|}{}       & \multicolumn{1}{c|}{512}  & -1.47 & -1.76 & \multicolumn{1}{c|}{-1.30} & -0.52 & -0.34 & \multicolumn{1}{c|}{-0.65} & -0.88 & -0.82  & \multicolumn{1}{c|}{-0.79} & -1.42    & -0.88   & -1.21   \\
\multicolumn{1}{|c|}{}       & \multicolumn{1}{c|}{2048} & -1.19 & -1.46 & \multicolumn{1}{c|}{-0.99} & -0.13 & -0.12 & \multicolumn{1}{c|}{-0.23} & -0.52 & -0.62  & \multicolumn{1}{c|}{-0.47} & -1.11    & -0.66   & -0.89   \\ \hline
\end{tabular}
\footnotesize{T: typical fixed model, F: frailty model, R: mixed effects model with random effects, Mean, Mode, and Median are the difference from true RMST.}
  \end{center}
\end{table}

\section{Actual data re-analysis}
\label{sec:adr}
We analyzed survival data of children in eight states of the Empowered Action Group (EAG) in India, collected from 2019 to 2021 through the Demographic and Health Surveys (DHS) program. Our primary interest was the difference in RMSTs of children under five among different maternal age groups at birth ("12-19 years old", "20-30 years old", "31+ years old"). The restricted time was set at 50 months, where the mortality rate exceeded 80\%. Kaplan-Meier plots were presented in Supplemental Figures \ref{App:KM_2030} and \ref{App:KM_31} in the Supplemental material.

\begin{figure}[H]
  \begin{center}
  \includegraphics[width=15cm]{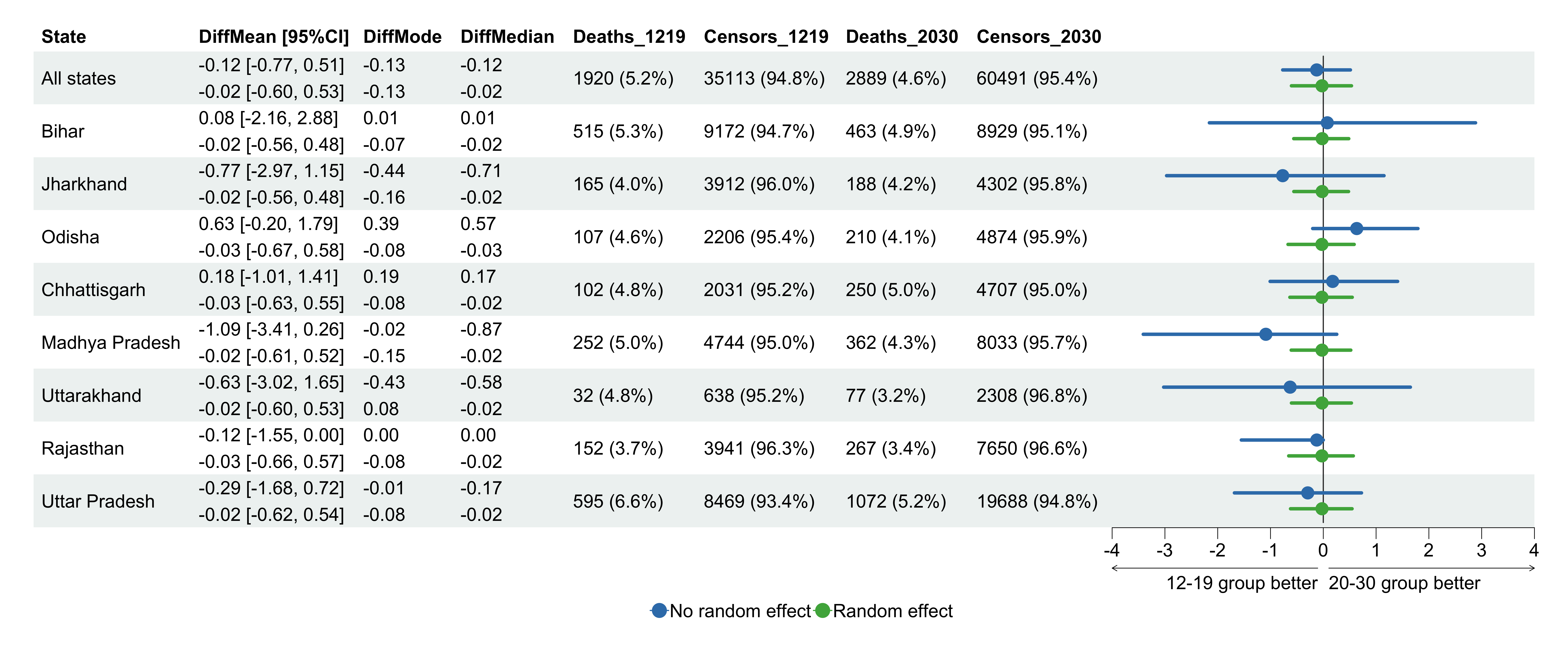}
  \caption{Forest plot of the difference in restricted mean survival times (12--19 group vs 20--30 group)}
  \label{Forest2030}
        \footnotesize{}
  \end{center}
\end{figure}

\begin{figure}[H]
  \begin{center}
  \includegraphics[width=15cm]{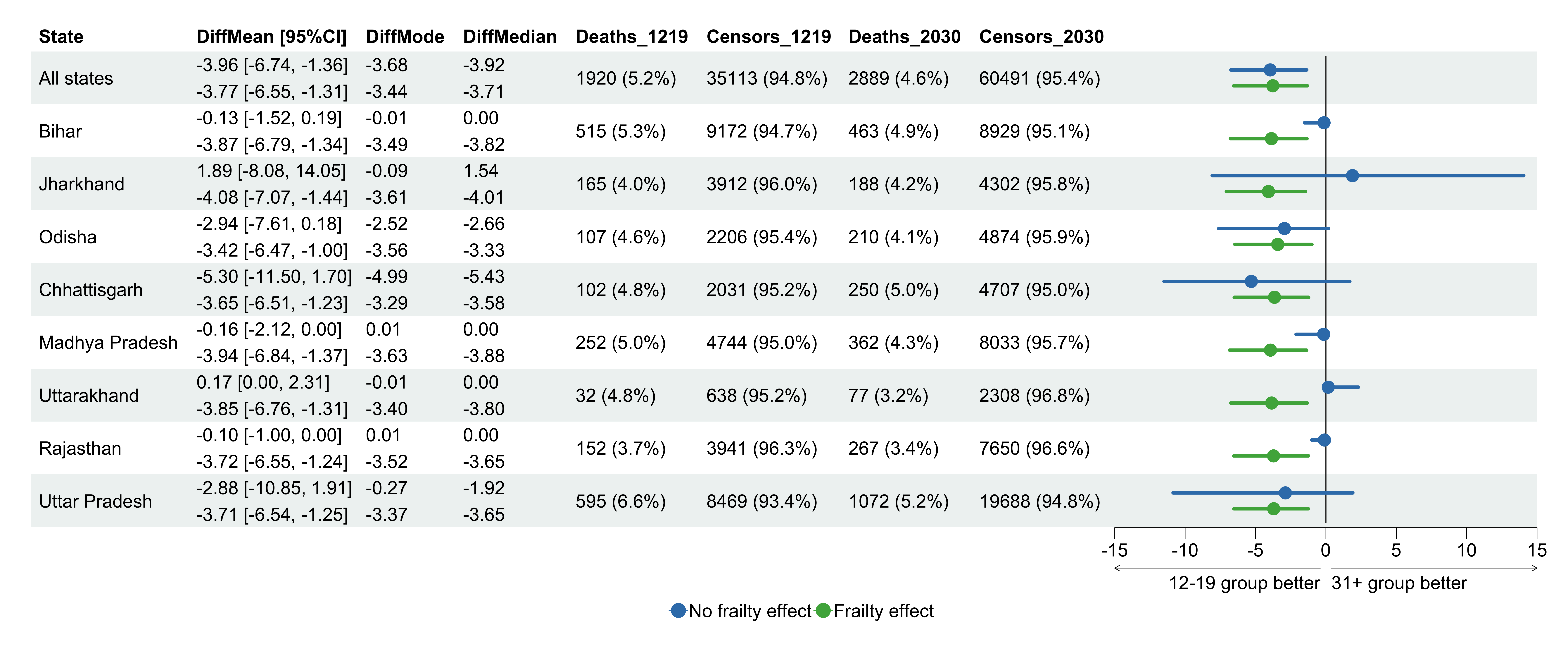}
  \caption{Forest plot of the difference in restricted mean survival times (12--19 group vs 31+ group)}
  \label{Forest31}
        \footnotesize{}
  \end{center}
\end{figure}

The covariates included sex, place of delivery (Respondent's home, Other home, Public sector, Government hospital, CS Govt health professional, Other public sector, Private hospital/clinic, CS private health facility, and Other), size of child at birth (Very large, Larger than average, Average, Smaller than average, Very small, and Don't know), and birth order. The clusters represented the eight states: Bihar, Jharkhand, Odisha, Chhattisgarh, Madhya Pradesh, Uttarakhand, Rajasthan, and Uttar Pradesh. The RMST differences between the two age groups across the eight states reveal that the trends varied by state (see Figures \ref{Forest2030} and \ref{Forest31}). We reanalyzed the data to determine if these differences were due to heterogeneity between the states or if they resulted from the sample size causing random variations in trends.

The MCMC specification included four chains, each with 2000 iterations and 1000 burn-ins. We used the brm functions from the \texttt{brms} package to analyze the exponential, Weibull, log-normal models, and mixed effects models. However, we developed custom Stan source code to analyze the exponential, Weibull, log-normal frailty models and all log-logistic models, as they could not be processed using the brm functions. These Stan files are included as supplemental material. Model selection was carried out using WAIC, with all WAIC results presented in Appendix \ref{App:DHS_WAIC} in the Supplemental material.

\begin{figure}[H]
  \begin{center}
  \includegraphics[width=15cm]{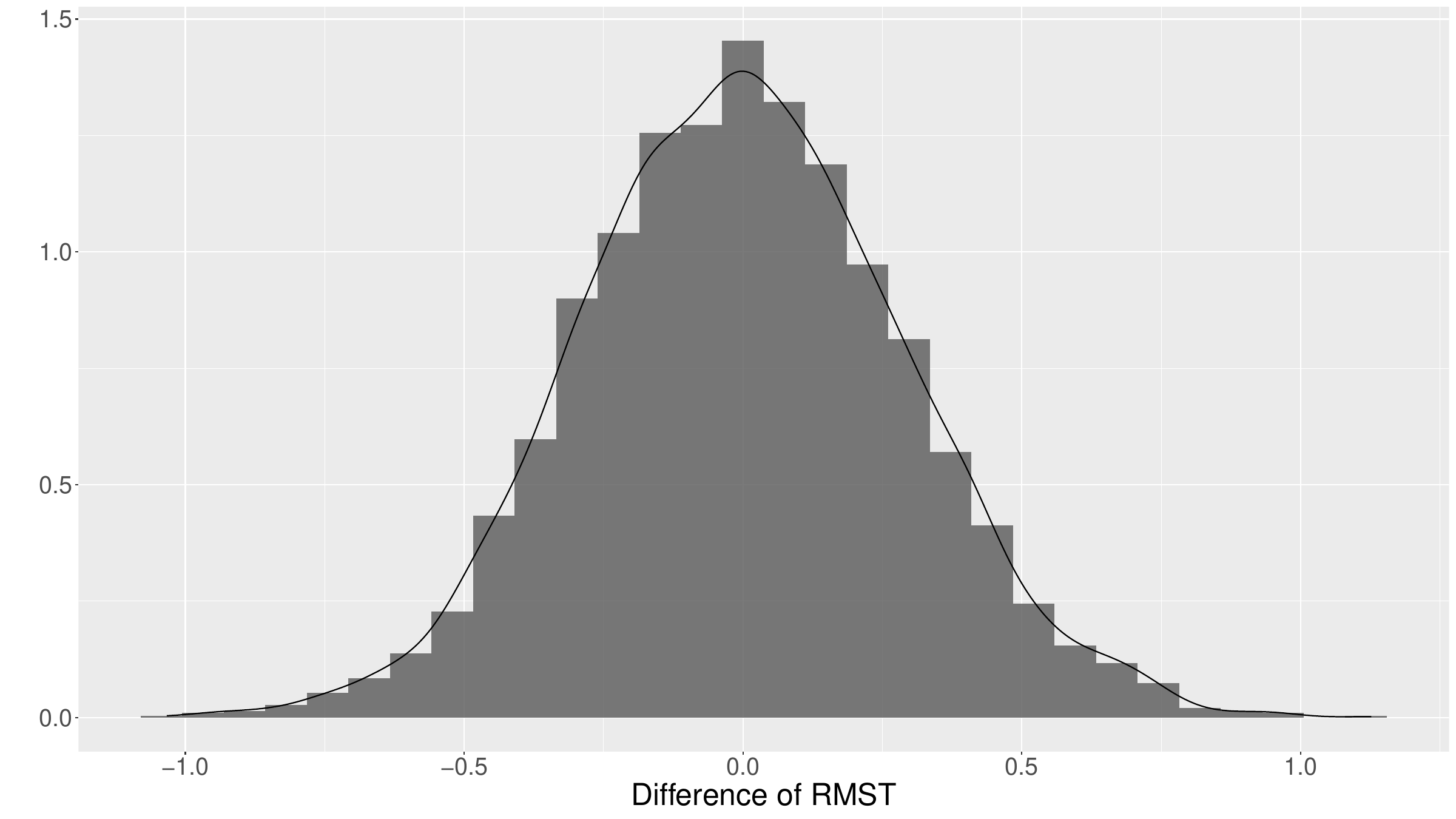}
  \caption{Histogram of the difference in restricted mean survival times (12--19 group vs 20--30 group)}
  \label{Hist2030}
  \footnotesize{}
  \end{center}
\end{figure}

For the subgroup ('12-19 years old' vs. '20-30 years old' groups), the WAIC for the mixed effects Weibull model was the lowest. Hence, the following discussion was based on the results of the mixed effects model. Other estimation results were summarized in Appendix \ref{App:Supp_DHS2030} in the Supplemental material. In the mixed-effects model, the RMST differences across states were similar due to the shrinkage effect, as the mean of the parameter $\phi$, representing the variance in the random effect distribution (indicating heterogeneity across clusters), was small, as shown in Figure \ref{Forest2030}. All estimation results were presented in Table \ref{t:DHS2030_W_n}, indicating that Place and Size affect the survival time of children under five. We obtained the RMST distribution, and its histogram is shown in Figure \ref{Hist2030}. From the RMST distribution, we calculated the probability of RMST under 0, -3, and -6 months. The probability of the RMST difference under 0 months was 0.507, while the probabilities under -3 and -6 months were both 0.000.

\begin{table}[H]
  \begin{center}
\caption{Results of mixed effects Weibull model (12--19 group vs 20--30 group)\label{t:DHS2030_W_n}}
\begin{tabular}{|c|c|c|c|c|c|c|c|c|}\hline
\multicolumn{2}{|c|}{Parameter} & Mode & Median & Mean & SE & 95\%CI & Rhat & ESS\\\hline
\multicolumn{2}{|c|}{Intercept} & 3.91 & 3.99 & 4.00 & 0.14 & [3.73, 4.28] & 1.01 & 622\\
\multicolumn{2}{|c|}{20--30 group} & 0.00 & 0.00 & 0.00 & 0.02 & [-0.04, 0.03] & 1.00 & 2658\\
\multicolumn{2}{|c|}{Sex} & 0.07 & 0.08 & 0.08 & 0.02 & [0.04, 0.11] & 1.00 & 2274\\
Place & Respondents home & 0.62 & 0.54 & 0.54 & 0.14 & [0.26, 0.79] & 1.01 & 551\\
Place & Other home & 0.51 & 0.58 & 0.60 & 0.27 & [0.10, 1.17] & 1.00 & 1204\\
Place & Public sector & 0.65 & 0.68 & 0.68 & 0.15 & [0.36, 0.95] & 1.00 & 646\\
Place & Government hospital & 0.57 & 0.58 & 0.58 & 0.14 & [0.30, 0.83] & 1.00 & 551\\
Place & CS Govt health professional & 0.63 & 0.57 & 0.57 & 0.13 & [0.29, 0.81] & 1.01 & 547\\
Place & Other public sector & 0.58 & 0.63 & 0.62 & 0.15 & [0.32, 0.89] & 1.00 & 602\\
Place & Private hospital/clinic & 0.45 & 0.41 & 0.40 & 0.14 & [0.12, 0.65] & 1.01 & 563\\
Place & CS private health facility & 0.52 & 0.62 & 0.62 & 0.17 & [0.29, 0.96] & 1.00 & 681\\
Size & Very large & 0.67 & 0.67 & 0.67 & 0.05 & [0.58, 0.76] & 1.01 & 974\\
Size & Larger than average & 0.73 & 0.71 & 0.70 & 0.05 & [0.61, 0.79] & 1.01 & 926\\
Size & Average & 0.78 & 0.77 & 0.76 & 0.04 & [0.69, 0.84] & 1.01 & 734\\
Size & Smaller than average & 0.54 & 0.54 & 0.53 & 0.04 & [0.45, 0.62] & 1.00 & 963\\
Size & Very small & 0.10 & 0.08 & 0.08 & 0.05 & [-0.01, 0.17] & 1.00 & 991\\
\multicolumn{2}{|c|}{Order} & -0.01 & -0.01 & -0.01 & 0.01 & [-0.02, 0.00] & 1.00 & 2864\\
\multicolumn{2}{|c|}{$k$} & 1.71 & 1.71 & 1.71 & 0.02 & [1.67, 1.75] & 1.00 & 1979\\
\multicolumn{2}{|c|}{$\phi$} & 0.08 & 0.10 & 0.11 & 0.04 & [0.05, 0.22] & 1.00 & 483\\
\multicolumn{2}{|c|}{RMST$_{12-19 group}$} & 37.99 & 37.35 & 37.33 & 2.27 & [32.73, 41.66] &  & \\
\multicolumn{2}{|c|}{RMST$_{20-30 group}$} & 37.14 & 37.34 & 37.31 & 2.28 & [32.78, 41.56] &  & \\
\multicolumn{2}{|c|}{RMST$_{diff}$} & -0.13 & -0.02 & -0.02 & 0.29 & [-0.60, 0.53] &  & \\\hline
\end{tabular}
  \footnotesize{95\%CI: 95\% credible interval, ESS: Effective sample size}
  \end{center}
\end{table}

\begin{figure}[H]
  \begin{center}
  \includegraphics[width=15cm]{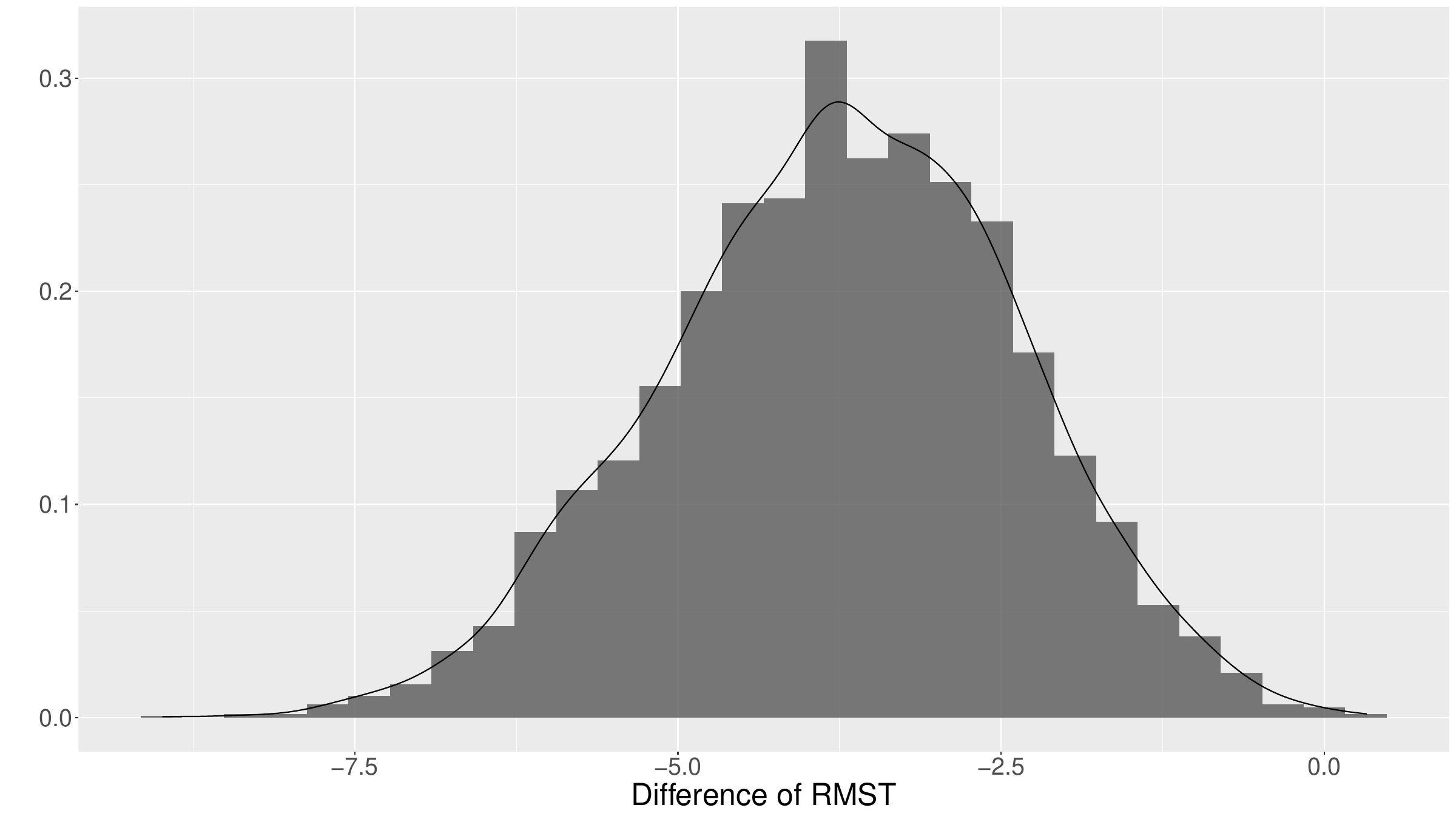}
  \caption{Histogram of the difference in restricted mean survival times (12--19 group vs 31+ group)}
  \label{Hist31}
        \footnotesize{}
  \end{center}
\end{figure}

For the subgroup ('12-19 years old' vs. '31+ years old' groups), the WAIC for the Weibull frailty model was the lowest. Hence, the following discussion is based on the results of the frailty model. Other estimation results were summarized in Appendix \ref{App:Supp_DHS31} in the Supplemental material. In the frailty model, the RMST differences across states were similar due to the shrinkage effect, as the mean of the parameter $\phi$, representing the variance in the gamma frailty distribution (indicating heterogeneity across clusters), was small, as shown in Figure \ref{Forest31}. All estimation results were presented in Table \ref{t:DHS31_W_f}, which implied that Place and Size affect the survival time of children under five. We obtained the RMST distribution, and its histogram is shown in Figure \ref{Hist31}. From the RMST distribution, we calculated the probability of RMST under 0, -3, and -6 months. The probability of the RMST difference under 0 months was 0.999, the probability under -3 months was 0.69375, and the probability under -6 months was 0.057.

\begin{table}[H]
  \begin{center}
\caption{Results of Weibull frailty model (12--19 group vs 31+ group)\label{t:DHS31_W_f}}
\begin{tabular}{|c|c|c|c|c|c|c|c|c|}\hline
\multicolumn{2}{|c|}{Parameter} & Mode & Median & Mean & SE & 95\%CI & Rhat & ESS\\\hline
\multicolumn{2}{|c|}{Intercept} & 3.98 & 4.02 & 4.03 & 0.19 & [3.68, 4.42] & 1.00 & 978\\
\multicolumn{2}{|c|}{31+ group} & -0.22 & -0.21 & -0.21 & 0.07 & [-0.34, -0.08] & 1.00 & 5178\\
\multicolumn{2}{|c|}{Sex} & 0.05 & 0.05 & 0.05 & 0.03 & [0.00, 0.10] & 1.00 & 4795\\
Place & Respondents home & 0.55 & 0.49 & 0.48 & 0.18 & [0.12, 0.82] & 1.00 & 1008\\
Place & Other home & 0.45 & 0.35 & 0.37 & 0.34 & [-0.24, 1.10] & 1.00 & 2087\\
Place & Public sector & 0.80 & 0.77 & 0.77 & 0.20 & [0.37, 1.17] & 1.00 & 1173\\
Place & Government hospital & 0.55 & 0.52 & 0.52 & 0.18 & [0.16, 0.86] & 1.00 & 1022\\
Place & CS Govt health professional & 0.58 & 0.55 & 0.54 & 0.18 & [0.19, 0.87] & 1.00 & 999\\
Place & Other public sector & 0.70 & 0.69 & 0.68 & 0.22 & [0.25, 1.11] & 1.00 & 1341\\
Place & Private hospital/clinic & 0.37 & 0.33 & 0.32 & 0.18 & [-0.04, 0.66] & 1.00 & 1024\\
Place & CS private health facility & 0.81 & 0.86 & 0.87 & 0.28 & [0.34, 1.42] & 1.00 & 1824\\
Size & Very large & 0.45 & 0.44 & 0.44 & 0.07 & [0.30, 0.57] & 1.00 & 2108\\
Size & Larger than average & 0.57 & 0.56 & 0.56 & 0.07 & [0.43, 0.69] & 1.00 & 2055\\
Size & Average & 0.63 & 0.63 & 0.63 & 0.06 & [0.52, 0.73] & 1.00 & 1754\\
Size & Smaller than average & 0.49 & 0.48 & 0.49 & 0.07 & [0.36, 0.61] & 1.00 & 2070\\
Size & Very small & 0.03 & 0.02 & 0.02 & 0.07 & [-0.11, 0.15] & 1.00 & 2183\\
\multicolumn{2}{|c|}{Order} & 0.00 & 0.00 & 0.00 & 0.01 & [-0.02, 0.01] & 1.00 & 5814\\
\multicolumn{2}{|c|}{$k$} & 1.82 & 1.83 & 1.83 & 0.04 & [1.76, 1.90] & 1.00 & 3477\\
\multicolumn{2}{|c|}{$\phi$} & 0.02 & 0.04 & 0.06 & 0.06 & [0.01, 0.22] & 1.00 & 1634\\
\multicolumn{2}{|c|}{RMST$_{12-19 group}$} & 38.70 & 38.54 & 38.38 & 3.03 & [31.98, 43.70] &  & \\
\multicolumn{2}{|c|}{RMST$_{31+ group}$} & 34.53 & 34.78 & 34.61 & 3.73 & [27.03, 41.36] &  & \\
\multicolumn{2}{|c|}{RMST$_{diff}$} & -3.44 & -3.71 & -3.77 & 1.36 & [-6.55, -1.31] &  & \\\hline
\end{tabular}
  \footnotesize{95\%CI: 95\% credible interval, ESS: Effective sample size}
  \end{center}
\end{table}

\section{Discussion}
In this paper, we proposed a Bayesian method for determining the distribution of RMST using posterior samples. We showed that the estimated RMST converged to the true value as the sufficient number of subjects when the estimated Bayesian parameters were consistent. We derived an explicit RMST equation by devising an integral of the survival function to obtain the distribution of RMST. Our proposed method allows to derive not only the mean and credible interval but also the mode, median, and probability of exceeding a certain value. In addition, we proposed two methods: one using random effects to account for heterogeneity among clusters, and the other using frailty. Because the frailty model cannot be analyzed with the existing \texttt{brms} package in R, we created a custom Stan file; see the supplemental material for the Stan files. Similarly, a log-logistic model could not be analyzed with \texttt{brms}, so a custom Stan code was created; see the supplemental material for the Stan files.

A limitation of our study is the necessity to assume a parametric model for survival time to obtain posterior samples. However, because there are several parametric models and many methods of model selection, selecting the model most appropriate for the data helps avoid incorrect analysis results. Additionally, a more flexible model can be built using the spline function proposed by Zhong and Schaubel (2022)~\cite{zhong2022restricted}. The RMST for each period between knots can be obtained using the formulas we have used, and the overall RMST can be derived by summing the RMSTs.

The simulation study confirms that the bias decreases, and the MSE becomes smaller as the sample size increases, as theory suggests. In Scenario A, the log-normal frailty model did not perform well due to a poor fit to the model, but in Scenario B, it performed adequately. The simulation results showed a small bias and MSE for models close enough to the true model. The importance of selecting an appropriate model by model selection for actual data analysis was suggested.

In the actual data analysis, we adjusted for differences between clusters by considering heterogeneity and obtained consistent results for each state. When using the exponential distribution from Scenario C, the estimation results were unstable, as with the log-normal frailty model in the real data analysis. The log-normal frailty model was found to increase frailty when the fit to the model was poor. A sensitivity analysis in Appendix \ref{App:LeukSurv} in the Supplemental material was performed to check the behavior of the log-normal frailty estimation and the behavior of the WAIC for log-logistic frailty, and no particular problems were found. The simulations for log-normal frailty showed that the estimation accuracy may be poor if the data were not generated from a log-normal base.

In conclusion, we derived explicit RMST formulas for the parametric model and the distribution of RMST. This allows us to calculate the mean, median, mode, and credible interval. We created the Stan code for the frailty and log-logistic models. Simulations confirmed that there were no performance problems, and we used the shrinkage effect to derive more accurate results for each cluster.
\bibliography{main.bib}
\bibliographystyle{unsrt}

\end{document}


\title{{\it Supplemental material of Bayesian Parametric Methods for Deriving Distribution of Restricted Mean Survival Time}}
\author{}
\date{}
\maketitle
\appendix
\section{Appendix}\label{App}
\subsection{Derivation of RMST}
\subsubsection{Exponential distribution}\label{App:exp_RMST}
For the exponential distribution $Exp(\lambda)$, we define the density function, survival function, and hazard function.
\begin{align}
f(t)=\lambda e^{-\lambda t}, S(t)=e^{-\lambda t}, h(t)=\lambda.\nonumber
\end{align}
The RMST can be calculated as follows:
\begin{align}
\int^\tau_0 S(t)dt&=\int^\tau_0 tf(t)dt+\tau S(\tau)\nonumber\\
&=\left[-te^{-\lambda t}\right]^t_0+\int^\tau_0 e^{-\lambda t}dt+\tau e^{-\lambda \tau}\nonumber\\
&=-\tau e^{-\lambda \tau}+\left[-\frac{e^{-\lambda t}}{\lambda}\right]^\tau_0+\tau e^{-\lambda \tau}\nonumber\\
&=\frac{1-e^{-\lambda \tau}}{\lambda}.\nonumber
\end{align}
In the exponential distribution, the survival function can be directly integrated as $\int^\tau_0 S(t)dt=\int^\tau_0 e^{-\lambda t}dt=\frac{1-e^{-\lambda t}}{\lambda}$.

\subsubsection{Weibull distribution}\label{App:Wei_RMST}
For the weibull distribution $W(\lambda,k)$ where the scale parameter $\lambda>0$ and the shape parameter $k>0$, we define the density function, survival function, and hazard function.
\begin{align}
f(t)=\lambda k t^{k-1}\exp(-\lambda t^k), S(t)=\exp(-\lambda t^k), h(t)=\lambda k t^{k-1}.\nonumber
\end{align}
The RMST is 
\begin{align}
\int^\tau_0 S(t)dt&=\int^\tau_0 t\lambda k t^{k-1}\exp(-\lambda t^k)dt+\tau \exp(-\lambda \tau^k)\nonumber\\
&(\lambda t^k=x\Leftrightarrow t=\lambda^{-\frac{1}{k}}x^{\frac{1}{k}}, \lambda k t^{k-1}dt=dx)\nonumber\\
&=\int^{\lambda\tau^k}_0 \lambda^{-\frac{1}{k}}x^{\frac{1}{k}}\exp(-x)dx+\tau \exp(-\lambda \tau^k)\nonumber\\
&=\lambda^{-\frac{1}{k}}\gamma\left(\lambda\tau^k;\frac{1}{k}+1\right)+\tau \exp(-\lambda \tau^k),\nonumber
\end{align}
where $\gamma(z;a)=\int^z_0 t^{a-1}e^{-t}dt$ is the incomplete gamma function.

\subsubsection{Log-logistic distribution}\label{App:LL_RMST}
For the log-logistic distribution $LL(\mu,k)$ where the parameters $\mu\in\R$, $k>0$, we define the density function, survival function, and hazard function.
\begin{align}
f(t)=\frac{e^\mu k t^{k-1}}{(1+e^\mu t^k)^2}, S(t)=\frac{1}{1+e^\mu t^k}, h(t)=\frac{e^\mu k t^{k-1}}{1+e^\mu t^k}\nonumber
\end{align}
The RMST is calculated as follows:
\begin{align}
\int^\tau_0 S(t)dt&=\int^\tau_0 t\frac{e^\mu k t^{k-1}}{(1+e^\mu t^k)^2}dt+\tau\frac{1}{1+e^\mu \tau^k}\nonumber\\
&(e^\mu t^k=x\Leftrightarrow t=e^{-\frac{\mu}{k}}x^\frac{1}{k}, dt=\frac{1}{k}e^{-\frac{\mu}{k}}x^{\frac{1}{k}-1}dx)\nonumber\\
&=\int^{e^\mu\tau^k}_0 \frac{e^{-\frac{\mu}{k}}x^\frac{1}{k}}{(1+x)^2}dx+\tau\frac{1}{1+e^\mu \tau^k}\nonumber\\
&=e^{-\frac{\mu}{k}}B\left(\frac{e^\mu\tau^k}{1+e^\mu\tau^k};1+\frac{1}{k},1-\frac{1}{k}\right)+\tau\frac{1}{1+e^\mu \tau^k},\label{eq:LL_RMST_last}
\end{align}
where $B(z;a,b)=\int^{z}_0 t^{a-1}(1-t)^{b-1}dt$ is the incomplete beta function. The following formula was used to derive the incomplete beta function in the equation \ref{eq:LL_RMST_last}.
\begin{align}
&\int^z_0\frac{x^{a-1}}{(1+x)^{a+b}}dx\nonumber\\
&\left(\frac{x}{1+x}=y\Leftrightarrow x=\frac{y}{1-y}, dx=\frac{1}{(1-y)^2}dy \right)\nonumber\\
=&\int^{\frac{z}{1+z}}_0 y^{a-1}(1-y)^{b-1}\nonumber\\
=&B\left(\frac{z}{1+z};a,b\right).\nonumber
\end{align}

\subsubsection{Log-normal distribution}\label{App:LN_RMST}
For the log-normal distribution, $LN(\mu,\sigma^2)$ where the parameters $\mu\in\R$, $\sigma^2>0$, we define the density function, survival function, and hazard function.
\begin{align}
&f(t)=\frac{1}{t\sqrt{2\pi\sigma^2}}\exp\left\{-\frac{(\log(t)-\mu)^2}{2\sigma^2}\right\}, S(t)=1-\Phi\left(\frac{\log(t)-\mu}{\sigma}\right), \nonumber\\
&h(t)=\frac{\frac{1}{t\sqrt{2\pi\sigma^2}}\exp\left\{-\frac{(\log(t)-\mu)^2}{2\sigma^2}\right\}}{1-\Phi\left(\frac{\log(t)-\mu}{\sigma}\right)}.\nonumber
\end{align}
The RMST is 
\begin{align}
\int^\tau_0 S(t)dt&=\int^\tau_0 t\frac{1}{t\sqrt{2\pi\sigma^2}}\exp\left\{-\frac{(\log(t)-\mu)^2}{2\sigma^2}\right\}dt+\tau\left(1-\Phi\left(\frac{\log(\tau)-\mu}{\sigma}\right)\right)\nonumber\\
&(\log(t)=x\sim\Nc(\mu,\si^2), dt=e^xdx)\nonumber\\
&=\int^{\log(\tau)}_{-\infty} \frac{1}{\sqrt{2\pi\sigma^2}}\exp\left\{-\frac{(x-\mu)^2}{2\sigma^2}\right\}\exp\{x\}dt+\tau\left(1-\Phi\left(\frac{\log(\tau)-\mu}{\sigma}\right)\right)\nonumber\\
&=\exp\left\{\mu+\frac{\si^2}{2}\right\}\int^{\log(\tau)}_{-\infty} \frac{1}{\sqrt{2\pi\sigma^2}}\exp\left\{-\frac{(x-\mu-\si^2)^2}{2\sigma^2}\right\}dt+\tau\left(1-\Phi\left(\frac{\log(\tau)-\mu}{\sigma}\right)\right)\nonumber\\
&\left(\frac{x-\mu-\si^2}{\si}=y\sim\Nc(-\si,1), dx=\si dy\right)\nonumber\\
&=\exp\left\{\mu+\frac{\si^2}{2}\right\}\int^{\frac{\log(\tau)-\mu-\si^2}{\si}}_{-\infty} \frac{1}{\sqrt{2\pi}}\exp\left\{-\frac{y^2}{2}\right\}dt+\tau\left(1-\Phi\left(\frac{\log(\tau)-\mu}{\sigma}\right)\right)\nonumber\\
&=\exp\left\{\mu+\frac{\si^2}{2}\right\}\Phi\left(\frac{\log(\tau)-\mu-\si^2}{\si}\right)+\tau\left(1-\Phi\left(\frac{\log(\tau)-\mu}{\sigma}\right)\right).\nonumber
\end{align}

\subsection{Distribution of RMST for each cluster in mixed effect models}\label{App:RMST_rand}
We demonstrate the computation of the RMST for each cluster, considering the heterogeneity via random effects.
\medskip
\\
{\bf [Mixed effects exponential model]}
The distribution of the RMST for the $i$-th cluster in each group is obtained below.
\begin{align}
RMST_{E,r}(\tau,x_1,\be_0^\ast,\be_1^\ast,u_i^\ast)=\frac{1-e^{-\exp\{\be_0^\ast+x_1\be_1^\ast+u_i^\ast\} \tau}}{\exp\{\be_0^\ast+x_1\be_1^\ast+u_i^\ast\}},\nonumber
\end{align}
where the $\ast$ denotes the posterior sample, and $x_1$ is 0 for the control group and 1 for the treatment group. The difference between the RMSTs is
\begin{align}
RMST_{E,r}(\tau,1,\be_0^\ast,\be_1^\ast,u_i^\ast)-RMST_{E,r}(\tau,0,\be_0^\ast,\be_1^\ast,u_i^\ast).\nonumber
\end{align}
If the model includes covariates, $\be_0^\ast$ and $\be_1^\ast$ are adjusted for the covariates.
\medskip
\\
{\bf [Mixed effects Weibull model]}
The distribution of the RMST for the $i$-th cluster in each group is obtained below.
\begin{align}
RMST_{W,r}(\tau,x_1,\be_0^\ast,\be_1^\ast,u_i^\ast)=&\exp\left\{-\frac{\be_0^\ast+x_1\be_1^\ast+u_i^\ast}{k^\ast}\right\}\gamma\left(\exp\{\be_0^\ast+x_1\be_1^\ast+u_i^\ast\}\tau^{k^\ast};\frac{1}{k^\ast}+1\right)\nonumber\\
&\hspace{1cm}+\tau \exp\left(-\exp\{\be_0^\ast+x_1\be_1^\ast+u_i^\ast\} \tau^{k^\ast}\right).\nonumber
\end{align}
 The difference between the RMSTs is
\begin{align}
RMST_{W,r}(\tau,1,\be_0^\ast,\be_1^\ast,u_i^\ast)-RMST_{W,r}(\tau,0,\be_0^\ast,\be_1^\ast,u_i^\ast).\nonumber
\end{align}
\medskip
\\
{\bf [Mixed effects log-logistic distribution]}
The distribution of the RMST for the $i$-th cluster in each group is obtained below.
\begin{align}
RMST_{LL,r}(\tau,x_1,\be_0^\ast,\be_1^\ast,u_i^\ast)=&e^{-\frac{\be_0^\ast+x_1\be_1^\ast+u_i^\ast}{k^\ast}}B\left(\frac{e^{\be_0^\ast+x_1\be_1^\ast+u_i^\ast}\tau^{k^\ast}}{1+e^{\be_0^\ast+x_1\be_1^\ast+u_i^\ast}\tau^{k^\ast}};1+\frac{1}{k^\ast},1-\frac{1}{k^\ast}\right)\nonumber\\
&\hspace{1cm}+\tau\frac{1}{1+e^{\be_0^\ast+x_1\be_1^\ast+u_i^\ast}\tau^{k^\ast}}.\nonumber
\end{align}
The difference between the RMSTs is
\begin{align}
RMST_{LL,r}(\tau,1,\be_0^\ast,\be_1^\ast,u_i^\ast)-RMST_{LL,r}(\tau,0,\be_0^\ast,\be_1^\ast,u_i^\ast).\nonumber
\end{align}
\medskip
\\
{\bf [Mixed effects log-normal distribution]}
The distribution of the RMST for the $i$-th cluster in each group is obtained below.
\begin{align}
RMST_{LN,r}(\tau,x_1,\be_0^\ast,\be_1^\ast,u_i^\ast)&=\exp\left\{\be_0^\ast+x_1\be_1^\ast+u_i^\ast+\frac{\si^{2\ast}}{2}\right\}\Phi\left(\frac{\log(\tau)-(\be_0^\ast+x_1\be_1^\ast+u_i^\ast-\si^{2\ast})}{\si^\ast}\right)\nonumber\\
&\hspace{1.5cm}+\tau\left(1-\Phi\left(\frac{\log(\tau)-(\be_0^\ast+x_1\be_1^\ast+u_i^\ast)}{\si^\ast}\right)\right).\nonumber
\end{align}
The difference between the RMSTs is
\begin{align}
RMST_{LN,r}(\tau,1,\be_0^\ast,\be_1^\ast,u_i^\ast)-RMST_{LN,r}(\tau,0,\be_0^\ast,\be_1^\ast,u_i^\ast).\nonumber
\end{align}

\subsubsection{Derivations of RMST for mixed effects models}\label{App:RMST_random_calc}
We consider adding random effects to the rate parameters or the scale parameters.
\medskip
\\
{\bf [Mixed effects exponential model]} The density and survival functions are conditioned on a random effect $u\sim\Nc(0,\phi)$, denoted as 
\begin{align}
f(t|u)=\lambda e^u e^{-\lambda e^u t}, S(t)=e^{-\lambda e^u t}.\nonumber
\end{align}
The RMST for the mixed effects exponential model is
\begin{align}
\int^\tau_0 S(t|u)dt=\frac{1-e^{-\lambda e^u \tau}}{\lambda e^u}.\nonumber
\end{align}
The mixed effects exponential model is equivalent to the exponential model with the log-normal frailty.
\medskip
\\
{\bf [Mixed effects Weibull model]} The density and survival functions are conditioned on a random effect $u\sim\Nc(0,\phi)$, denoted as 
\begin{align}
f(t|u)=\lambda e^u k t^{k-1}\exp(-\lambda e^u t^k), S(t)=\exp(-\lambda e^u t^k).\nonumber
\end{align}
The RMST is
\begin{align}
\int^\tau_0 S(t|u)dt=\lambda^{-\frac{1}{k}}e^{-\frac{u}{k}}\gamma\left(\lambda e^u\tau^k;\frac{1}{k}+1\right)+\tau \exp(-\lambda e^u\tau^k),\nonumber
\end{align}
where $\gamma(z;a)=\int^z_0 t^{a-1}e^{-t}dt$ is an incomplete gamma function. The mixed effects Weibull model is equivalent to the Weibull model with the log-normal frailty.
\medskip
\\
{\bf [Mixed effects log-logistic model]} The density and survival functions are conditioned on a random effect $u\sim\Nc(0,\phi)$, denoted as
\begin{align}
f(t|u)=\frac{e^{\mu+u} k t^{k-1}}{(1+e^{\mu+u} t^k)^2}, S(t|u)=\frac{1}{1+e^{\mu+u}t^k}.\nonumber
\end{align}
The RMST is
\begin{align}
\int^\tau_0 S(t|u)dt=e^{-\frac{\mu+u}{k}}B\left(\frac{e^{\mu+u}\tau^k}{1+e^{\mu+u}\tau^k};1+\frac{1}{k},1-\frac{1}{k}\right)+\tau\frac{1}{1+e^{\mu+u} \tau^k},\nonumber
\end{align}
where $B(z;a,b)$ is an incomplete beta function.
\medskip
\\
{\bf [Mixed effects log-normal model]} The density and survival functions are conditioned on a random effect $u\sim\Nc(0,\phi)$, denoted as
\begin{align}
f(t|u)=\frac{1}{t\sqrt{2\pi\si^2}}\exp\left\{-\frac{(\log(t)-\mu-u)^2}{2\si^2}\right\}, S(t|u)=1-\Phi\left(\frac{\log(t)-\mu-u}{\si}\right).\nonumber
\end{align}
The RMST is
\begin{align}
\int^\tau_0 S(t|u)dt=\exp\left\{\mu+\frac{\si^2}{2}\right\}\Phi\left(\frac{\log(\tau)-\mu-u-\si^2}{\si}\right)+\tau\left(1-\Phi\left(\frac{\log(\tau)-\mu-u}{\sigma}\right)\right).\nonumber
\end{align}

\subsection{Distribution of RMST for frailty models}\label{App:RMST_frail}
The derivations of the RMSTs for the frailty models are shown in the Appendix \ref{App:RMST_frail_calc}.
\medskip
\\
{\bf [Exponential frailty model]} The distribution of the RMST with frailty term for the $i$-th cluster in each group is obtained below.
\begin{align}
RMST_{E,f}(\tau,x_1,\be_0^\ast,\be_1^\ast,v_i^\ast)=\frac{1-e^{-v_i^\ast\exp\{\be_0^\ast+x_1\be_1^\ast\} \tau}}{v_i^\ast\exp\{\be_0^\ast+x_1\be_1^\ast\}}.\nonumber
\end{align}
The difference between the RMSTs is
\begin{align}
RMST_{E,f}(\tau,1,\be_0^\ast,\be_1^\ast,v_i^\ast)-RMST_{E,f}(\tau,0,\be_0^\ast,\be_1^\ast,v_i^\ast).\nonumber
\end{align}
\medskip
\\
{\bf [Weibull frailty model]} The distribution of the RMST with frailty term for the $i$-th cluster in each group is obtained below.
\begin{align}
RMST_{W,f}(\tau,x_1,\be_0^\ast,\be_1^\ast,v_i^\ast)=&(v_i^\ast)^{-\frac{1}{k}}\exp\left\{-\frac{\be_0^\ast+x_1\be_1^\ast}{k^\ast}\right\}\gamma\left(v_i^\ast\exp\{\be_0^\ast+x_1\be_1^\ast\}\tau^{k^\ast};\frac{1}{k^\ast}+1\right)\nonumber\\
&\hspace{1cm}+\tau \exp\left(-v_i^\ast\exp\{\be_0^\ast+x_1\be_1^\ast\} \tau^{k^\ast}\right).\nonumber
\end{align}
The difference between the RMSTs is
\begin{align}
RMST_{W,f}(\tau,1,\be_0^\ast,\be_1^\ast,v_i^\ast)-RMST_{W,f}(\tau,0,\be_0^\ast,\be_1^\ast,v_i^\ast).\nonumber
\end{align}
\medskip
\\
{\bf [Log-logistic frailty model]} The distribution of the RMST with frailty term for the $i$-th cluster in each group is obtained below.
\begin{align}
RMST_{LL,f}(\tau,x_1,\be_0^\ast,\be_1^\ast,v_i^\ast)=&v_i^\ast e^{-\frac{\be_0^\ast+x_1\be_1^\ast}{k^\ast}}B\left(\frac{e^{\be_0^\ast+x_1\be_1^\ast}\tau^{k^\ast}}{1+e^{\be_0^\ast+x_1\be_1^\ast}\tau^{k^\ast}};1+\frac{1}{k^\ast},v_i^\ast-\frac{1}{k^\ast}\right)\nonumber\\
&\hspace{1cm}+\tau\left(\frac{1}{1+e^{\be_0^\ast+x_1\be_1^\ast}\tau^{k^\ast}}\right)^{v_i^\ast}.\nonumber
\end{align}
The difference between the RMSTs is
\begin{align}
RMST_{LL,f}(\tau,1,\be_0^\ast,\be_1^\ast,v_i^\ast)-RMST_{LL,f}(\tau,0,\be_0^\ast,\be_1^\ast,v_i^\ast).\nonumber
\end{align}
\medskip
\\
{\bf [Log-normal frailty model]} The distribution of the RMST with frailty term for the $i$-th cluster in each group is obtained below.
\begin{align}
RMST_{LN,f}(\tau,x_1,\be_0^\ast,\be_1^\ast,v_i^\ast)&=\exp\left\{\be_0^\ast+x_1\be_1^\ast+\frac{\si^{2\ast}}{2}\right\}\frac{1}{v_i^\ast}\nonumber\\
&\hspace{1cm}\times\left(1-\left(1-\Phi\left(\frac{\log(\tau)-(\be_0^\ast+x_1\be_1^\ast-\si^{2\ast})}{\si^\ast}\right)\right)^{v_i^\ast}\right)\nonumber\\
&\hspace{1.5cm}+\tau\left(1-\Phi\left(\frac{\log(\tau)-(\be_0^\ast+x_1\be_1^\ast)}{\si^\ast}\right)\right)^{v_i^\ast}.\nonumber
\end{align}
The difference between the RMSTs is
\begin{align}
RMST_{LN,f}(\tau,1,\be_0^\ast,\be_1^\ast,v_i^\ast)-RMST_{LN,f}(\tau,0,\be_0^\ast,\be_1^\ast,v_i^\ast).\nonumber
\end{align}

\subsubsection{Derivations of RMST for frailty models}\label{App:RMST_frail_calc}
We derive the RMST using frailty models before incorporating covariates.
\medskip
\\
{\bf [Exponential frailty model]} The RMST with frailty term is
\begin{align}
\int^\tau_0 S_f(t|v)dt&=\int^\tau_0 tf_f(t|v)dt+\tau S_f(\tau|v)\nonumber\\
&=\left[-te^{-v\lambda t}\right]^t_0+\int^\tau_0 e^{-v\lambda t}dt+\tau e^{-v\lambda \tau}\nonumber\\
&=-\tau e^{-v\lambda \tau}+\left[-\frac{e^{-v\lambda t}}{v\lambda}\right]^\tau_0+\tau e^{-v\lambda \tau}\nonumber\\
&=\frac{1-e^{-v\lambda \tau}}{v\lambda}.\nonumber
\end{align}
\medskip
\\
{\bf [Weibull frailty model]} The RMST with frailty term is
\begin{align}
\int^\tau_0 S_f(t|v)dt&=\int^\tau_0 t v\lambda k t^{k-1}\exp(-v\lambda t^k)dt+\tau \exp(-v\lambda \tau^k)\nonumber\\
&(v\lambda t^k=x\Leftrightarrow t=v^{-\frac{1}{k}}\lambda^{-\frac{1}{k}}x^{\frac{1}{k}}, v\lambda k t^{k-1}dt=dx)\nonumber\\
&=\int^{v\lambda\tau^k}_0 v^{-\frac{1}{k}}\lambda^{-\frac{1}{k}}x^{\frac{1}{k}}\exp(-x)dx+\tau \exp(-v\lambda \tau^k)\nonumber\\
&=v^{-\frac{1}{k}}\lambda^{-\frac{1}{k}}\gamma\left(v\lambda\tau^k;\frac{1}{k}+1\right)+\tau \exp(-v\lambda \tau^k).\nonumber
\end{align}
\medskip
\\
{\bf [Log-logistic frailty model]} The RMST with frailty term is
\begin{align}
\int^\tau_0 S_f(t|v)dt&=\int^\tau_0 t v\frac{e^\mu k t^{k-1}}{(1+e^\mu t^k)^{v+1}}dt+\tau\left(\frac{1}{1+e^\mu \tau^k}\right)^{v}\nonumber\\
&(e^\mu t^k=x\Leftrightarrow t=e^{-\frac{\mu}{k}}x^\frac{1}{k}, dt=\frac{1}{k}e^{-\frac{\mu}{k}}x^{\frac{1}{k}-1}dx)\nonumber\\
&=\int^{e^\mu\tau^k}_0 v\frac{e^{-\frac{\mu}{k}}x^\frac{1}{k}}{(1+x)^{v+1}}dx+\tau\left(\frac{1}{1+e^\mu \tau^k}\right)^{v}\nonumber\\
&=v e^{-\frac{\mu}{k}}B\left(\frac{e^\mu \tau^k}{1+e^\mu\tau^k};1+\frac{1}{k},v-\frac{1}{k}\right)+\tau\left(\frac{1}{1+e^\mu \tau^k}\right)^{v}.\nonumber
\end{align}
\medskip
\\
{\bf [Log-normal frailty model]} The RMST with frailty term is
\begin{align}
\int^\tau_0 S_f(t|v)dt&=\int^\tau_0 t v\frac{1}{t\sqrt{2\pi\sigma^2}}\exp\left\{-\frac{(\log(t)-\mu)^2}{2\sigma^2}\right\}\left(1-\Phi\left(\frac{\log(t)-\mu}{\sigma}\right)\right)^{v-1}dt\nonumber\\
&\hspace{1cm}+\tau\left(1-\Phi\left(\frac{\log(\tau)-\mu}{\sigma}\right)\right)^{v}\nonumber\\
&(\log(t)=x\sim\Nc(\mu,\si^2), dt=e^xdx)\nonumber\\
&=\int^{\log(\tau)}_{-\infty} \frac{1}{\sqrt{2\pi\sigma^2}}\exp\left\{-\frac{(x-\mu)^2}{2\sigma^2}\right\}\exp\{x\}\left(1-\Phi\left(\frac{x-\mu}{\sigma}\right)\right)^{v-1}dx\nonumber\\
&\hspace{1cm}+\tau\left(1-\Phi\left(\frac{\log(\tau)-\mu}{\sigma}\right)\right)^{v}\nonumber\\
&=\exp\left\{\mu+\frac{\si^2}{2}\right\}\int^{\log(\tau)}_{-\infty} \frac{1}{\sqrt{2\pi\sigma^2}}\exp\left\{-\frac{(x-\mu-\si^2)^2}{2\sigma^2}\right\}\left(1-\Phi\left(\frac{x-\mu}{\sigma}\right)\right)^{v-1}dx\nonumber\\
&\hspace{1cm}+\tau\left(1-\Phi\left(\frac{\log(\tau)-\mu}{\sigma}\right)\right)^{v}\nonumber\\
&\left(\frac{x-\mu-\si^2}{\si}=y\sim\Nc(-\si,1), dx=\si dy\right)\nonumber\\
&=\exp\left\{\mu+\frac{\si^2}{2}\right\}\int^{\frac{\log(\tau)-\mu-\si^2}{\si}}_{-\infty} \frac{1}{\sqrt{2\pi}}\exp\left\{-\frac{y^2}{2}\right\}\left(1-\Phi\left(y+\si\right)\right)^{v-1}dy\nonumber\\
&\hspace{1cm}+\tau\left(1-\Phi\left(\frac{\log(\tau)-\mu}{\sigma}\right)\right)^{v}\nonumber\\
&\approx\exp\left\{\mu+\frac{\si^2}{2}\right\}\int^{\frac{\log(\tau)-\mu-\si^2}{\si}}_{-\infty} \frac{1}{\sqrt{2\pi}}\exp\left\{-\frac{y^2}{2}\right\}\left(1-\Phi\left(y\right)\right)^{v-1}dy\nonumber\\
&\hspace{1cm}+\tau\left(1-\Phi\left(\frac{\log(\tau)-\mu}{\sigma}\right)\right)^{v}\label{eq:appro}\\
&=\exp\left\{\mu+\frac{\si^2}{2}\right\}\frac{1}{v}\left(1-\left(1-\Phi\left(\frac{\log(\tau)-\mu-\si^2}{\si}\right)\right)^{v}\right)\nonumber\\
&\hspace{1cm}+\tau\left(1-\Phi\left(\frac{\log(\tau)-\mu}{\sigma}\right)\right)^{v}.\nonumber
\end{align}
The second-to-last approximate equation $\ref{eq:appro}$ is the result of modifying the integral because it is not solvable. To obtain a better approximation, the Monte Carlo integral can be applied, but this may be computationally time-consuming since it is applied to each posterior sample.

\subsection{Another representation of the Weibull distribution}\label{App:Ano_W}
We demonstrate the RMST in the weibull distribution $W(\lambda,k)$, where a scale parameter $\lambda>0$ and the shape parameter $k>0$. We define the density function, survival function, and hazard function.
\begin{align}
f(t)=\frac{k}{\lambda}\left(\frac{t}{\lambda}\right)^{k-1}\exp\left\{-\left(\frac{t}{\lambda}\right)^k\right\}, S(t)=\exp\left\{-\left(\frac{t}{\lambda}\right)^k\right\}, h(t)=\frac{k}{\lambda}\left(\frac{t}{\lambda}\right)^{k-1}.\nonumber
\end{align}
The RMST is calculated as follows:
\begin{align}
\int^\tau_0 S(t)dt&=\int^\tau_0 t\frac{k}{\lambda}\left(\frac{t}{\lambda}\right)^{k-1}\exp\left\{-\left(\frac{t}{\lambda}\right)^k\right\}dt+\tau \exp\left\{-\left(\frac{\tau}{\lambda}\right)^k\right\}\nonumber\\
&\left(\left(\frac{t}{\lambda}\right)^k=x\Leftrightarrow t=\lambda x^{\frac{1}{k}}, \frac{k}{\lambda}\left(\frac{t}{\lambda}\right)^{k-1}dt=dx\right)\nonumber\\
&=\int^{\left(\frac{\tau}{\lambda}\right)^k}_0 \lambda x^{\frac{1}{k}}\exp(-x)dx+\tau \exp\left\{-\left(\frac{\tau}{\lambda}\right)^k\right\}\nonumber\\
&=\lambda\gamma\left(\left(\frac{\tau}{\lambda}\right)^k;\frac{1}{k}+1\right)+\tau \exp\left\{-\left(\frac{\tau}{\lambda}\right)^k\right\},\nonumber
\end{align}
where $\gamma(z;a)=\int^z_0 t^{a-1}e^{-t}dt$ is an incomplete gamma function.

Next we consider the Weibull model. We transform the parameter $\lambda$ to account for covariates.
\begin{align}
\la_{ij}=\exp\{\x_{ij}^T\bbe\}.\nonumber
\end{align}
The hazard function with frailty term is
\begin{align}
h(t_{ij}|v_i)=v_i\frac{k}{\exp(\x_{ij}^T\bbe)}\left(\frac{t}{\exp(\x_{ij}^T\bbe)}\right)^{k-1}.\nonumber
\end{align}

For the Weibull distribution, the distribution of the RMST for each group is obtained below.
\begin{align}
RMST_{W}(\tau,x_1,\be_0^\ast,\be_1^\ast)&=\exp\left\{\be_0^\ast+x_1\be_1^\ast\right\}\gamma\left(\left(\frac{\tau}{\exp\left\{\be_0^\ast+x_1\be_1^\ast\right\}}\right)^{k^\ast};\frac{1}{k^\ast}+1\right)\nonumber\\
&\hspace{3cm}+\tau \exp\left\{-\left(\frac{\tau}{\exp\left\{\be_0^\ast+x_1\be_1^\ast\right\}}\right)^{k^\ast}\right\}.\nonumber
\end{align}
The difference between the RMSTs is
\begin{align}
RMST_W(\tau,1,\be_0^\ast,\be_1^\ast)-RMST_W(\tau,0,\be_0^\ast,\be_1^\ast).\nonumber
\end{align}

The RMSTs for the mixed effects Weibull model are
\begin{align}
RMST_{W,r}(\tau,x_1,\be_0^\ast,\be_1^\ast,u_i^\ast)&=\exp\left\{\be_0^\ast+x_1\be_1^\ast+u_i^\ast\right\}\gamma\left(\left(\frac{\tau}{\exp\left\{\be_0^\ast+x_1\be_1^\ast+u_i^\ast\right\}}\right)^{k^\ast};\frac{1}{k^\ast}+1\right)\nonumber\\
&\hspace{3cm}+\tau \exp\left\{-\left(\frac{\tau}{\exp\left\{\be_0^\ast+x_1\be_1^\ast+u_i^\ast\right\}}\right)^{k^\ast}\right\}.\nonumber
\end{align}

For the frailty model, the hazard function is 
\begin{align}
h_f(t|v)=v\frac{k}{\lambda}\left(\frac{t}{\lambda}\right)^{k-1}.\nonumber
\end{align}
The density function and survival function are
\begin{align}
f_f(t|v)=v\frac{k}{\lambda}\left(\frac{t}{\lambda}\right)^{k-1}\exp\left\{-v\left(\frac{t}{\lambda}\right)^k\right\}, S(t|v)=\exp\left\{-v\left(\frac{t}{\lambda}\right)^k\right\}.\nonumber
\end{align}

The RMST is calculated as follows
\begin{align}
\int^\tau_0 S(t|v)dt&=\int^\tau_0 vt\frac{k}{\lambda}\left(\frac{t}{\lambda}\right)^{k-1}\exp\left\{-v\left(\frac{t}{\lambda}\right)^k\right\}dt+\tau \exp\left\{-v\left(\frac{\tau}{\lambda}\right)^k\right\}\nonumber\\
&\left(v\left(\frac{t}{\lambda}\right)^k=x\Leftrightarrow t=v^{\frac{1}{k}}\lambda x^{\frac{1}{k}}, v\frac{k}{\lambda}\left(\frac{t}{\lambda}\right)^{k-1}dt=dx\right)\nonumber\\
&=\int^{v\left(\frac{\tau}{\lambda}\right)^k}_0 v^{\frac{1}{k}}\lambda x^{\frac{1}{k}}\exp(-x)dx+\tau \exp\left\{-v\left(\frac{\tau}{\lambda}\right)^k\right\}\nonumber\\
&=v^{\frac{1}{k}}\lambda\gamma\left(v\left(\frac{\tau}{\lambda}\right)^k;\frac{1}{k}+1\right)+\tau \exp\left\{-v\left(\frac{\tau}{\lambda}\right)^k\right\}.\nonumber
\end{align}

The RMST for the Weibull frailty model is
\begin{align}
RMST_{W,f}(\tau,x_1,\be_0^\ast,\be_1^\ast,v_i^\ast)&=(v_i^\ast)^{\frac{1}{k^\ast}}\exp\left\{\be_0^\ast+x_1\be_1^\ast\right\}\gamma\left(v_i^\ast\left(\frac{\tau}{\exp\left\{\be_0^\ast+x_1\be_1^\ast\right\}}\right)^{k^\ast};\frac{1}{k^\ast}+1\right)\nonumber\\
&\hspace{3cm}+\tau\exp\left\{-v_i^\ast\left(\frac{\tau}{\exp\left\{\be_0^\ast+x_1\be_1^\ast\right\}}\right)^{k^\ast}\right\}.\nonumber
\end{align}

\subsection{Another representation of the log-logistic distribution}\label{App:Ano_LL}
We demonstrate the RMST in a log-logistic distribution $LL(\al,k)$, where parameters $\al>0$ and $k>0$. We define the density function, survival function, and hazard function.
\begin{align}
f(t)=\frac{\frac{k}{\al}\left(\frac{t}{\al}\right)^{k-1}}{\left(1+\left(\frac{t}{\al}\right)^k\right)^2}, S(t)=\frac{1}{1+\left(\frac{t}{\al}\right)^k}, h(t)=\frac{\frac{k}{\al}\left(\frac{t}{\al}\right)^{k-1}}{1+\left(\frac{t}{\al}\right)^k}.\nonumber
\end{align}
The RMST is calculated as follows:
\begin{align}
\int^\tau_0 S(t)dt&=\int^\tau_0 t\frac{\frac{k}{\al}\left(\frac{t}{\al}\right)^{k-1}}{\left(1+\left(\frac{t}{\al}\right)^k\right)^2}dt+\tau\frac{1}{1+\left(\frac{\tau}{\al}\right)^k}\nonumber\\
&\left(\left(\frac{t}{\al}\right)^k=x\Leftrightarrow t=\al x^\frac{1}{k}, \frac{k}{\al}\left(\frac{t}{\al}\right)^{k-1}dt=dx\right)\nonumber\\
&=\int^{\left(\frac{\tau}{\al}\right)^k}_0 \frac{\al x^\frac{1}{k}}{(1+x)^2}dx+\tau\frac{1}{1+\left(\frac{\tau}{\al}\right)^k}\nonumber\\
&=\al B\left(\frac{\left(\frac{\tau}{\al}\right)^k}{1+\left(\frac{\tau}{\al}\right)^k};1+\frac{1}{k},1-\frac{1}{k}\right)+\tau\frac{1}{1+\left(\frac{\tau}{\al}\right)^k},\nonumber
\end{align}
where $B(z;a,b)$ is an incomplete beta function.
We transform the parameter $\al$ to account for covariates.
\begin{align}
\al_{ij}=\exp(\x_{ij}^T\bbe).\nonumber
\end{align}
The hazard function with frailty term is
\begin{align}
h(t_{ij}|v_i)=v_i\frac{\frac{k}{\exp(\x_{ij}^T\bbe)}\left(\frac{t}{\exp(\x_{ij}^T\bbe)}\right)^{k-1}}{1+\left(\frac{t}{\exp(\x_{ij}^T\bbe)}\right)^k},\nonumber
\end{align}
\begin{align}
S(t_{ij}|v_i)=\left(\frac{1}{1+\left(\frac{t}{\exp(\x_{ij}^T\bbe)}\right)^k}\right)^{v_i},\nonumber
\end{align}
\begin{align}
f(t_{ij}|v_i)=v_i\frac{k}{\exp(\x_{ij}^T\bbe)}\left(\frac{t}{\exp(\x_{ij}^T\bbe)}\right)^{k-1}\left(\frac{1}{1+\left(\frac{t}{\exp(\x_{ij}^T\bbe)}\right)^k}\right)^{v_i+1}.\nonumber
\end{align}
For log-logistic distribution, the distribution of RMST for each group is obtained below
\begin{align}
RMST_{LL}(\tau,x_1,\be_0^\ast,\be_1^\ast)&=\exp(\be_0^\ast+x_1\be_1^\ast) B\left(\frac{\left(\frac{\tau}{\exp(\be_0^\ast+x_1\be_1^\ast)}\right)^{k^\ast}}{1+\left(\frac{\tau}{\exp(\be_0^\ast+x_1\be_1^\ast)}\right)^{k^\ast}};1+\frac{1}{k^\ast},1-\frac{1}{k^\ast}\right)\nonumber\\
&\hspace{3cm}+\tau\frac{1}{1+\left(\frac{\tau}{\exp(\be_0^\ast+x_1\be_1^\ast)}\right)^{k^\ast}}.\nonumber
\end{align}
The difference between the RMSTs is
\begin{align}
RMST_{LL}(\tau,1,\be_0^\ast,\be_1^\ast)-RMST_{LL}(\tau,0,\be_0^\ast,\be_1^\ast).\nonumber
\end{align}
The RMST for the mixed effects log-logistic model is
\begin{align}
RMST_{LL,r}(\tau,x_1,\be_0^\ast,\be_1^\ast,u_i^\ast)&=\exp(\be_0^\ast+x_1\be_1^\ast+u_i^\ast) B\left(\frac{\left(\frac{\tau}{\exp(\be_0^\ast+x_1\be_1^\ast+u_i^\ast)}\right)^{k^\ast}}{1+\left(\frac{\tau}{\exp(\be_0^\ast+x_1\be_1^\ast+u_i^\ast)}\right)^{k^\ast}};1+\frac{1}{k^\ast},1-\frac{1}{k^\ast}\right)\nonumber\\
&\hspace{3cm}+\tau\frac{1}{1+\left(\frac{\tau}{\exp(\be_0^\ast+x_1\be_1^\ast+u_i^\ast)}\right)^{k^\ast}}.\nonumber
\end{align}
For frailty model, the hazard function is 
\begin{align}
h_f(t|v)=v\frac{\frac{k}{\al}\left(\frac{t}{\al}\right)^{k-1}}{1+\left(\frac{t}{\al}\right)^k}.\nonumber
\end{align}
The density function and survival function are
\begin{align}
f_f(t|v)=v\frac{k}{\al}\left(\frac{t}{\al}\right)^{k-1}\left(\frac{1}{1+\left(\frac{t}{\al}\right)^k}\right)^{v+1}, S(t|v)=\left(\frac{1}{1+\left(\frac{t}{\al}\right)^k}\right)^{v}.\nonumber
\end{align}
The RMST is calculated as follows
\begin{align}
\int^\tau_0 S(t|v)dt&=\int^\tau_0 vt\frac{k}{\al}\left(\frac{t}{\al}\right)^{k-1}\left(\frac{1}{1+\left(\frac{t}{\al}\right)^k}\right)^{v+1}dt+\tau\left(\frac{1}{1+\left(\frac{t}{\al}\right)^k}\right)^{v}\nonumber\\
&\left(\left(\frac{t}{\al}\right)^k=x\Leftrightarrow t=\al x^{\frac{1}{k}}, \frac{k}{\al}\left(\frac{t}{\al}\right)^{k-1}dt=dx\right)\nonumber\\
&=\int^{\left(\frac{\tau}{\al}\right)^k}_0 \al vx^{\frac{1}{k}}\left(\frac{1}{1+x}\right)^{v+1}dx+\tau\left(\frac{1}{1+\left(\frac{t}{\al}\right)^k}\right)^{v}\nonumber\\
&=\al vB\left(\frac{\left(\frac{\tau}{\al}\right)^k}{1+\left(\frac{\tau}{\al}\right)^k};1+\frac{1}{k},v-\frac{1}{k}\right)+\tau\left(\frac{1}{1+\left(\frac{t}{\al}\right)^k}\right)^{v}.\nonumber
\end{align}

The RMST for the log-logistic frailty model is
\begin{align}
RMST_{LL,f}(\tau,x_1,\be_0^\ast,\be_1^\ast,v_i^\ast)&=v\left(\be_0^\ast+x_1\be_1^\ast\right)B\left(\frac{\left(\frac{\tau}{\left(\be_0^\ast+x_1\be_1^\ast\right)}\right)^k}{1+\left(\frac{\tau}{\left(\be_0^\ast+x_1\be_1^\ast\right)}\right)^k};1+\frac{1}{k},v_i^\ast-\frac{1}{k}\right)\nonumber\\
&\hspace{3cm}+\tau\left(\frac{1}{1+\left(\frac{t}{\left(\be_0^\ast+x_1\be_1^\ast\right)}\right)^k}\right)^{v_i^\ast}.\nonumber
\end{align}

\subsection{Posterior probability of mixed effect models}
\subsubsection{Mixed effects exponential model}\label{App:exp_rand}
The density function and survival function, conditional on the random effect $u_i$ for $i$-th cluster, are
\begin{align}
&f_r(t_{ij}|u_i)=\exp\{\x_{ij}^T\bbe+u_i\}\exp\left\{-\exp(\x_{ij}^T\bbe+u_i)t_{ij}\right\},\nonumber\\
&S_r(t_{ij}|u_i)=\exp\left\{-\exp(\x_{ij}^T\bbe+u_i)t_{ij}\right\}.\nonumber
\end{align}
The posterior probability is 
\begin{align}
\pi_r(\bbe,\u,\phi|\t,\X)\propto L_r(\t|\X,\bbe,\u,\phi)g_r(\u|\phi)\pi(\bbe|\c)\pi(\phi|\xi),\nonumber
\end{align}
where
\begin{align}
L_r(\t|\X,\bbe,\u,\phi)&=\prod^M_{i=1}\prod^{n_i}_{j=1}\{f_r(t_{ij}|v_i)\}^{\de_{ij}}\{S_r(t_{ij}|v_i)\}^{1-\de_{ij}}\nonumber\\
&=\prod^M_{i=1}\prod^{n_i}_{j=1}\left[\exp\{\x_{ij}^T\bbe+u_i\}\exp\left\{-\exp(\x_{ij}^T\bbe+u_i)t_{ij}\right\}\right]^{\de_{ij}}\nonumber\\
&\hspace{2cm}\times\left[\exp\left\{-\exp(\x_{ij}^T\bbe+u_i)t_{ij}\right\}\right]^{1-\de_{ij}},\nonumber
\end{align}
the variable $\de_{ij}$  is a censoring indicator that takes the value 1 if an event occurs and 0 if censored, $g_r(\u|\phi)$ is the density function of $\u$,
$\bbe\sim\Nc(0,\bSi_\c)$, $\bSi_\c=\diag(c_1,c_2,\ldots,c_p)$, $u_i\sim \Nc\left(0,\phi^2\right)$, $\phi\sim \Uc(0,\xi)$, and $\xi$ is a hyperparameter.

\subsubsection{Mixed effects Weibull model}\label{App:wei_rand}
The density function and survival function, conditional on the random effect $u_i$ for $i$-th cluster, are
\begin{align}
&f_r(t_{ij}|u_i)=\exp(\x_{ij}^T\bbe+u_i)k t_{ij}^{k-1}\exp\left\{-\exp(\x_{ij}^T\bbe+u_i)t_{ij}^k\right\},\nonumber
&S_r(t_{ij}|u_i)=\exp\left\{ - \exp(\x_{ij}^T\bbe+u_i) t_{ij}^{k}\right\}.\nonumber
\end{align}
The posterior probability is 
\begin{align}
\pi_r(\bbe,\u,\phi,k|\t,\X)\propto L_r(\t|\X,\bbe,\u,\phi,k)g_r(\u|\phi)\pi(\bbe|\c)\pi(\phi|\xi)\pi(k|a,b),\nonumber
\end{align}
where
\begin{align}
L_r(\t|\X,\bbe,\u,\phi,k)&=\prod^M_{i=1}\prod^{n_i}_{j=1}\{f_r(t_{ij}|u_i)\}^{\de_{ij}}\{S_r(t_{ij}|u_i)\}^{1-\de_{ij}}\nonumber\\
&=\prod^M_{i=1}\prod^{n_i}_{j=1}\left[\exp(\x_{ij}^T\bbe+u_i)k t_{ij}^{k-1}\exp\left\{-\exp(\x_{ij}^T\bbe+u_i)t_{ij}^k\right\}\right]^{\de_{ij}}\nonumber\\
&\hspace{2cm}\times\left[\exp\left\{ - \exp(\x_{ij}^T\bbe+u_i) t_{ij}^{k}\right\}\right]^{1-\de_{ij}},\nonumber
\end{align}
$\bbe\sim\Nc(0,\bSi_\c)$, $\bSi_\c=\diag(c_1,c_2,\ldots,c_p)$, $u_i\sim \Nc\left(0,\phi^2\right)$, $\phi\sim \Uc(0,\xi)$, and $k\sim Gamma\left(a,b\right)$.

\subsubsection{Mixed effects log-logistic model}\label{App:LL_rand}
The density function and survival function, conditional on the random effect $u_i$ for $i$-th cluster, are
\begin{align}
f_r(t_{ij}|u_i)=\frac{\exp\{\x_{ij}^T\bbe+u_i\}k t_{ij}^{k-1}}{\left(1+\exp(\x_{ij}^T\bbe+u_i) t^k\right)^{2}},S_r(t_{ij}|u_i)=\frac{1}{1+\exp(\x_{ij}^T\bbe+u_i) t^k}.\nonumber
\end{align}
The posterior probability is 
\begin{align}
\pi_r(\bbe,\u,\phi,k|\t,\X)\propto L_r(\t|\X,\bbe,\u,\phi,k)g_r(\u|\phi)\pi(\bbe|\c)\pi(\phi|\xi)\pi(k|a,b),\nonumber
\end{align}
where
\begin{align}
L_r(\t|\X,\bbe,\u,\phi,k)&=\prod^M_{i=1}\prod^{n_i}_{j=1}\{f_r(t_{ij}|u_i)\}^{\de_{ij}}\{S_r(t_{ij}|u_i)\}^{1-\de_{ij}}\nonumber\\
&=\prod^M_{i=1}\prod^{n_i}_{j=1}\left[\frac{\exp\{\x_{ij}^T\bbe+u_i\}k t_{ij}^{k-1}}{\left(1+\exp(\x_{ij}^T\bbe+u_i) t^k\right)^{2}}\right]^{\de_{ij}}\left[\frac{1}{1+\exp(\x_{ij}^T\bbe+u_i) t^k}\right]^{1-\de_{ij}},\nonumber
\end{align}
$\bbe\sim\Nc(0,\bSi_\c)$, $\bSi_\c=\diag(c_1,c_2,\ldots,c_p)$, $u_i\sim \Nc\left(0,\phi^2\right)$, $\phi\sim \Uc(0,\xi)$, and $k\sim Gamma\left(a,b\right)$.

\subsubsection{Mixed effects log-normal model}\label{App:LN_rand}
The density function and survival function, conditional on the random effect $u_i$ for $i$-th cluster, are
\begin{align}
&f_r(t_{ij}|u_i)=\frac{1}{t\sqrt{2\pi\sigma^2}}\exp\left[-\frac{\{\log(t)-(\x_{ij}^T\bbe+u_i)\}^2}{2\sigma^2}\right],\nonumber\\
&S_r(t_{ij}|u_i)=1-\Phi\left(\frac{\log(t)-(\x_{ij}^T\bbe+u_i)}{\sigma}\right).\nonumber
\end{align}
The posterior probability is 
\begin{align}
\pi_r(\bbe,\u,\phi,\si^2|\t,\X)\propto L_r(\t|\X,\bbe,\u,\phi,\si^2)g_r(\u|\phi)\pi(\bbe|\c)\pi(\phi|\xi)\pi(\si^2|a,b),\nonumber
\end{align}
where
\begin{align}
L_r(\t|\X,\bbe,\u,\phi,\si^2)&=\prod^M_{i=1}\prod^{n_i}_{j=1}\{f_r(t_{ij}|u_i)\}^{\de_{ij}}\{S_r(t_{ij}|u_i)\}^{1-\de_{ij}}\nonumber\\
&=\prod^M_{i=1}\prod^{n_i}_{j=1}\left[\frac{1}{t\sqrt{2\pi\sigma^2}}\exp\left[-\frac{\{\log(t)-(\x_{ij}^T\bbe+u_i)\}^2}{2\sigma^2}\right]\right]^{\de_{ij}}\nonumber\\
&\hspace{2cm}\times\left[1-\Phi\left(\frac{\log(t)-(\x_{ij}^T\bbe+u_i)}{\sigma}\right)\right]^{1-\de_{ij}},\nonumber
\end{align}
$\bbe\sim\Nc(0,\bSi_\c)$, $\bSi_\c=\diag(c_1,c_2,\ldots,c_p)$, $u_i\sim \Nc\left(0,\phi^2\right)$, $\phi\sim \Uc(0,\xi)$, and $\si^2\sim \Uc(0,s)$.

\subsection{Posterior probability of frailty models}
\subsubsection{Exponential frailty model}\label{App:exp_frail}
The survival function $S_f(t_{ij}|v_i)$ conditional on the frailty $u_i$ for $i$-th cluster is expressed as
\begin{align}
S_f(t_{ij}|v_i)=\exp\left\{-\int^{t_{ij}}_0 h_f(s|v_i)ds\right\}=\exp\left\{-\int^{t_{ij}}_0 v_i\exp\{\x_{ij}^T\bbe\}ds\right\}=\exp\left\{-v_i\exp\{\x_{ij}^T\bbe\}t_{ij}\right\}.\nonumber
\end{align}
The density function $f_f(t_{ij}|v_i)$ conditional on the frailty $u_i$ for $i$-th cluster is
\begin{align}
f_f(t_{ij}|v_i)=\left.-\frac{d}{dt}S_f(t|v_i)\right|_{t=t_{ij}}=v_i\exp\{\x_{ij}^T\bbe\}\exp\left\{-v_i\exp\{\x_{ij}^T\bbe\}t_{ij}\right\}.\nonumber
\end{align}
The posterior probability is 
\begin{align}
\pi_f(\bbe,\v,\phi|\t,\X)\propto L_f(\t|\X,\bbe,\v,\phi)g_f(\v|\phi)\pi(\bbe|\c)\pi(\phi|\xi),\nonumber
\end{align}
where
\begin{align}
L_f(\t|\X,\bbe,\v,\phi)&=\prod^M_{i=1}\prod^{n_i}_{j=1}\{f_f(t_{ij}|v_i)\}^{\de_{ij}}\{S_f(t_{ij}|v_i)\}^{1-\de_{ij}}\nonumber\\
&=\prod^M_{i=1}\prod^{n_i}_{j=1}\left[v_j\exp\{\x_{ij}^T\bbe\}\exp\left\{-v_j\exp\{\x_{ij}^T\bbe\}t\right\}\right]^{\de_{ij}}\left[\exp\left\{-v_j\exp\{\x_{ij}^T\bbe\}t\right\}\right]^{1-\de_{ij}},\nonumber
\end{align}
$\bbe\sim\Nc(0,\bSi_\c)$, $\bSi_\c=\diag(c_1,c_2,\ldots,c_p)$, $v_i\sim Gamma\left(\frac{1}{\phi}, \frac{1}{\phi}\right)$ , $\phi\sim \Uc(0,\xi)$.

\subsubsection{Weibull frailty model}\label{App:wei_frail}
The survival function conditional on the frailty $u_i$ for $i$-th cluster is 
\begin{align}
S_f(t_{ij}|v_i)=\exp\left\{- \int^{t_{ij}}_0v_i\exp\{\x_{ij}^T\bbe\} k s^{k-1}ds\right\}=\exp\left\{ - v_i\exp\{\x_{ij}^T\bbe\} t_{ij}^{k}\right\}.\nonumber
\end{align}
The density function conditional on the frailty $u_i$ for $i$-th cluster is 
\begin{align}
f_f(t_{ij}|v_i)=\left.-\frac{d}{dt}S_f(t|v_i)\right|_{t=t_{ij}}=v_i\exp\{\x_{ij}^T\bbe\}k t_{ij}^{k-1}\exp\left\{-v_i\exp\{\x_{ij}^T\bbe\}t_{ij}^k\right\}.\nonumber
\end{align}
The posterior probability is 
\begin{align}
\pi_f(\bbe,\v,\phi,k|\t,\X)\propto L_f(\t|\X,\bbe,\v,\phi,k)g_f(\v|\phi)\pi(\bbe|\c)\pi(\phi|\xi)\pi(k|a,b),\nonumber
\end{align}
where
\begin{align}
L_f(\t|\X,\bbe,\v,\phi,k)&=\prod^M_{i=1}\prod^{n_i}_{j=1}\{f_f(t_{ij}|v_i)\}^{\de_{ij}}\{S_f(t_{ij}|v_i)\}^{1-\de_{ij}}\nonumber\\
&=\prod^M_{i=1}\prod^{n_i}_{j=1}\left[v_j\exp\{\x_{ij}^T\bbe\}kt^{k-1}\exp\left\{-v_j\exp\{\x_{ij}^T\bbe\}t^k\right\}\right]^{\de_{ij}}\nonumber\\
&\hspace{2cm}\times\left[\exp\left\{-v_j\exp\{\x_{ij}^T\bbe\}t^k\right\}\right]^{1-\de_{ij}},\nonumber
\end{align}
$\bbe\sim\Nc(0,\bSi_\c)$, $\bSi_\c=\diag(c_1,c_2,\ldots,c_p)$, $v_i\sim Gamma\left(\frac{1}{\phi},\frac{1}{\phi}\right)$, $\phi\sim \Uc(0,\xi)$, and $k\sim Gamma\left(a,b\right)$.

\subsubsection{Log-logistic frailty model}\label{App:LL_frail}
The following formula is used to calculate the survival function conditional on the frailty.
\begin{align}
S(t)=e^{-\int^t_0 h(s)ds}\Leftrightarrow -\int^t_0 h(s)ds=\log\left(S(t)\right)\nonumber
\end{align}
The survival function conditional on the frailty $u_i$ for $i$-th cluster is 
\begin{align}
S_f(t_{ij}|v_i)=\exp\left\{-\int^{t_{ij}}_0h_f(s|v_i)ds\right\}=\exp\left\{ - v_i\int^{t_{ij}}_0h(s)ds\right\}=\exp\left\{\log\left(S(t)\right)^{v_i}\right\}=\left(\frac{1}{1+e^{\x_{ij}^T\bbe} t^k}\right)^{v_i}.\nonumber
\end{align}
The density function conditional on the frailty $u_i$ for $i$-th cluster is 
\begin{align}
f_f(t_{ij}|v_i)=\left.-\frac{d}{dt}S(t|v_i)\right|_{t=t_{ij}}=v_i\exp\{\x_{ij}^T\bbe\}k t_{ij}^{k-1}\left(\frac{1}{1+e^{\x_{ij}^T\bbe} t^k}\right)^{v_i+1}.\nonumber
\end{align}
The posterior probability is
\begin{align}
\pi_f(\bbe,\v,\phi,k|\t,\X)\propto L_f(\t|\X,\bbe,\v,\phi,k)g_f(\v|\phi)\pi(\bbe|\c)\pi(\phi|\xi)\pi(k|a,b),\nonumber
\end{align}
where
\begin{align}
L_f(\t|\X,\bbe,\v,\phi,k)&=\prod^M_{i=1}\prod^{n_i}_{j=1}\{f(t_{ij}|v_i)\}^{\de_{ij}}\{S(t_{ij}|v_i)\}^{1-\de_{ij}}\nonumber\\
&=\prod^M_{i=1}\prod^{n_i}_{j=1}\left[v_i\exp\{\x_{ij}^T\bbe\}k t_{ij}^{k-1}\left(\frac{1}{1+e^{\x_{ij}^T\bbe} t^k}\right)^{v_i+1}\right]^{\de_{ij}}\left[\left(\frac{1}{1+e^{\x_{ij}^T\bbe} t^k}\right)^{v_i}\right]^{1-\de_{ij}},\nonumber
\end{align}
$\bbe\sim\Nc(0,\bSi_\c)$, $\bSi_\c=\diag(c_1,c_2,\ldots,c_p)$, $v_i\sim Gamma\left(\frac{1}{\phi},\frac{1}{\phi}\right)$, and $\phi\sim \Uc(0,\xi), k\sim Gamma\left(a,b\right)$.

\subsubsection{Log-normal frailty model}\label{App:LN_frail}
The survival function conditional on the frailty $u_i$ for $i$-th cluster is 
\begin{align}
S_f(t_{ij}|v_i)=\left(S(t)\right)^{v_i}=\left(1-\Phi\left(\frac{\log(t)-\x_{ij}^T\bbe}{\sigma}\right)\right)^{v_i}.\nonumber
\end{align}
The density function conditional on the frailty $u_i$ for $i$-th cluster is 
\begin{align}
f_f(t_{ij}|v_i)=\left.-\frac{d}{dt}S(t|v_i)\right|_{t=t_{ij}}=v_i\frac{1}{t\sqrt{2\pi\sigma^2}}\exp\left\{-\frac{(\log(t)-\x_{ij}^T\bbe)^2}{2\sigma^2}\right\}\left(1-\Phi\left(\frac{\log(t)-\x_{ij}^T\bbe}{\sigma}\right)\right)^{v_i-1}.\nonumber
\end{align}
The posterior probability is
\begin{align}
\pi_f(\bbe,\v,\phi,\si^2|\t,\X)\propto L_f(\t|\X,\bbe,\v,\phi,\si^2)g_f(\v|\phi)\pi(\bbe|\c)\pi(\phi|\xi)\pi(\si^2|a,b),\nonumber
\end{align}
where
\begin{align}
L_f(\t|\X,\bbe,\v,\phi,\si^2)&=\prod^M_{i=1}\prod^{n_i}_{j=1}\{f(t_{ij}|v_i)\}^{\de_{ij}}\{S(t_{ij}|v_i)\}^{1-\de_{ij}}\nonumber\\
&=\prod^M_{i=1}\prod^{n_i}_{j=1}\left[v_i\frac{1}{t\sqrt{2\pi\sigma^2}}\exp\left\{-\frac{(\log(t)-\x_{ij}^T\bbe)^2}{2\sigma^2}\right\}\left(1-\Phi\left(\frac{\log(t)-\x_{ij}^T\bbe}{\sigma}\right)\right)^{v_i-1}\right]^{\de_{ij}}\nonumber\\
&\hspace{2.5cm}\times\left[\left(1-\Phi\left(\frac{\log(t)-\x_{ij}^T\bbe}{\sigma}\right)\right)^{v_i}\right]^{1-\de_{ij}},\nonumber
\end{align}
$\bbe\sim\Nc(0,\bSi_\c)$, $\bSi_\c=\diag(c_1,c_2,\ldots,c_p)$, $v_i\sim Gamma\left(\frac{1}{\phi},\frac{1}{\phi}\right)$, $\phi\sim \Uc(0,\xi)$, and $\si^2\sim \Uc(0,s)$.

\subsection{Supplemental information for simulation}
\subsubsection{Survival probability in each scenario}
\begin{figure}[H]
  \begin{center}
  \includegraphics[width=15cm]{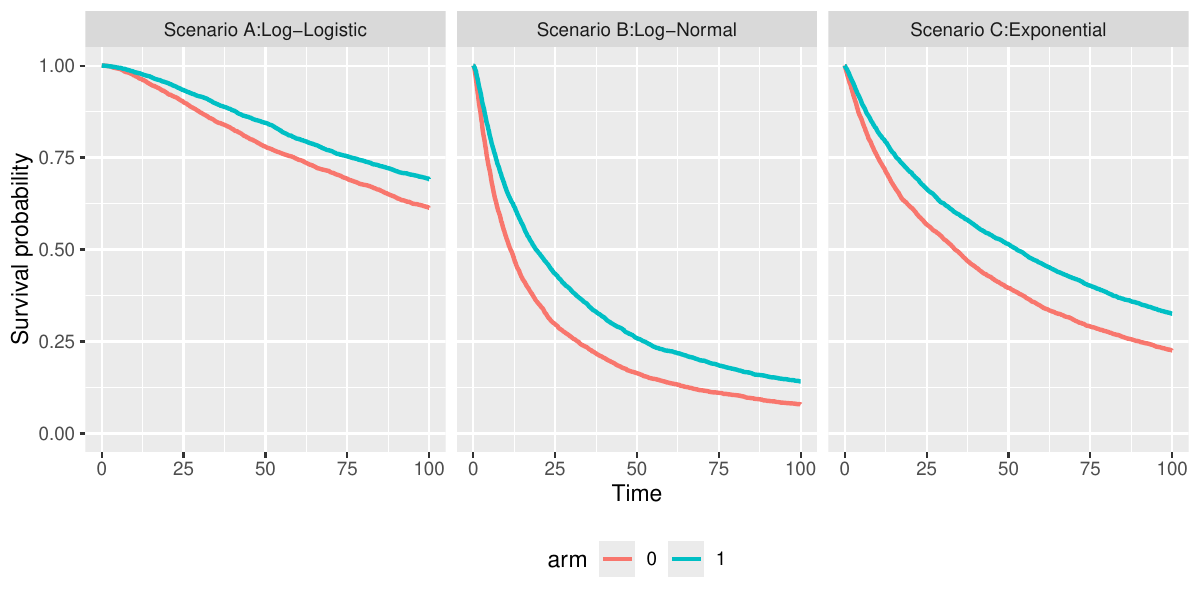}
  \caption{Survival probabilities of simulation.}
  \label{Sim:data}
        \footnotesize{}
  \end{center}
\end{figure}

\subsubsection{Simulation specifications and results of scenario C}
Scenario C is a basic proportional hazard assumption based on exponential distribution.
The parameter is given as 
\begin{align*}
    \lambda_{ij} &= \exp( \beta_0 + \beta_1 x_{1ij} + \beta_2 x_{2ij} + v_i),
\end{align*}
where $\beta_0=-4.5$ is a baseline effect, $\beta_1=0.5$ is a treatment effect, $\beta_2=1$ is an effect of the covariate, and other parameters are set as well as scenario A.
We consider $4$ clusters for the simulation. 
The theoretical RMSTs in Scenario C are $45.85$ and $60.37$ by equation \eqref{eq:RMST-E}, respectively. Thus, the true difference between the RMSTs is $-14.52$.

The simulation results were in Table \ref{t:sim-C} for scenario C in the Appendix. In scenario C, where the exponential model with random effect is true, the exponential and Weibull models had small biases and MSEs. The log-logistic and log-normal models had some biases in the scenario. The log-normal model with frailty effects included a significant bias, and consequently, the MSE was also significant.
In summary, bias and MSE were larger when inferences were made using models that deviated from the true model. 
It should be noted that the frailty effect may obtain extreme accuracy, like the log-normal model in scenario C.
Therefore, an appropriate model with or without frailty or random effects should be selected for estimating RMST with small bias and MSE.
\begin{table}[H]
  \begin{center}
\caption{Simulation results of scenario C: exponential model with random effects \label{t:sim-C}}
\begin{tabular}{|cccccccccccccc|}
\hline
\multicolumn{1}{|c|}{}       & \multicolumn{1}{c|}{}     & \multicolumn{3}{c|}{Exponential}           & \multicolumn{3}{c|}{Log-Logistic}          & \multicolumn{3}{c|}{Log-Normal}             & \multicolumn{3}{c|}{Weibull} \\
\multicolumn{1}{|c|}{}       & \multicolumn{1}{c|}{n}    & T     & F     & \multicolumn{1}{c|}{M}     & T     & F     & \multicolumn{1}{c|}{M}     & T     & F      & \multicolumn{1}{c|}{M}     & T        & F       & M       \\ \hline
\multicolumn{14}{|l|}{Scenario C: Exponential model with random effect}                                                                                                                                                         \\ \hline
\multicolumn{1}{|c|}{Mean}   & \multicolumn{1}{c|}{64}   & 0.94  & 1.82  & \multicolumn{1}{c|}{0.88}  & 1.64  & 1.37  & \multicolumn{1}{c|}{1.72}  & 2.52  & 2.97   & \multicolumn{1}{c|}{2.52}  & 1.15     & 0.97    & 1.08    \\
\multicolumn{1}{|c|}{(Bias)}       & \multicolumn{1}{c|}{512}  & 0.50  & 0.92  & \multicolumn{1}{c|}{0.26}  & 1.39  & 2.60  & \multicolumn{1}{c|}{1.28}  & 2.19  & 10.46  & \multicolumn{1}{c|}{2.11}  & 0.71     & 0.37    & 0.28    \\
\multicolumn{1}{|c|}{}       & \multicolumn{1}{c|}{2048} & 0.94  & 1.59  & \multicolumn{1}{c|}{1.00}  & 1.86  & 3.71  & \multicolumn{1}{c|}{1.99}  & 2.53  & 14.25  & \multicolumn{1}{c|}{2.60}  & 1.08     & 1.08    & 1.00    \\ \hline
\multicolumn{1}{|c|}{MSE}    & \multicolumn{1}{c|}{64}   & 81.75 & 73.34 & \multicolumn{1}{c|}{83.53} & 87.73 & 92.44 & \multicolumn{1}{c|}{87.10} & 86.94 & 89.72  & \multicolumn{1}{c|}{88.25} & 80.19    & 83.11   & 80.80   \\
\multicolumn{1}{|c|}{}       & \multicolumn{1}{c|}{512}  & 11.00 & 10.03 & \multicolumn{1}{c|}{10.27} & 12.84 & 15.08 & \multicolumn{1}{c|}{11.91} & 15.43 & 113.58 & \multicolumn{1}{c|}{14.85} & 10.89    & 10.13   & 9.95    \\
\multicolumn{1}{|c|}{}       & \multicolumn{1}{c|}{2048} & 4.13  & 5.27  & \multicolumn{1}{c|}{3.75}  & 6.48  & 15.55 & \multicolumn{1}{c|}{6.75}  & 9.27  & 203.06 & \multicolumn{1}{c|}{9.41}  & 4.39     & 3.77    & 3.68    \\ \hline
\multicolumn{1}{|c|}{Mode}   & \multicolumn{1}{c|}{64}   & 0.74  & 2.27  & \multicolumn{1}{c|}{1.28}  & 1.28  & 1.40  & \multicolumn{1}{c|}{1.64}  & 1.99  & 3.50   & \multicolumn{1}{c|}{2.35}  & 1.00     & 1.22    & 1.02    \\
\multicolumn{1}{|c|}{}       & \multicolumn{1}{c|}{512}  & 0.40  & 0.48  & \multicolumn{1}{c|}{0.19}  & 1.37  & 2.70  & \multicolumn{1}{c|}{1.22}  & 2.06  & 11.84  & \multicolumn{1}{c|}{2.03}  & 0.57     & 0.35    & 0.35    \\
\multicolumn{1}{|c|}{}       & \multicolumn{1}{c|}{2048} & 0.94  & 1.12  & \multicolumn{1}{c|}{0.73}  & 1.83  & 3.71  & \multicolumn{1}{c|}{1.90}  & 2.48  & 14.40  & \multicolumn{1}{c|}{2.55}  & 1.04     & 0.85    & 0.78    \\ \hline
\multicolumn{1}{|c|}{Median} & \multicolumn{1}{c|}{64}   & 0.81  & 1.94  & \multicolumn{1}{c|}{0.84}  & 1.53  & 1.33  & \multicolumn{1}{c|}{1.68}  & 2.39  & 3.06   & \multicolumn{1}{c|}{2.44}  & 1.06     & 0.99    & 1.05    \\
\multicolumn{1}{|c|}{}       & \multicolumn{1}{c|}{512}  & 0.49  & 0.77  & \multicolumn{1}{c|}{0.21}  & 1.38  & 2.62  & \multicolumn{1}{c|}{1.25}  & 2.17  & 10.96  & \multicolumn{1}{c|}{2.09}  & 0.69     & 0.29    & 0.24    \\
\multicolumn{1}{|c|}{}       & \multicolumn{1}{c|}{2048} & 0.94  & 1.31  & \multicolumn{1}{c|}{0.91}  & 1.86  & 3.71  & \multicolumn{1}{c|}{1.92}  & 2.53  & 14.30  & \multicolumn{1}{c|}{2.56}  & 1.07     & 0.93    & 0.91    \\ \hline
\end{tabular}
\footnotesize{T: typical fixed model, F: frailty effects model, R: random effects model, Mean, Mode, and Median are the difference from true RMST.}
  \end{center}
\end{table}

\subsection{Supplemental information for DHS data analysis}
\subsubsection{Kaplan-Meier plot of DHS data}
\begin{figure}[H]
  \begin{center}
  \includegraphics[width=15cm]{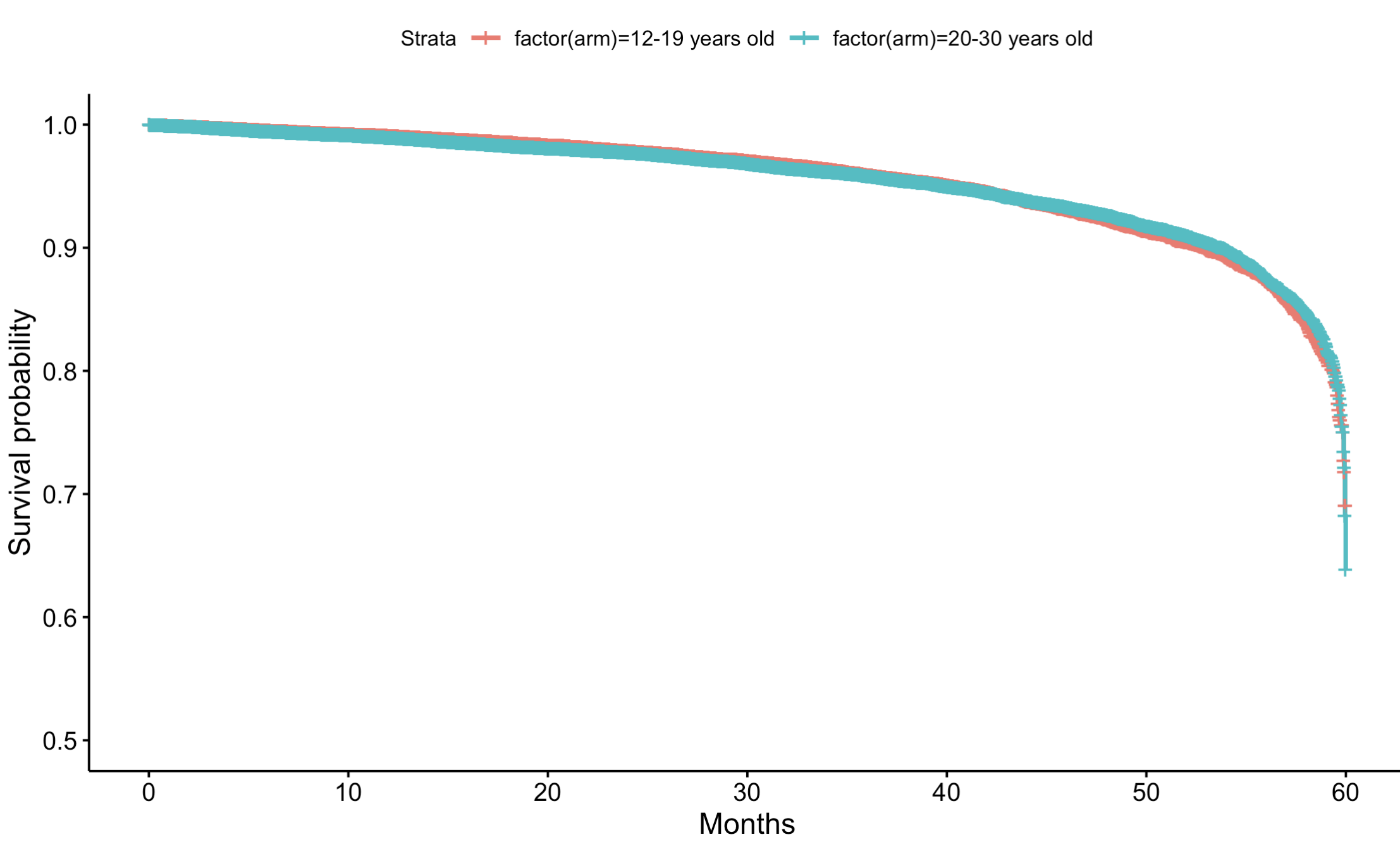}
  \caption{Kaplan-Meier plot (12--19 group vs 20--30 group)}
  \label{App:KM_2030}
        \footnotesize{}
  \end{center}
\end{figure}
\begin{figure}[H]
  \begin{center}
  \includegraphics[width=15cm]{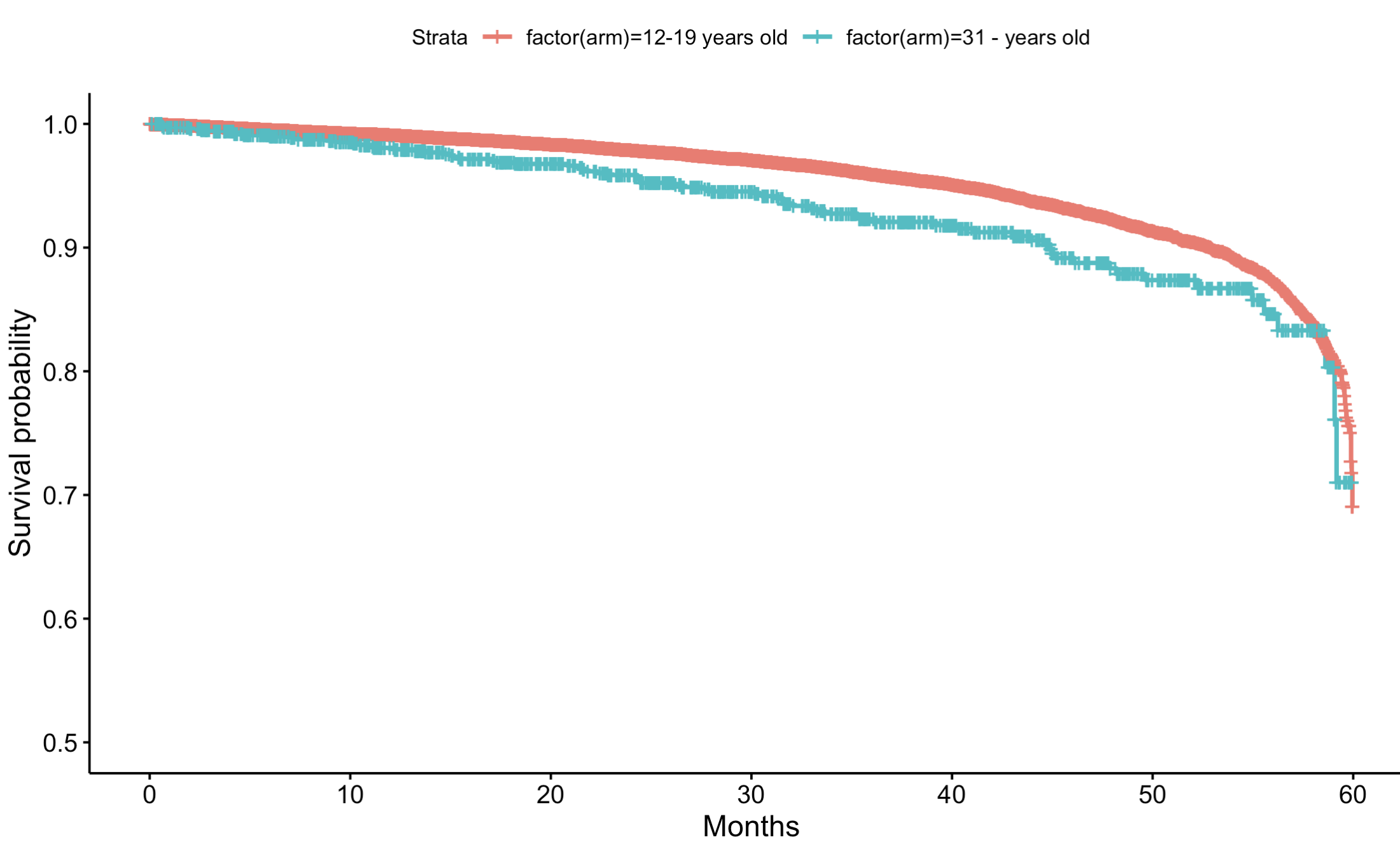}
  \caption{Kaplan-Meier plot (12--19 group vs 31+ group)}
  \label{App:KM_31}
        \footnotesize{}
  \end{center}
\end{figure}

\subsubsection{Estimation results (12--19 group vs 20--30 group)}\label{App:Supp_DHS2030}
\begin{table}[H]
  \begin{center}
\caption{Results of exponential model (12--19 group vs 20--30 group)\label{t:DHS2030_E}}
\begin{tabular}{|c|c|c|c|c|c|c|c|c|}\hline
\multicolumn{2}{|c|}{Parameter} & Mode & Median & Mean & SE & 95\%CI & Rhat & ESS\\\hline
\multicolumn{2}{|c|}{Intercept} & 4.44 & 4.42 & 4.43 & 0.25 & [3.97, 4.95] & 1.00 & 788\\
\multicolumn{2}{|c|}{20--30 group} & 0.02 & 0.01 & 0.01 & 0.03 & [-0.04, 0.07] & 1.00 & 3653\\
\multicolumn{2}{|c|}{Sex} & 0.12 & 0.13 & 0.13 & 0.03 & [0.07, 0.19] & 1.00 & 4055\\
Place & Respondents home & 0.89 & 0.80 & 0.79 & 0.24 & [0.28, 1.23] & 1.00 & 758\\
Place & Other home & 0.92 & 0.88 & 0.89 & 0.46 & [0.00, 1.83] & 1.00 & 1780\\
Place & Public sector & 1.09 & 1.06 & 1.06 & 0.27 & [0.50, 1.57] & 1.00 & 850\\
Place & Government hospital & 0.99 & 0.94 & 0.93 & 0.24 & [0.42, 1.37] & 1.00 & 723\\
Place & CS Govt health professional & 0.86 & 0.90 & 0.90 & 0.24 & [0.39, 1.34] & 1.00 & 738\\
Place & Other public sector & 0.95 & 0.98 & 0.97 & 0.27 & [0.41, 1.51] & 1.00 & 854\\
Place & Private hospital/clinic & 0.58 & 0.61 & 0.61 & 0.24 & [0.10, 1.06] & 1.00 & 741\\
Place & CS private health facility & 0.96 & 0.95 & 0.94 & 0.29 & [0.33, 1.49] & 1.00 & 963\\
Size & Very large & 1.13 & 1.15 & 1.15 & 0.09 & [0.98, 1.32] & 1.00 & 1643\\
Size & Larger than average & 1.23 & 1.22 & 1.22 & 0.08 & [1.07, 1.38] & 1.00 & 1554\\
Size & Average & 1.33 & 1.32 & 1.32 & 0.07 & [1.19, 1.45] & 1.00 & 1271\\
Size & Smaller than average & 0.96 & 0.94 & 0.94 & 0.08 & [0.78, 1.10] & 1.00 & 1453\\
Size & Very small & 0.15 & 0.16 & 0.16 & 0.08 & [0.00, 0.32] & 1.00 & 1581\\
\multicolumn{2}{|c|}{Order} & -0.03 & -0.03 & -0.03 & 0.01 & [-0.05, -0.01] & 1.00 & 4215\\
\multicolumn{2}{|c|}{RMST$_{12-19 group}$} & 37.76 & 37.58 & 37.45 & 2.46 & [32.33, 42.11] &  & \\
\multicolumn{2}{|c|}{RMST$_{20-30 group}$} & 38.13 & 37.71 & 37.58 & 2.43 & [32.59, 42.16] &  & \\
\multicolumn{2}{|c|}{RMST$_{diff}$} & 0.05 & 0.13 & 0.13 & 0.30 & [-0.43, 0.72] &  & \\\hline
\end{tabular}
  \footnotesize{95\%CI: 95\% credible interval, ESS: Effective sample size}
  \end{center}
\end{table}

\begin{table}[H]
  \begin{center}
\caption{Results of mixed effects exponential model (12--19 group vs 20--30 group)\label{t:DHS2030_E_n}}
\begin{tabular}{|c|c|c|c|c|c|c|c|c|}\hline
\multicolumn{2}{|c|}{Parameter} & Mode & Median & Mean & SE & 95\%CI & Rhat & ESS\\\hline
\multicolumn{2}{|c|}{Intercept} & 4.35 & 4.43 & 4.44 & 0.26 & [3.96, 4.97] & 1.00 & 1230\\
\multicolumn{2}{|c|}{20--30 group} & 0.01 & 0.03 & 0.03 & 0.03 & [-0.04, 0.09] & 1.00 & 3905\\
\multicolumn{2}{|c|}{Sex} & 0.14 & 0.13 & 0.13 & 0.03 & [0.08, 0.19] & 1.00 & 3528\\
Place & Respondents home & 0.85 & 0.84 & 0.82 & 0.24 & [0.32, 1.28] & 1.01 & 1083\\
Place & Other home & 0.82 & 0.92 & 0.93 & 0.47 & [0.06, 1.89] & 1.00 & 2359\\
Place & Public sector & 1.08 & 1.08 & 1.07 & 0.27 & [0.52, 1.58] & 1.00 & 1231\\
Place & Government hospital & 0.93 & 0.95 & 0.94 & 0.24 & [0.44, 1.39] & 1.01 & 1085\\
Place & CS Govt health professional & 0.89 & 0.92 & 0.91 & 0.24 & [0.42, 1.35] & 1.01 & 1105\\
Place & Other public sector & 1.08 & 1.02 & 1.01 & 0.27 & [0.45, 1.53] & 1.00 & 1300\\
Place & Private hospital/clinic & 0.59 & 0.66 & 0.65 & 0.24 & [0.15, 1.10] & 1.01 & 1080\\
Place & CS private health facility & 1.20 & 1.03 & 1.02 & 0.29 & [0.42, 1.58] & 1.00 & 1385\\
Size & Very large & 1.15 & 1.16 & 1.16 & 0.08 & [1.00, 1.33] & 1.00 & 2092\\
Size & Larger than average & 1.22 & 1.22 & 1.22 & 0.08 & [1.07, 1.38] & 1.00 & 1949\\
Size & Average & 1.34 & 1.32 & 1.32 & 0.07 & [1.19, 1.45] & 1.00 & 1720\\
Size & Smaller than average & 0.97 & 0.94 & 0.94 & 0.08 & [0.79, 1.09] & 1.00 & 1944\\
Size & Very small & 0.16 & 0.17 & 0.17 & 0.08 & [0.00, 0.33] & 1.00 & 2105\\
\multicolumn{2}{|c|}{Order} & -0.02 & -0.02 & -0.02 & 0.01 & [-0.04, 0.00] & 1.00 & 4496\\
\multicolumn{2}{|c|}{$\phi$} & 0.17 & 0.17 & 0.18 & 0.07 & [0.09, 0.36] & 1.01 & 1099\\
\multicolumn{2}{|c|}{RMST$_{12-19 group}$} & 36.86 & 37.62 & 37.55 & 2.58 & [32.18, 42.23] &  & \\
\multicolumn{2}{|c|}{RMST$_{20-30 group}$} & 38.70 & 37.89 & 37.80 & 2.55 & [32.43, 42.44] &  & \\
\multicolumn{2}{|c|}{RMST$_{diff}$} & 0.19 & 0.25 & 0.25 & 0.32 & [-0.36, 0.89] &  & \\\hline
\end{tabular}
  \footnotesize{95\%CI: 95\% credible interval, ESS: Effective sample size}
  \end{center}
\end{table}

\begin{table}[H]
  \begin{center}
\caption{Results of exponential frailty model (12--19 group vs 20--30 group)\label{t:DHS2030_E_f}}
\begin{tabular}{|c|c|c|c|c|c|c|c|c|}\hline
\multicolumn{2}{|c|}{Parameter} & Mode & Median & Mean & SE & 95\%CI & Rhat & ESS\\\hline
\multicolumn{2}{|c|}{Intercept} & 4.39 & 4.44 & 4.45 & 0.26 & [3.96, 5.00] & 1.00 & 888\\
\multicolumn{2}{|c|}{20--30 group} & 0.02 & 0.02 & 0.02 & 0.03 & [-0.03, 0.08] & 1.00 & 4368\\
\multicolumn{2}{|c|}{Sex} & 0.13 & 0.13 & 0.13 & 0.03 & [0.08, 0.19] & 1.00 & 4676\\
Place & Respondents home & 0.82 & 0.84 & 0.83 & 0.25 & [0.30, 1.27] & 1.00 & 849\\
Place & Other home & 0.76 & 0.93 & 0.94 & 0.47 & [0.08, 1.91] & 1.00 & 2227\\
Place & Public sector & 1.12 & 1.08 & 1.07 & 0.27 & [0.51, 1.59] & 1.00 & 907\\
Place & Government hospital & 0.96 & 0.95 & 0.94 & 0.25 & [0.42, 1.39] & 1.00 & 833\\
Place & CS Govt health professional & 0.86 & 0.93 & 0.92 & 0.24 & [0.40, 1.36] & 1.00 & 838\\
Place & Other public sector & 1.06 & 1.02 & 1.02 & 0.27 & [0.47, 1.54] & 1.00 & 991\\
Place & Private hospital/clinic & 0.63 & 0.66 & 0.65 & 0.25 & [0.14, 1.10] & 1.00 & 835\\
Place & CS private health facility & 1.04 & 1.03 & 1.03 & 0.30 & [0.42, 1.60] & 1.00 & 1165\\
Size & Very large & 1.15 & 1.16 & 1.16 & 0.08 & [1.00, 1.33] & 1.00 & 1812\\
Size & Larger than average & 1.22 & 1.22 & 1.22 & 0.08 & [1.07, 1.37] & 1.00 & 1824\\
Size & Average & 1.33 & 1.31 & 1.31 & 0.07 & [1.18, 1.44] & 1.00 & 1576\\
Size & Smaller than average & 0.91 & 0.93 & 0.93 & 0.08 & [0.78, 1.09] & 1.00 & 1797\\
Size & Very small & 0.16 & 0.17 & 0.17 & 0.08 & [0.00, 0.33] & 1.00 & 1940\\
\multicolumn{2}{|c|}{Order} & -0.02 & -0.02 & -0.02 & 0.01 & [-0.04, 0.00] & 1.00 & 4882\\
\multicolumn{2}{|c|}{$\phi$} & 0.02 & 0.03 & 0.05 & 0.07 & [0.01, 0.19] & 1.00 & 1595\\
\multicolumn{2}{|c|}{RMST$_{12-19 group}$} & 38.25 & 37.80 & 37.66 & 2.61 & [32.21, 42.43] &  & \\
\multicolumn{2}{|c|}{RMST$_{20-30 group}$} & 38.60 & 38.06 & 37.91 & 2.57 & [32.59, 42.59] &  & \\
\multicolumn{2}{|c|}{RMST$_{diff}$} & 0.22 & 0.24 & 0.25 & 0.30 & [-0.32, 0.88] &  & \\\hline
\end{tabular}
  \footnotesize{95\%CI: 95\% credible interval, ESS: Effective sample size}
  \end{center}
\end{table}

\begin{table}[H]
  \begin{center}
\caption{Results of Weibull model (12--19 group vs 20--30 group)\label{t:DHS2030_W}}
\begin{tabular}{|c|c|c|c|c|c|c|c|c|}\hline
\multicolumn{2}{|c|}{Parameter} & Mode & Median & Mean & SE & 95\%CI & Rhat & ESS\\\hline
\multicolumn{2}{|c|}{Intercept} & 3.97 & 3.99 & 3.99 & 0.15 & [3.71, 4.28] & 1.00 & 415\\
\multicolumn{2}{|c|}{20--30 group} & 0.00 & -0.01 & -0.01 & 0.02 & [-0.04, 0.03] & 1.00 & 2078\\
\multicolumn{2}{|c|}{Sex} & 0.08 & 0.08 & 0.08 & 0.02 & [0.04, 0.11] & 1.00 & 1939\\
Place & Respondents home & 0.52 & 0.52 & 0.51 & 0.14 & [0.23, 0.79] & 1.00 & 377\\
Place & Other home & 0.54 & 0.56 & 0.57 & 0.27 & [0.09, 1.12] & 1.00 & 906\\
Place & Public sector & 0.67 & 0.67 & 0.66 & 0.16 & [0.34, 0.98] & 1.00 & 427\\
Place & Government hospital & 0.58 & 0.58 & 0.57 & 0.14 & [0.30, 0.85] & 1.00 & 371\\
Place & CS Govt health professional & 0.55 & 0.56 & 0.56 & 0.14 & [0.29, 0.83] & 1.00 & 374\\
Place & Other public sector & 0.58 & 0.60 & 0.60 & 0.16 & [0.29, 0.91] & 1.00 & 465\\
Place & Private hospital/clinic & 0.33 & 0.37 & 0.37 & 0.14 & [0.09, 0.65] & 1.00 & 377\\
Place & CS private health facility & 0.62 & 0.58 & 0.57 & 0.17 & [0.24, 0.92] & 1.00 & 504\\
Size & Very large & 0.65 & 0.66 & 0.66 & 0.05 & [0.57, 0.75] & 1.00 & 1019\\
Size & Larger than average & 0.71 & 0.70 & 0.70 & 0.05 & [0.61, 0.79] & 1.00 & 907\\
Size & Average & 0.77 & 0.77 & 0.76 & 0.04 & [0.69, 0.84] & 1.00 & 815\\
Size & Smaller than average & 0.54 & 0.53 & 0.53 & 0.05 & [0.44, 0.62] & 1.00 & 957\\
Size & Very small & 0.07 & 0.07 & 0.07 & 0.05 & [-0.03, 0.16] & 1.00 & 1106\\
\multicolumn{2}{|c|}{Order} & -0.02 & -0.02 & -0.02 & 0.01 & [-0.03, -0.01] & 1.00 & 2833\\
\multicolumn{2}{|c|}{$k$} & 1.70 & 1.70 & 1.70 & 0.02 & [1.66, 1.75] & 1.00 & 1049\\
\multicolumn{2}{|c|}{RMST$_{12-19 group}$} & 38.80 & 37.34 & 37.21 & 2.39 & [32.15, 41.57] &  & \\
\multicolumn{2}{|c|}{RMST$_{20-30 group}$} & 37.34 & 37.26 & 37.09 & 2.41 & [31.90, 41.44] &  & \\
\multicolumn{2}{|c|}{RMST$_{diff}$} & -0.13 & -0.12 & -0.12 & 0.31 & [-0.77, 0.51] &  & \\\hline
\end{tabular}
  \footnotesize{95\%CI: 95\% credible interval, ESS: Effective sample size}
  \end{center}
\end{table}

\begin{table}[H]
  \begin{center}
\caption{Results of Weibull frailty model (12--19 group vs 20--30 group)\label{t:DHS2030_W_f}}
\begin{tabular}{|c|c|c|c|c|c|c|c|c|}\hline
\multicolumn{2}{|c|}{Parameter} & Mode & Median & Mean & SE & 95\%CI & Rhat & ESS\\\hline
\multicolumn{2}{|c|}{Intercept} & 4.00 & 3.99 & 4.00 & 0.15 & [3.71, 4.30] & 1.00 & 489\\
\multicolumn{2}{|c|}{20--30 group} & 0.00 & 0.00 & 0.00 & 0.02 & [-0.04, 0.03] & 1.00 & 2376\\
\multicolumn{2}{|c|}{Sex} & 0.08 & 0.08 & 0.08 & 0.02 & [0.04, 0.11] & 1.00 & 2132\\
Place & Respondents home & 0.52 & 0.53 & 0.53 & 0.14 & [0.24, 0.80] & 1.00 & 476\\
Place & Other home & 0.46 & 0.58 & 0.59 & 0.27 & [0.08, 1.14] & 1.00 & 1090\\
Place & Public sector & 0.69 & 0.67 & 0.66 & 0.16 & [0.34, 0.96] & 1.00 & 534\\
Place & Government hospital & 0.67 & 0.57 & 0.57 & 0.14 & [0.28, 0.83] & 1.00 & 482\\
Place & CS Govt health professional & 0.51 & 0.56 & 0.56 & 0.14 & [0.27, 0.82] & 1.00 & 462\\
Place & Other public sector & 0.69 & 0.62 & 0.61 & 0.16 & [0.28, 0.92] & 1.00 & 515\\
Place & Private hospital/clinic & 0.38 & 0.39 & 0.39 & 0.14 & [0.09, 0.66] & 1.00 & 471\\
Place & CS private health facility & 0.60 & 0.61 & 0.61 & 0.18 & [0.27, 0.96] & 1.00 & 638\\
Size & Very large & 0.66 & 0.67 & 0.67 & 0.05 & [0.57, 0.77] & 1.00 & 878\\
Size & Larger than average & 0.72 & 0.70 & 0.70 & 0.05 & [0.61, 0.79] & 1.01 & 868\\
Size & Average & 0.77 & 0.76 & 0.76 & 0.04 & [0.68, 0.84] & 1.01 & 727\\
Size & Smaller than average & 0.52 & 0.53 & 0.53 & 0.05 & [0.44, 0.62] & 1.01 & 788\\
Size & Very small & 0.08 & 0.07 & 0.07 & 0.05 & [-0.02, 0.17] & 1.00 & 912\\
\multicolumn{2}{|c|}{Order} & -0.01 & -0.01 & -0.01 & 0.01 & [-0.02, 0.00] & 1.00 & 3125\\
\multicolumn{2}{|c|}{$k$} & 1.70 & 1.71 & 1.71 & 0.02 & [1.67, 1.75] & 1.00 & 2106\\
\multicolumn{2}{|c|}{$\phi$} & 0.02 & 0.04 & 0.05 & 0.06 & [0.01, 0.20] & 1.00 & 511\\
\multicolumn{2}{|c|}{RMST$_{12-19 group}$} & 37.33 & 37.37 & 37.33 & 2.51 & [32.20, 41.80] &  & \\
\multicolumn{2}{|c|}{RMST$_{20-30 group}$} & 36.53 & 37.37 & 37.31 & 2.51 & [32.10, 41.79] &  & \\
\multicolumn{2}{|c|}{RMST$_{diff}$} & 0.03 & -0.01 & -0.02 & 0.30 & [-0.62, 0.60] &  & \\\hline
\end{tabular}
  \footnotesize{95\%CI: 95\% credible interval, ESS: Effective sample size}
  \end{center}
\end{table}

\begin{table}[H]
  \begin{center}
\caption{Results of log-normal model (12--19 group vs 20--30 group)\label{t:DHS2030_LN}}
\begin{tabular}{|c|c|c|c|c|c|c|c|c|}\hline
\multicolumn{2}{|c|}{Parameter} & Mode & Median & Mean & SE & 95\%CI & Rhat & ESS\\\hline
\multicolumn{2}{|c|}{Intercept} & 4.44 & 4.48 & 4.49 & 0.23 & [4.04, 4.96] & 1.00 & 824\\
\multicolumn{2}{|c|}{20--30 group} & -0.01 & -0.02 & -0.02 & 0.02 & [-0.06, 0.02] & 1.00 & 3736\\
\multicolumn{2}{|c|}{Sex} & 0.10 & 0.11 & 0.11 & 0.02 & [0.06, 0.15] & 1.00 & 3710\\
Place & Respondents home & 0.71 & 0.62 & 0.62 & 0.23 & [0.14, 1.06] & 1.00 & 854\\
Place & Other home & 0.54 & 0.60 & 0.61 & 0.38 & [-0.12, 1.36] & 1.00 & 1651\\
Place & Public sector & 0.83 & 0.78 & 0.78 & 0.24 & [0.28, 1.26] & 1.00 & 915\\
Place & Government hospital & 0.78 & 0.73 & 0.72 & 0.23 & [0.25, 1.16] & 1.00 & 857\\
Place & CS Govt health professional & 0.79 & 0.70 & 0.70 & 0.23 & [0.23, 1.14] & 1.00 & 837\\
Place & Other public sector & 0.80 & 0.78 & 0.77 & 0.25 & [0.27, 1.24] & 1.00 & 927\\
Place & Private hospital/clinic & 0.54 & 0.47 & 0.46 & 0.23 & [-0.01, 0.90] & 1.00 & 842\\
Place & CS private health facility & 0.72 & 0.69 & 0.69 & 0.26 & [0.19, 1.20] & 1.00 & 1005\\
Size & Very large & 0.87 & 0.87 & 0.87 & 0.07 & [0.73, 1.02] & 1.01 & 1836\\
Size & Larger than average & 0.93 & 0.93 & 0.93 & 0.07 & [0.79, 1.07] & 1.01 & 1767\\
Size & Average & 1.02 & 1.02 & 1.02 & 0.06 & [0.89, 1.14] & 1.01 & 1577\\
Size & Smaller than average & 0.73 & 0.71 & 0.71 & 0.07 & [0.58, 0.85] & 1.00 & 1711\\
Size & Very small & 0.05 & 0.07 & 0.07 & 0.08 & [-0.07, 0.22] & 1.00 & 1896\\
\multicolumn{2}{|c|}{Order} & -0.02 & -0.02 & -0.02 & 0.01 & [-0.03, 0.00] & 1.00 & 4360\\
\multicolumn{2}{|c|}{$\si$} & 1.54 & 1.54 & 1.54 & 0.02 & [1.50, 1.57] & 1.00 & 2287\\
\multicolumn{2}{|c|}{RMST$_{12-19 group}$} & 40.63 & 40.39 & 40.28 & 1.88 & [36.31, 43.79] &  & \\
\multicolumn{2}{|c|}{RMST$_{20-30 group}$} & 40.57 & 40.18 & 40.11 & 1.90 & [36.11, 43.71] &  & \\
\multicolumn{2}{|c|}{RMST$_{diff}$} & -0.16 & -0.17 & -0.17 & 0.19 & [-0.55, 0.19] &  & \\\hline
\end{tabular}
  \footnotesize{95\%CI: 95\% credible interval, ESS: Effective sample size}
  \end{center}
\end{table}

\begin{table}[H]
  \begin{center}
\caption{Results of mixed effects log-normal model (12--19 group vs 20--30 group)\label{t:DHS2030_LN_n}}
\begin{tabular}{|c|c|c|c|c|c|c|c|c|}\hline
\multicolumn{2}{|c|}{Parameter} & Mode & Median & Mean & SE & 95\%CI & Rhat & ESS\\\hline
\multicolumn{2}{|c|}{Intercept} & 4.51 & 4.48 & 4.49 & 0.23 & [4.04, 4.97] & 1.00 & 1119\\
\multicolumn{2}{|c|}{20--30 group} & -0.02 & -0.01 & -0.01 & 0.02 & [-0.06, 0.03] & 1.00 & 3885\\
\multicolumn{2}{|c|}{Sex} & 0.12 & 0.11 & 0.11 & 0.02 & [0.07, 0.15] & 1.00 & 4243\\
Place & Respondents home & 0.58 & 0.65 & 0.65 & 0.22 & [0.22, 1.09] & 1.00 & 1074\\
Place & Other home & 0.53 & 0.64 & 0.66 & 0.37 & [-0.04, 1.39] & 1.00 & 2022\\
Place & Public sector & 0.73 & 0.79 & 0.79 & 0.24 & [0.33, 1.25] & 1.00 & 1125\\
Place & Government hospital & 0.70 & 0.73 & 0.73 & 0.22 & [0.29, 1.16] & 1.00 & 1083\\
Place & CS Govt health professional & 0.76 & 0.72 & 0.71 & 0.22 & [0.29, 1.15] & 1.00 & 1089\\
Place & Other public sector & 0.84 & 0.80 & 0.80 & 0.24 & [0.32, 1.26] & 1.00 & 1144\\
Place & Private hospital/clinic & 0.53 & 0.50 & 0.50 & 0.22 & [0.07, 0.92] & 1.00 & 1100\\
Place & CS private health facility & 0.73 & 0.75 & 0.75 & 0.25 & [0.26, 1.24] & 1.00 & 1299\\
Size & Very large & 0.89 & 0.89 & 0.89 & 0.08 & [0.74, 1.03] & 1.00 & 1815\\
Size & Larger than average & 0.93 & 0.93 & 0.93 & 0.07 & [0.79, 1.07] & 1.00 & 1830\\
Size & Average & 1.01 & 1.02 & 1.02 & 0.06 & [0.90, 1.14] & 1.00 & 1624\\
Size & Smaller than average & 0.74 & 0.72 & 0.72 & 0.07 & [0.58, 0.85] & 1.00 & 1914\\
Size & Very small & 0.04 & 0.07 & 0.08 & 0.08 & [-0.08, 0.23] & 1.00 & 2100\\
\multicolumn{2}{|c|}{Order} & -0.02 & -0.01 & -0.01 & 0.01 & [-0.03, 0.00] & 1.00 & 4405\\
\multicolumn{2}{|c|}{$\si$} & 1.54 & 1.53 & 1.53 & 0.02 & [1.50, 1.57] & 1.00 & 3455\\
\multicolumn{2}{|c|}{$\phi$} & 0.10 & 0.12 & 0.13 & 0.05 & [0.07, 0.25] & 1.00 & 1151\\
\multicolumn{2}{|c|}{RMST$_{12-19 group}$} & 40.62 & 40.40 & 40.31 & 1.88 & [36.43, 43.83] &  & \\
\multicolumn{2}{|c|}{RMST$_{20-30 group}$} & 40.24 & 40.29 & 40.21 & 1.88 & [36.27, 43.79] &  & \\
\multicolumn{2}{|c|}{RMST$_{diff}$} & -0.09 & -0.10 & -0.10 & 0.19 & [-0.48, 0.26] &  & \\\hline
\end{tabular}
  \footnotesize{95\%CI: 95\% credible interval, ESS: Effective sample size}
  \end{center}
\end{table}

\begin{table}[H]
  \begin{center}
\caption{Results of log-normal frailty model (12--19 group vs 20--30 group)\label{t:DHS2030_LN_f}}
\begin{tabular}{|c|c|c|c|c|c|c|c|c|}\hline
\multicolumn{2}{|c|}{Parameter} & Mode & Median & Mean & SE & 95\%CI & Rhat & ESS\\\hline
\multicolumn{2}{|c|}{Intercept} & 19.26 & 19.11 & 19.10 & 1.19 & [16.79, 21.42] & 1.00 & 790\\
\multicolumn{2}{|c|}{20--30 group} & 0.00 & 0.00 & 0.00 & 0.02 & [-0.04, 0.03] & 1.00 & 3775\\
\multicolumn{2}{|c|}{Sex} & 0.08 & 0.08 & 0.08 & 0.02 & [0.05, 0.12] & 1.00 & 3934\\
Place & Respondents home & 0.60 & 0.57 & 0.56 & 0.15 & [0.25, 0.85] & 1.00 & 1039\\
Place & Other home & 0.63 & 0.61 & 0.62 & 0.28 & [0.11, 1.18] & 1.00 & 1938\\
Place & Public sector & 0.80 & 0.71 & 0.71 & 0.17 & [0.36, 1.02] & 1.00 & 1171\\
Place & Government hospital & 0.58 & 0.62 & 0.61 & 0.16 & [0.30, 0.90] & 1.00 & 1022\\
Place & CS Govt health professional & 0.59 & 0.60 & 0.60 & 0.15 & [0.28, 0.89] & 1.00 & 1030\\
Place & Other public sector & 0.70 & 0.67 & 0.66 & 0.17 & [0.32, 0.99] & 1.00 & 1141\\
Place & Private hospital/clinic & 0.39 & 0.43 & 0.42 & 0.16 & [0.10, 0.71] & 1.00 & 1007\\
Place & CS private health facility & 0.61 & 0.65 & 0.65 & 0.19 & [0.27, 1.02] & 1.00 & 1319\\
Size & Very large & 0.72 & 0.72 & 0.72 & 0.05 & [0.61, 0.83] & 1.00 & 1505\\
Size & Larger than average & 0.75 & 0.75 & 0.75 & 0.05 & [0.65, 0.85] & 1.00 & 1565\\
Size & Average & 0.82 & 0.82 & 0.82 & 0.04 & [0.73, 0.90] & 1.00 & 1225\\
Size & Smaller than average & 0.59 & 0.57 & 0.57 & 0.05 & [0.47, 0.67] & 1.00 & 1509\\
Size & Very small & 0.07 & 0.08 & 0.08 & 0.06 & [-0.03, 0.18] & 1.00 & 1676\\
\multicolumn{2}{|c|}{Order} & -0.01 & -0.01 & -0.01 & 0.01 & [-0.02, 0.00] & 1.00 & 3479\\
\multicolumn{2}{|c|}{$\si$} & 3.27 & 3.27 & 3.27 & 0.11 & [3.05, 3.49] & 1.00 & 849\\
\multicolumn{2}{|c|}{$\phi$} & 131492 & 446651 & 841037 & 1300743 & [55426, 4005760] & 1.00 & 1007\\
\multicolumn{2}{|c|}{RMST$_{12-19 group}$} & 50.00 & 50.00 & 50.00 & 0.00 & [50.00, 50.00] &  & \\
\multicolumn{2}{|c|}{RMST$_{20-30 group}$} & 50.00 & 50.00 & 50.00 & 0.00 & [50.00, 50.00] &  & \\
\multicolumn{2}{|c|}{RMST$_{diff}$} & 0.00 & 0.00 & 0.00 & 0.00 & [0.00, 0.00] &  & \\\hline
\end{tabular}
  \footnotesize{95\%CI: 95\% credible interval, ESS: Effective sample size}
  \end{center}
\end{table}

The estimated results of the log-normal frailty model differ from the others; therefore, we examined the behavior using different data in Appendix \ref{App:LeukSurv}.

\begin{table}[H]
  \begin{center}
\caption{Results of log-logistic model (12--19 group vs 20--30 group)\label{t:DHS2030_LL}}
\begin{tabular}{|c|c|c|c|c|c|c|c|c|}\hline
\multicolumn{2}{|c|}{Parameter} & Mode & Median & Mean & SE & 95\%CI & Rhat & ESS\\\hline
\multicolumn{2}{|c|}{Intercept} & 4.01 & 4.01 & 4.02 & 0.16 & [3.73, 4.35] & 1.01 & 1059\\
\multicolumn{2}{|c|}{20--30 group} & -0.01 & -0.01 & -0.01 & 0.02 & [-0.04, 0.03] & 1.00 & 4416\\
\multicolumn{2}{|c|}{Sex} & 0.08 & 0.08 & 0.08 & 0.02 & [0.04, 0.11] & 1.00 & 4233\\
Place & Respondents home & 0.54 & 0.53 & 0.53 & 0.16 & [0.21, 0.81] & 1.00 & 997\\
Place & Other home & 0.55 & 0.55 & 0.56 & 0.29 & [0.01, 1.14] & 1.00 & 1979\\
Place & Public sector & 0.68 & 0.68 & 0.67 & 0.17 & [0.32, 0.99] & 1.00 & 1119\\
Place & Government hospital & 0.62 & 0.60 & 0.59 & 0.16 & [0.27, 0.88] & 1.00 & 1001\\
Place & CS Govt health professional & 0.56 & 0.58 & 0.58 & 0.16 & [0.26, 0.86] & 1.00 & 1003\\
Place & Other public sector & 0.62 & 0.62 & 0.62 & 0.17 & [0.26, 0.95] & 1.01 & 1197\\
Place & Private hospital/clinic & 0.45 & 0.40 & 0.39 & 0.16 & [0.07, 0.68] & 1.00 & 1009\\
Place & CS private health facility & 0.63 & 0.59 & 0.59 & 0.19 & [0.21, 0.93] & 1.00 & 1274\\
Size & Very large & 0.68 & 0.68 & 0.68 & 0.05 & [0.58, 0.78] & 1.00 & 1728\\
Size & Larger than average & 0.71 & 0.72 & 0.72 & 0.05 & [0.63, 0.82] & 1.00 & 1679\\
Size & Average & 0.79 & 0.78 & 0.78 & 0.04 & [0.70, 0.87] & 1.00 & 1451\\
Size & Smaller than average & 0.54 & 0.55 & 0.55 & 0.05 & [0.45, 0.65] & 1.00 & 1604\\
Size & Very small & 0.05 & 0.06 & 0.06 & 0.05 & [-0.04, 0.16] & 1.00 & 1938\\
\multicolumn{2}{|c|}{Order} & -0.02 & -0.02 & -0.02 & 0.01 & [-0.03, -0.01] & 1.00 & 4461\\
\multicolumn{2}{|c|}{$k$} & 1.73 & 1.73 & 1.73 & 0.02 & [1.69, 1.77] & 1.00 & 2020\\
\multicolumn{2}{|c|}{RMST$_{12-19 group}$} & 39.76 & 39.64 & 39.62 & 1.95 & [35.80, 43.30] &  & \\
\multicolumn{2}{|c|}{RMST$_{20-30 group}$} & 39.85 & 39.54 & 39.53 & 1.96 & [35.66, 43.23] &  & \\
\multicolumn{2}{|c|}{RMST$_{diff}$} & -0.02 & -0.09 & -0.09 & 0.22 & [-0.53, 0.34] &  & \\\hline
\end{tabular}
  \footnotesize{95\%CI: 95\% credible interval, ESS: Effective sample size}
  \end{center}
\end{table}

\begin{table}[H]
  \begin{center}
\caption{Results of mixed effects log-logistic model (12--19 group vs 20--30 group)\label{t:DHS2030_LL_n}}
\begin{tabular}{|c|c|c|c|c|c|c|c|c|}\hline
\multicolumn{2}{|c|}{Parameter} & Mode & Median & Mean & SE & 95\%CI & Rhat & ESS\\\hline
\multicolumn{2}{|c|}{Intercept} & 4.00 & 4.02 & 4.03 & 0.16 & [3.73, 4.36] & 1.01 & 1278\\
\multicolumn{2}{|c|}{20--30 group} & 0.00 & 0.00 & 0.00 & 0.02 & [-0.04, 0.03] & 1.00 & 3898\\
\multicolumn{2}{|c|}{Sex} & 0.08 & 0.08 & 0.08 & 0.02 & [0.04, 0.11] & 1.00 & 4324\\
Place & Respondents home & 0.61 & 0.56 & 0.55 & 0.15 & [0.24, 0.83] & 1.01 & 1214\\
Place & Other home & 0.46 & 0.58 & 0.60 & 0.28 & [0.04, 1.18] & 1.00 & 2242\\
Place & Public sector & 0.71 & 0.70 & 0.69 & 0.17 & [0.34, 1.01] & 1.00 & 1350\\
Place & Government hospital & 0.62 & 0.61 & 0.60 & 0.15 & [0.29, 0.89] & 1.00 & 1231\\
Place & CS Govt health professional & 0.57 & 0.60 & 0.59 & 0.15 & [0.28, 0.87] & 1.01 & 1220\\
Place & Other public sector & 0.68 & 0.65 & 0.64 & 0.17 & [0.30, 0.97] & 1.00 & 1394\\
Place & Private hospital/clinic & 0.45 & 0.43 & 0.42 & 0.15 & [0.11, 0.70] & 1.00 & 1228\\
Place & CS private health facility & 0.59 & 0.64 & 0.64 & 0.18 & [0.27, 0.99] & 1.00 & 1543\\
Size & Very large & 0.69 & 0.69 & 0.69 & 0.05 & [0.58, 0.79] & 1.00 & 1944\\
Size & Larger than average & 0.72 & 0.72 & 0.72 & 0.05 & [0.63, 0.82] & 1.00 & 1785\\
Size & Average & 0.77 & 0.78 & 0.78 & 0.04 & [0.70, 0.87] & 1.00 & 1599\\
Size & Smaller than average & 0.55 & 0.55 & 0.55 & 0.05 & [0.45, 0.65] & 1.00 & 1749\\
Size & Very small & 0.07 & 0.07 & 0.07 & 0.05 & [-0.04, 0.17] & 1.00 & 2246\\
\multicolumn{2}{|c|}{Order} & -0.01 & -0.01 & -0.01 & 0.01 & [-0.02, 0.00] & 1.00 & 4957\\
\multicolumn{2}{|c|}{$k$} & 1.74 & 1.73 & 1.73 & 0.02 & [1.69, 1.77] & 1.00 & 3740\\
\multicolumn{2}{|c|}{$\phi$} & 0.08 & 0.10 & 0.11 & 0.05 & [0.05, 0.23] & 1.00 & 1239\\
\multicolumn{2}{|c|}{RMST$_{12-19 group}$} & 39.48 & 39.74 & 39.70 & 1.96 & [35.73, 43.31] &  & \\
\multicolumn{2}{|c|}{RMST$_{20-30 group}$} & 39.67 & 39.71 & 39.68 & 1.95 & [35.69, 43.33] &  & \\
\multicolumn{2}{|c|}{RMST$_{diff}$} & 0.03 & -0.02 & -0.03 & 0.23 & [-0.48, 0.42] &  & \\\hline
\end{tabular}
  \footnotesize{95\%CI: 95\% credible interval, ESS: Effective sample size}
  \end{center}
\end{table}

\begin{table}[H]
  \begin{center}
\caption{Results of log-logistic frailty model (12--19 group vs 20--30 group)\label{t:DHS2030_LL_f}}
\begin{tabular}{|c|c|c|c|c|c|c|c|c|}\hline
\multicolumn{2}{|c|}{Parameter} & Mode & Median & Mean & SE & 95\%CI & Rhat & ESS\\\hline
\multicolumn{2}{|c|}{Intercept} & 3.17 & 3.18 & 3.18 & 0.21 & [2.76, 3.59] & 1.00 & 671\\
\multicolumn{2}{|c|}{20--30 group} & 0.00 & 0.00 & 0.00 & 0.02 & [-0.04, 0.03] & 1.00 & 3295\\
\multicolumn{2}{|c|}{Sex} & 0.07 & 0.08 & 0.08 & 0.02 & [0.04, 0.11] & 1.00 & 2520\\
Place & Respondents home & 0.58 & 0.55 & 0.55 & 0.14 & [0.27, 0.81] & 1.00 & 880\\
Place & Other home & 0.56 & 0.60 & 0.61 & 0.27 & [0.10, 1.17] & 1.00 & 1821\\
Place & Public sector & 0.71 & 0.69 & 0.69 & 0.16 & [0.39, 0.98] & 1.00 & 991\\
Place & Government hospital & 0.57 & 0.60 & 0.59 & 0.14 & [0.31, 0.85] & 1.00 & 864\\
Place & CS Govt health professional & 0.59 & 0.58 & 0.58 & 0.14 & [0.31, 0.84] & 1.00 & 870\\
Place & Other public sector & 0.62 & 0.64 & 0.63 & 0.16 & [0.32, 0.94] & 1.00 & 1024\\
Place & Private hospital/clinic & 0.43 & 0.42 & 0.41 & 0.14 & [0.13, 0.68] & 1.00 & 886\\
Place & CS private health facility & 0.70 & 0.64 & 0.63 & 0.18 & [0.28, 0.96] & 1.00 & 1112\\
Size & Very large & 0.69 & 0.67 & 0.67 & 0.05 & [0.57, 0.77] & 1.00 & 1762\\
Size & Larger than average & 0.70 & 0.71 & 0.71 & 0.05 & [0.61, 0.80] & 1.00 & 1586\\
Size & Average & 0.76 & 0.77 & 0.77 & 0.04 & [0.68, 0.84] & 1.00 & 1360\\
Size & Smaller than average & 0.52 & 0.54 & 0.54 & 0.05 & [0.44, 0.63] & 1.00 & 1516\\
Size & Very small & 0.08 & 0.07 & 0.07 & 0.05 & [-0.02, 0.17] & 1.00 & 1700\\
\multicolumn{2}{|c|}{Order} & -0.01 & -0.01 & -0.01 & 0.01 & [-0.02, 0.00] & 1.00 & 3336\\
\multicolumn{2}{|c|}{$k$} & 1.71 & 1.71 & 1.71 & 0.02 & [1.66, 1.75] & 1.00 & 2485\\
\multicolumn{2}{|c|}{$\phi$} & 7.22 & 9.51 & 11.58 & 8.24 & [3.00, 32.40] & 1.00 & 997\\
\multicolumn{2}{|c|}{RMST$_{12-19 group}$} & 26.85 & 27.13 & 27.16 & 3.36 & [20.60, 33.57] &  & \\
\multicolumn{2}{|c|}{RMST$_{20-30 group}$} & 26.75 & 27.13 & 27.14 & 3.36 & [20.47, 33.55] &  & \\
\multicolumn{2}{|c|}{RMST$_{diff}$} & 0.07 & -0.02 & -0.02 & 0.28 & [-0.57, 0.53] &  & \\\hline
\end{tabular}
  \footnotesize{95\%CI: 95\% credible interval, ESS: Effective sample size}
  \end{center}
\end{table}

\subsubsection{Estimation results (12--19 group vs 31+ group)}\label{App:Supp_DHS31}
\begin{table}[H]
  \begin{center}
\caption{Results of exponential model (12--19 group vs 31-- group)\label{t:DHS31_E}}
\begin{tabular}{|c|c|c|c|c|c|c|c|c|}\hline
\multicolumn{2}{|c|}{Parameter} & Mode & Median & Mean & SE & 95\%CI & Rhat & ESS\\\hline
\multicolumn{2}{|c|}{Intercept} & 4.46 & 4.45 & 4.47 & 0.34 & [3.86, 5.18] & 1.00 & 908\\
\multicolumn{2}{|c|}{31+ group} & -0.37 & -0.37 & -0.37 & 0.12 & [-0.60, -0.13] & 1.00 & 4149\\
\multicolumn{2}{|c|}{Sex} & 0.09 & 0.10 & 0.10 & 0.04 & [0.01, 0.19] & 1.00 & 4255\\
Place & Respondents home & 0.89 & 0.84 & 0.83 & 0.33 & [0.15, 1.45] & 1.00 & 825\\
Place & Other home & 0.75 & 0.60 & 0.61 & 0.61 & [-0.48, 1.87] & 1.00 & 1817\\
Place & Public sector & 1.31 & 1.39 & 1.39 & 0.37 & [0.63, 2.09] & 1.00 & 970\\
Place & Government hospital & 1.02 & 0.97 & 0.96 & 0.33 & [0.27, 1.58] & 1.00 & 801\\
Place & CS Govt health professional & 1.05 & 1.01 & 1.00 & 0.33 & [0.32, 1.61] & 1.01 & 790\\
Place & Other public sector & 1.26 & 1.30 & 1.29 & 0.40 & [0.48, 2.05] & 1.00 & 1003\\
Place & Private hospital/clinic & 0.61 & 0.58 & 0.57 & 0.33 & [-0.11, 1.18] & 1.01 & 800\\
Place & CS private health facility & 1.34 & 1.49 & 1.50 & 0.50 & [0.51, 2.50] & 1.00 & 1414\\
Size & Very large & 0.82 & 0.83 & 0.83 & 0.12 & [0.59, 1.06] & 1.00 & 1750\\
Size & Larger than average & 1.09 & 1.05 & 1.05 & 0.12 & [0.82, 1.29] & 1.00 & 1635\\
Size & Average & 1.19 & 1.17 & 1.17 & 0.10 & [0.98, 1.36] & 1.00 & 1406\\
Size & Smaller than average & 0.92 & 0.92 & 0.91 & 0.12 & [0.68, 1.15] & 1.00 & 1721\\
Size & Very small & 0.10 & 0.07 & 0.07 & 0.13 & [-0.18, 0.32] & 1.00 & 1846\\
\multicolumn{2}{|c|}{Order} & -0.01 & -0.01 & -0.01 & 0.01 & [-0.04, 0.02] & 1.00 & 4436\\
\multicolumn{2}{|c|}{RMST$_{12-19 group}$} & 37.98 & 37.90 & 37.71 & 3.26 & [30.91, 43.56] &  & \\
\multicolumn{2}{|c|}{RMST$_{31+ group}$} & 34.19 & 33.73 & 33.60 & 4.13 & [25.25, 41.10] &  & \\
\multicolumn{2}{|c|}{RMST$_{diff}$} & -3.90 & -4.05 & -4.11 & 1.58 & [-7.38, -1.23] &  & \\\hline
\end{tabular}
  \footnotesize{95\%CI: 95\% credible interval, ESS: Effective sample size}
  \end{center}
\end{table}

\begin{table}[H]
  \begin{center}
\caption{Results of mixed effects exponential model (12--19 group vs 31+ group)\label{t:DHS31_E_n}}
\begin{tabular}{|c|c|c|c|c|c|c|c|c|}\hline
\multicolumn{2}{|c|}{Parameter} & Mode & Median & Mean & SE & 95\%CI & Rhat & ESS\\\hline
\multicolumn{2}{|c|}{Intercept} & 4.58 & 4.49 & 4.50 & 0.34 & [3.85, 5.17] & 1.00 & 1131\\
\multicolumn{2}{|c|}{31+ group} & -0.38 & -0.36 & -0.36 & 0.12 & [-0.59, -0.12] & 1.00 & 5588\\
\multicolumn{2}{|c|}{Sex} & 0.11 & 0.10 & 0.10 & 0.04 & [0.01, 0.18] & 1.00 & 5349\\
Place & Respondents home & 0.94 & 0.86 & 0.85 & 0.32 & [0.22, 1.45] & 1.00 & 1057\\
Place & Other home & 0.45 & 0.62 & 0.65 & 0.64 & [-0.51, 1.96] & 1.00 & 2017\\
Place & Public sector & 1.56 & 1.39 & 1.38 & 0.38 & [0.64, 2.09] & 1.00 & 1260\\
Place & Government hospital & 0.82 & 0.96 & 0.96 & 0.32 & [0.32, 1.56] & 1.00 & 1056\\
Place & CS Govt health professional & 1.04 & 1.00 & 1.00 & 0.32 & [0.36, 1.60] & 1.00 & 1050\\
Place & Other public sector & 1.37 & 1.27 & 1.27 & 0.40 & [0.50, 2.06] & 1.00 & 1390\\
Place & Private hospital/clinic & 0.50 & 0.62 & 0.61 & 0.32 & [-0.03, 1.20] & 1.00 & 1044\\
Place & CS private health facility & 1.44 & 1.56 & 1.57 & 0.52 & [0.61, 2.63] & 1.00 & 1807\\
Size & Very large & 0.84 & 0.83 & 0.83 & 0.12 & [0.59, 1.08] & 1.00 & 2446\\
Size & Larger than average & 1.09 & 1.06 & 1.05 & 0.12 & [0.82, 1.28] & 1.00 & 2195\\
Size & Average & 1.18 & 1.16 & 1.16 & 0.10 & [0.97, 1.34] & 1.00 & 1851\\
Size & Smaller than average & 0.88 & 0.91 & 0.91 & 0.12 & [0.67, 1.14] & 1.00 & 2265\\
Size & Very small & 0.07 & 0.08 & 0.09 & 0.13 & [-0.17, 0.33] & 1.00 & 2395\\
\multicolumn{2}{|c|}{Order} & 0.00 & 0.00 & 0.00 & 0.01 & [-0.03, 0.03] & 1.00 & 5932\\
\multicolumn{2}{|c|}{$\phi$} & 0.15 & 0.17 & 0.19 & 0.08 & [0.09, 0.40] & 1.00 & 1223\\
\multicolumn{2}{|c|}{RMST$_{12-19 group}$} & 39.08 & 38.28 & 37.97 & 3.32 & [30.74, 43.51] &  & \\
\multicolumn{2}{|c|}{RMST$_{31+ group}$} & 35.00 & 34.36 & 34.08 & 4.16 & [25.26, 41.36] &  & \\
\multicolumn{2}{|c|}{RMST$_{diff}$} & -3.99 & -3.83 & -3.88 & 1.54 & [-7.25, -1.08] &  & \\\hline
\end{tabular}
  \footnotesize{95\%CI: 95\% credible interval, ESS: Effective sample size}
  \end{center}
\end{table}

\begin{table}[H]
  \begin{center}
\caption{Results of exponential frailty model (12--19 group vs 31+ group)\label{t:DHS31_E_f}}
\begin{tabular}{|c|c|c|c|c|c|c|c|c|}\hline
\multicolumn{2}{|c|}{Parameter} & Mode & Median & Mean & SE & 95\%CI & Rhat & ESS\\\hline
\multicolumn{2}{|c|}{Intercept} & 4.45 & 4.49 & 4.51 & 0.35 & [3.86, 5.27] & 1.00 & 865\\
\multicolumn{2}{|c|}{31+ group} & -0.37 & -0.36 & -0.36 & 0.12 & [-0.58, -0.12] & 1.00 & 3971\\
\multicolumn{2}{|c|}{Sex} & 0.09 & 0.10 & 0.10 & 0.04 & [0.01, 0.19] & 1.00 & 4744\\
Place & Respondents home & 0.91 & 0.87 & 0.85 & 0.32 & [0.15, 1.44] & 1.01 & 752\\
Place & Other home & 0.46 & 0.60 & 0.63 & 0.62 & [-0.49, 1.94] & 1.00 & 1789\\
Place & Public sector & 1.37 & 1.40 & 1.39 & 0.37 & [0.59, 2.09] & 1.00 & 933\\
Place & Government hospital & 0.99 & 0.98 & 0.96 & 0.32 & [0.25, 1.56] & 1.01 & 749\\
Place & CS Govt health professional & 1.13 & 1.02 & 1.00 & 0.32 & [0.29, 1.59] & 1.01 & 751\\
Place & Other public sector & 1.36 & 1.29 & 1.27 & 0.40 & [0.45, 2.04] & 1.00 & 976\\
Place & Private hospital/clinic & 0.66 & 0.63 & 0.61 & 0.33 & [-0.11, 1.20] & 1.01 & 758\\
Place & CS private health facility & 1.52 & 1.56 & 1.57 & 0.51 & [0.61, 2.62] & 1.00 & 1308\\
Size & Very large & 0.85 & 0.84 & 0.83 & 0.12 & [0.59, 1.08] & 1.00 & 1921\\
Size & Larger than average & 1.09 & 1.06 & 1.06 & 0.12 & [0.81, 1.29] & 1.00 & 1813\\
Size & Average & 1.14 & 1.17 & 1.17 & 0.10 & [0.97, 1.36] & 1.00 & 1428\\
Size & Smaller than average & 0.89 & 0.92 & 0.92 & 0.12 & [0.68, 1.15] & 1.00 & 1685\\
Size & Very small & 0.07 & 0.09 & 0.09 & 0.13 & [-0.16, 0.34] & 1.00 & 1881\\
\multicolumn{2}{|c|}{Order} & 0.00 & 0.00 & 0.00 & 0.01 & [-0.03, 0.03] & 1.00 & 4536\\
\multicolumn{2}{|c|}{$\phi$} & 0.02 & 0.04 & 0.06 & 0.06 & [0.01, 0.21] & 1.00 & 1235\\
\multicolumn{2}{|c|}{RMST$_{12-19 group}$} & 37.92 & 38.27 & 38.11 & 3.28 & [30.90, 44.08] &  & \\
\multicolumn{2}{|c|}{RMST$_{31+ group}$} & 34.31 & 34.37 & 34.27 & 4.12 & [25.71, 42.05] &  & \\
\multicolumn{2}{|c|}{RMST$_{diff}$} & -4.11 & -3.81 & -3.84 & 1.52 & [-7.06, -1.08] &  & \\\hline
\end{tabular}
  \footnotesize{95\%CI: 95\% credible interval, ESS: Effective sample size}
  \end{center}
\end{table}

\begin{table}[H]
  \begin{center}
\caption{Results of Weibull model (12--19 group vs 31+ group)\label{t:DHS31_W}}
\begin{tabular}{|c|c|c|c|c|c|c|c|c|}\hline
\multicolumn{2}{|c|}{Parameter} & Mode & Median & Mean & SE & 95\%CI & Rhat & ESS\\\hline
\multicolumn{2}{|c|}{Intercept} & 4.01 & 4.01 & 4.02 & 0.19 & [3.68, 4.42] & 1.00 & 1363\\
\multicolumn{2}{|c|}{31+ group} & -0.22 & -0.22 & -0.22 & 0.07 & [-0.35, -0.08] & 1.00 & 2631\\
\multicolumn{2}{|c|}{Sex} & 0.05 & 0.05 & 0.05 & 0.02 & [0.00, 0.10] & 1.00 & 2900\\
Place & Respondents home & 0.44 & 0.47 & 0.47 & 0.18 & [0.09, 0.79] & 1.00 & 1341\\
Place & Other home & 0.35 & 0.33 & 0.34 & 0.34 & [-0.28, 1.04] & 1.00 & 2649\\
Place & Public sector & 0.75 & 0.77 & 0.77 & 0.21 & [0.34, 1.15] & 1.01 & 1898\\
Place & Government hospital & 0.56 & 0.52 & 0.52 & 0.18 & [0.15, 0.85] & 1.00 & 1310\\
Place & CS Govt health professional & 0.55 & 0.55 & 0.54 & 0.18 & [0.17, 0.86] & 1.00 & 1295\\
Place & Other public sector & 0.70 & 0.70 & 0.70 & 0.22 & [0.26, 1.11] & 1.00 & 1774\\
Place & Private hospital/clinic & 0.33 & 0.31 & 0.30 & 0.18 & [-0.08, 0.63] & 1.01 & 1378\\
Place & CS private health facility & 0.76 & 0.83 & 0.83 & 0.28 & [0.29, 1.38] & 1.00 & 2328\\
Size & Very large & 0.43 & 0.44 & 0.44 & 0.07 & [0.30, 0.57] & 1.00 & 2547\\
Size & Larger than average & 0.56 & 0.56 & 0.56 & 0.07 & [0.43, 0.69] & 1.00 & 2530\\
Size & Average & 0.63 & 0.63 & 0.63 & 0.05 & [0.52, 0.74] & 1.00 & 2229\\
Size & Smaller than average & 0.49 & 0.48 & 0.48 & 0.07 & [0.35, 0.61] & 1.00 & 2266\\
Size & Very small & 0.05 & 0.02 & 0.02 & 0.07 & [-0.12, 0.15] & 1.00 & 2751\\
\multicolumn{2}{|c|}{Order} & -0.01 & -0.01 & -0.01 & 0.01 & [-0.02, 0.01] & 1.00 & 2858\\
\multicolumn{2}{|c|}{$k$} & 1.82 & 1.83 & 1.83 & 0.04 & [1.76, 1.90] & 1.00 & 2280\\
\multicolumn{2}{|c|}{RMST$_{12-19 group}$} & 38.80 & 38.41 & 38.26 & 2.98 & [32.08, 43.72] &  & \\
\multicolumn{2}{|c|}{RMST$_{31+ group}$} & 34.90 & 34.44 & 34.31 & 3.72 & [27.19, 41.21] &  & \\
\multicolumn{2}{|c|}{RMST$_{diff}$} & -3.68 & -3.92 & -3.96 & 1.40 & [-6.74, -1.36] &  & \\\hline
\end{tabular}
  \footnotesize{95\%CI: 95\% credible interval, ESS: Effective sample size}
  \end{center}
\end{table}

\begin{table}[H]
  \begin{center}
\caption{Results of mixed effects Weibull model (12--19 group vs 31+ group)\label{t:DHS31_W_n}}
\begin{tabular}{|c|c|c|c|c|c|c|c|c|}\hline
\multicolumn{2}{|c|}{Parameter} & Mode & Median & Mean & SE & 95\%CI & Rhat & ESS\\\hline
\multicolumn{2}{|c|}{Intercept} & 4.09 & 4.03 & 4.04 & 0.19 & [3.69, 4.43] & 1.00 & 1059\\
\multicolumn{2}{|c|}{31+ group} & -0.22 & -0.21 & -0.21 & 0.07 & [-0.34, -0.08] & 1.00 & 4577\\
\multicolumn{2}{|c|}{Sex} & 0.05 & 0.05 & 0.05 & 0.02 & [0.00, 0.10] & 1.00 & 4768\\
Place & Respondents home & 0.48 & 0.49 & 0.48 & 0.18 & [0.10, 0.81] & 1.00 & 950\\
Place & Other home & 0.38 & 0.34 & 0.35 & 0.34 & [-0.25, 1.05] & 1.00 & 2002\\
Place & Public sector & 0.83 & 0.77 & 0.76 & 0.21 & [0.34, 1.14] & 1.00 & 1142\\
Place & Government hospital & 0.53 & 0.53 & 0.52 & 0.18 & [0.14, 0.85] & 1.00 & 956\\
Place & CS Govt health professional & 0.59 & 0.55 & 0.54 & 0.18 & [0.16, 0.86] & 1.00 & 940\\
Place & Other public sector & 0.63 & 0.68 & 0.68 & 0.22 & [0.24, 1.09] & 1.00 & 1304\\
Place & Private hospital/clinic & 0.29 & 0.33 & 0.32 & 0.18 & [-0.06, 0.65] & 1.00 & 961\\
Place & CS private health facility & 0.86 & 0.86 & 0.87 & 0.28 & [0.33, 1.43] & 1.00 & 1719\\
Size & Very large & 0.44 & 0.44 & 0.44 & 0.07 & [0.30, 0.58] & 1.00 & 1917\\
Size & Larger than average & 0.56 & 0.56 & 0.56 & 0.07 & [0.43, 0.69] & 1.00 & 1870\\
Size & Average & 0.62 & 0.63 & 0.63 & 0.06 & [0.52, 0.73] & 1.00 & 1552\\
Size & Smaller than average & 0.50 & 0.48 & 0.48 & 0.07 & [0.35, 0.62] & 1.00 & 1855\\
Size & Very small & 0.02 & 0.02 & 0.02 & 0.07 & [-0.11, 0.16] & 1.00 & 2102\\
\multicolumn{2}{|c|}{Order} & 0.00 & 0.00 & 0.00 & 0.01 & [-0.02, 0.01] & 1.00 & 5911\\
\multicolumn{2}{|c|}{$k$} & 1.82 & 1.82 & 1.83 & 0.03 & [1.76, 1.89] & 1.00 & 3812\\
\multicolumn{2}{|c|}{$\phi$} & 0.08 & 0.09 & 0.10 & 0.04 & [0.05, 0.21] & 1.00 & 1389\\
\multicolumn{2}{|c|}{RMST$_{12-19 group}$} & 39.57 & 38.76 & 38.59 & 3.00 & [32.22, 43.90] &  & \\
\multicolumn{2}{|c|}{RMST$_{31+ group}$} & 35.80 & 34.97 & 34.88 & 3.75 & [27.03, 41.86] &  & \\
\multicolumn{2}{|c|}{RMST$_{diff}$} & -3.88 & -3.66 & -3.71 & 1.40 & [-6.55, -1.18] &  & \\\hline
\end{tabular}
  \footnotesize{95\%CI: 95\% credible interval, ESS: Effective sample size}
  \end{center}
\end{table}

\begin{table}[H]
  \begin{center}
\caption{Results of log-normal model (12--19 group vs 31+ group)\label{t:DHS31_LN}}
\begin{tabular}{|c|c|c|c|c|c|c|c|c|}\hline
\multicolumn{2}{|c|}{Parameter} & Mode & Median & Mean & SE & 95\%CI & Rhat & ESS\\\hline
\multicolumn{2}{|c|}{Intercept} & 4.41 & 4.44 & 4.45 & 0.29 & [3.92, 5.04] & 1.00 & 980\\
\multicolumn{2}{|c|}{31-- group} & -0.34 & -0.33 & -0.33 & 0.09 & [-0.51, -0.15] & 1.00 & 3741\\
\multicolumn{2}{|c|}{Sex} & 0.07 & 0.07 & 0.07 & 0.03 & [0.00, 0.13] & 1.00 & 4148\\
Place & Respondents home & 0.59 & 0.65 & 0.64 & 0.28 & [0.08, 1.18] & 1.00 & 978\\
Place & Other home & 0.19 & 0.33 & 0.34 & 0.49 & [-0.59, 1.32] & 1.00 & 1937\\
Place & Public sector & 1.07 & 1.04 & 1.04 & 0.31 & [0.42, 1.64] & 1.00 & 1086\\
Place & Government hospital & 0.65 & 0.71 & 0.71 & 0.28 & [0.13, 1.23] & 1.00 & 984\\
Place & CS Govt health professional & 0.81 & 0.75 & 0.75 & 0.28 & [0.18, 1.27] & 1.00 & 973\\
Place & Other public sector & 0.97 & 0.90 & 0.90 & 0.32 & [0.27, 1.49] & 1.00 & 1154\\
Place & Private hospital/clinic & 0.41 & 0.43 & 0.42 & 0.28 & [-0.16, 0.95] & 1.00 & 1001\\
Place & CS private health facility & 1.01 & 0.94 & 0.94 & 0.37 & [0.22, 1.65] & 1.00 & 1415\\
Size & Very large & 0.58 & 0.57 & 0.57 & 0.10 & [0.37, 0.76] & 1.00 & 1615\\
Size & Larger than average & 0.73 & 0.76 & 0.76 & 0.10 & [0.56, 0.95] & 1.00 & 1598\\
Size & Average & 0.87 & 0.84 & 0.84 & 0.09 & [0.67, 1.01] & 1.00 & 1409\\
Size & Smaller than average & 0.69 & 0.66 & 0.66 & 0.10 & [0.46, 0.85] & 1.00 & 1596\\
Size & Very small & -0.02 & 0.00 & 0.00 & 0.11 & [-0.21, 0.21] & 1.00 & 1748\\
\multicolumn{2}{|c|}{Order} & -0.01 & -0.01 & -0.01 & 0.01 & [-0.03, 0.01] & 1.00 & 4046\\
\multicolumn{2}{|c|}{$\si$} & 1.42 & 1.43 & 1.43 & 0.02 & [1.38, 1.48] & 1.00 & 2188\\
\multicolumn{2}{|c|}{RMST$_{12-19 group}$} & 41.05 & 40.67 & 40.58 & 2.40 & [35.60, 44.96] &  & \\
\multicolumn{2}{|c|}{RMST$_{31+ group}$} & 37.32 & 37.64 & 37.57 & 2.91 & [31.65, 43.06] &  & \\
\multicolumn{2}{|c|}{RMST$_{diff}$} & -2.81 & -2.98 & -3.01 & 0.97 & [-4.98, -1.26] &  & \\\hline
\end{tabular}
  \footnotesize{95\%CI: 95\% credible interval, ESS: Effective sample size}
  \end{center}
\end{table}

\begin{table}[H]
  \begin{center}
\caption{Results of mixed effects log-normal model (12--19 group vs 31+ group)\label{t:DHS31_LN_n}}
\begin{tabular}{|c|c|c|c|c|c|c|c|c|}\hline
\multicolumn{2}{|c|}{Parameter} & Mode & Median & Mean & SE & 95\%CI & Rhat & ESS\\\hline
\multicolumn{2}{|c|}{Intercept} & 4.44 & 4.47 & 4.47 & 0.30 & [3.90, 5.09] & 1.00 & 1041\\
\multicolumn{2}{|c|}{31+ group} & -0.31 & -0.32 & -0.32 & 0.09 & [-0.49, -0.15] & 1.00 & 4358\\
\multicolumn{2}{|c|}{Sex} & 0.07 & 0.07 & 0.07 & 0.03 & [0.00, 0.13] & 1.00 & 4494\\
Place & Respondents home & 0.76 & 0.66 & 0.66 & 0.29 & [0.06, 1.21] & 1.00 & 978\\
Place & Other home & 0.27 & 0.36 & 0.37 & 0.48 & [-0.53, 1.32] & 1.00 & 1830\\
Place & Public sector & 1.00 & 1.03 & 1.03 & 0.32 & [0.40, 1.64] & 1.00 & 1038\\
Place & Government hospital & 0.75 & 0.71 & 0.70 & 0.29 & [0.11, 1.26] & 1.00 & 985\\
Place & CS Govt health professional & 0.84 & 0.75 & 0.75 & 0.29 & [0.16, 1.29] & 1.00 & 973\\
Place & Other public sector & 0.87 & 0.86 & 0.86 & 0.33 & [0.19, 1.49] & 1.00 & 1160\\
Place & Private hospital/clinic & 0.50 & 0.45 & 0.45 & 0.29 & [-0.14, 1.00] & 1.00 & 979\\
Place & CS private health facility & 1.16 & 1.00 & 1.00 & 0.38 & [0.28, 1.76] & 1.00 & 1430\\
Size & Very large & 0.55 & 0.57 & 0.57 & 0.10 & [0.36, 0.76] & 1.00 & 1879\\
Size & Larger than average & 0.76 & 0.76 & 0.76 & 0.10 & [0.56, 0.95] & 1.00 & 1772\\
Size & Average & 0.85 & 0.84 & 0.84 & 0.09 & [0.67, 1.00] & 1.00 & 1569\\
Size & Smaller than average & 0.66 & 0.67 & 0.67 & 0.10 & [0.47, 0.86] & 1.00 & 1840\\
Size & Very small & 0.03 & 0.00 & 0.00 & 0.11 & [-0.20, 0.21] & 1.00 & 2041\\
\multicolumn{2}{|c|}{Order} & 0.00 & 0.00 & 0.00 & 0.01 & [-0.02, 0.02] & 1.00 & 4585\\
\multicolumn{2}{|c|}{$\si$} & 1.44 & 1.43 & 1.43 & 0.02 & [1.38, 1.48] & 1.00 & 3286\\
\multicolumn{2}{|c|}{$\phi$} & 0.13 & 0.12 & 0.13 & 0.05 & [0.06, 0.27] & 1.00 & 1172\\
\multicolumn{2}{|c|}{RMST$_{12-19 group}$} & 40.69 & 40.90 & 40.76 & 2.50 & [35.51, 45.27] &  & \\
\multicolumn{2}{|c|}{RMST$_{31+ group}$} & 39.25 & 38.00 & 37.89 & 3.02 & [31.78, 43.53] &  & \\
\multicolumn{2}{|c|}{RMST$_{diff}$} & -2.66 & -2.83 & -2.87 & 0.95 & [-4.81, -1.17] &  & \\\hline
\end{tabular}
  \footnotesize{95\%CI: 95\% credible interval, ESS: Effective sample size}
  \end{center}
\end{table}

\begin{table}[H]
  \begin{center}
\caption{Results of log-normal frailty model (12--19 group vs 31+ group)\label{t:DHS31_LN_f}}
\begin{tabular}{|c|c|c|c|c|c|c|c|c|}\hline
\multicolumn{2}{|c|}{Parameter} & Mode & Median & Mean & SE & 95\%CI & Rhat & ESS\\\hline
\multicolumn{2}{|c|}{Intercept} & 12.37 & 12.49 & 12.53 & 0.90 & [10.81, 14.43] & 1.00 & 713\\
\multicolumn{2}{|c|}{31+ group} & -0.25 & -0.25 & -0.25 & 0.07 & [-0.38, -0.10] & 1.00 & 3585\\
\multicolumn{2}{|c|}{Sex} & 0.05 & 0.06 & 0.06 & 0.03 & [0.00, 0.11] & 1.00 & 3723\\
Place & Respondents home & 0.59 & 0.55 & 0.54 & 0.20 & [0.11, 0.91] & 1.00 & 959\\
Place & Other home & 0.32 & 0.37 & 0.39 & 0.39 & [-0.33, 1.19] & 1.00 & 2026\\
Place & Public sector & 0.89 & 0.87 & 0.86 & 0.23 & [0.40, 1.30] & 1.00 & 1136\\
Place & Government hospital & 0.55 & 0.59 & 0.58 & 0.20 & [0.16, 0.96] & 1.00 & 966\\
Place & CS Govt health professional & 0.63 & 0.62 & 0.61 & 0.20 & [0.19, 0.98] & 1.00 & 964\\
Place & Other public sector & 0.81 & 0.76 & 0.75 & 0.25 & [0.25, 1.24] & 1.00 & 1244\\
Place & Private hospital/clinic & 0.37 & 0.37 & 0.37 & 0.20 & [-0.07, 0.73] & 1.00 & 975\\
Place & CS private health facility & 0.94 & 0.91 & 0.92 & 0.30 & [0.33, 1.53] & 1.00 & 1467\\
Size & Very large & 0.49 & 0.49 & 0.48 & 0.08 & [0.33, 0.63] & 1.00 & 1863\\
Size & Larger than average & 0.64 & 0.62 & 0.62 & 0.08 & [0.47, 0.77] & 1.00 & 1858\\
Size & Average & 0.68 & 0.69 & 0.69 & 0.06 & [0.56, 0.82] & 1.00 & 1645\\
Size & Smaller than average & 0.56 & 0.54 & 0.54 & 0.08 & [0.39, 0.69] & 1.00 & 1807\\
Size & Very small & 0.03 & 0.02 & 0.02 & 0.08 & [-0.14, 0.18] & 1.00 & 1745\\
\multicolumn{2}{|c|}{Order} & 0.00 & 0.00 & 0.00 & 0.01 & [-0.02, 0.02] & 1.00 & 3913\\
\multicolumn{2}{|c|}{$k$} & 2.50 & 2.50 & 2.51 & 0.10 & [2.31, 2.72] & 1.00 & 835\\
\multicolumn{2}{|c|}{$\si$} & 1187 & 2298 & 3696 & 4759 & [433, 15736] & 1.00 & 890\\
\multicolumn{2}{|c|}{RMST$_{12-19 group}$} & 50.00 & 50.00 & 50.00 & 0.00 & [50.00, 50.00] &  & \\
\multicolumn{2}{|c|}{RMST$_{31+ group}$} & 50.00 & 50.00 & 50.00 & 0.00 & [50.00, 50.00] &  & \\
\multicolumn{2}{|c|}{RMST$_{diff}$} & 0.00 & 0.00 & 0.00 & 0.00 & [0.00, 0.00] &  & \\\hline
\end{tabular}
  \footnotesize{95\%CI: 95\% credible interval, ESS: Effective sample size}
  \end{center}
\end{table}

The estimated results of the log-normal frailty model differ from the others; therefore, we examined the behavior using different data in Appendix \ref{App:LeukSurv}.

\begin{table}[H]
  \begin{center}
\caption{Results of log-logistic model (12--19 group vs 31+ group)\label{t:DHS31_LL}}
\begin{tabular}{|c|c|c|c|c|c|c|c|c|}\hline
\multicolumn{2}{|c|}{Parameter} & Mode & Median & Mean & SE & 95\%CI & Rhat & ESS\\\hline
\multicolumn{2}{|c|}{Intercept} & 4.05 & 4.03 & 4.04 & 0.21 & [3.65, 4.48] & 1.00 & 931\\
\multicolumn{2}{|c|}{31+ group} & -0.22 & -0.23 & -0.23 & 0.07 & [-0.37, -0.10] & 1.00 & 4101\\
\multicolumn{2}{|c|}{Sex} & 0.05 & 0.05 & 0.05 & 0.03 & [0.00, 0.10] & 1.00 & 3482\\
Place & Respondents home & 0.48 & 0.52 & 0.51 & 0.20 & [0.09, 0.88] & 1.00 & 939\\
Place & Other home & 0.33 & 0.36 & 0.37 & 0.37 & [-0.34, 1.10] & 1.00 & 1873\\
Place & Public sector & 0.88 & 0.82 & 0.82 & 0.22 & [0.35, 1.24] & 1.00 & 1096\\
Place & Government hospital & 0.66 & 0.57 & 0.57 & 0.20 & [0.14, 0.94] & 1.00 & 925\\
Place & CS Govt health professional & 0.58 & 0.60 & 0.59 & 0.20 & [0.17, 0.95] & 1.00 & 921\\
Place & Other public sector & 0.77 & 0.75 & 0.74 & 0.24 & [0.26, 1.22] & 1.00 & 1169\\
Place & Private hospital/clinic & 0.29 & 0.35 & 0.34 & 0.20 & [-0.09, 0.71] & 1.00 & 926\\
Place & CS private health facility & 0.85 & 0.87 & 0.87 & 0.30 & [0.28, 1.49] & 1.00 & 1528\\
Size & Very large & 0.43 & 0.45 & 0.45 & 0.07 & [0.30, 0.59] & 1.00 & 1729\\
Size & Larger than average & 0.59 & 0.58 & 0.58 & 0.07 & [0.44, 0.72] & 1.00 & 1620\\
Size & Average & 0.63 & 0.64 & 0.65 & 0.06 & [0.53, 0.77] & 1.00 & 1446\\
Size & Smaller than average & 0.49 & 0.50 & 0.50 & 0.07 & [0.35, 0.64] & 1.00 & 1616\\
Size & Very small & 0.02 & 0.00 & 0.00 & 0.08 & [-0.15, 0.16] & 1.00 & 1756\\
\multicolumn{2}{|c|}{Order} & -0.01 & -0.01 & -0.01 & 0.01 & [-0.02, 0.01] & 1.00 & 4171\\
\multicolumn{2}{|c|}{$k$} & 1.86 & 1.86 & 1.86 & 0.04 & [1.79, 1.93] & 1.00 & 2427\\
\multicolumn{2}{|c|}{RMST$_{12-19 group}$} & 40.93 & 40.40 & 40.26 & 2.56 & [34.88, 45.00] &  & \\
\multicolumn{2}{|c|}{RMST$_{31+ group}$} & 37.31 & 37.22 & 37.09 & 3.13 & [30.89, 43.06] &  & \\
\multicolumn{2}{|c|}{RMST$_{diff}$} & -3.43 & -3.15 & -3.17 & 1.08 & [-5.37, -1.18] &  & \\\hline
\end{tabular}
  \footnotesize{95\%CI: 95\% credible interval, ESS: Effective sample size}
  \end{center}
\end{table}

\begin{table}[H]
  \begin{center}
\caption{Results of mixed effects log-logistic model (12--19 group vs 31+ group)\label{t:DHS31_LL_n}}
\begin{tabular}{|c|c|c|c|c|c|c|c|c|}\hline
\multicolumn{2}{|c|}{Parameter} & Mode & Median & Mean & SE & 95\%CI & Rhat & ESS\\\hline
\multicolumn{2}{|c|}{Intercept} & 4.00 & 4.05 & 4.06 & 0.21 & [3.66, 4.49] & 1.00 & 1133\\
\multicolumn{2}{|c|}{31+ group} & -0.24 & -0.23 & -0.22 & 0.07 & [-0.36, -0.08] & 1.00 & 4336\\
\multicolumn{2}{|c|}{Sex} & 0.05 & 0.05 & 0.05 & 0.03 & [0.00, 0.10] & 1.00 & 4488\\
Place & Respondents home & 0.55 & 0.54 & 0.53 & 0.20 & [0.12, 0.89] & 1.00 & 1042\\
Place & Other home & 0.29 & 0.37 & 0.39 & 0.36 & [-0.31, 1.11] & 1.00 & 1932\\
Place & Public sector & 0.83 & 0.83 & 0.83 & 0.23 & [0.37, 1.25] & 1.00 & 1204\\
Place & Government hospital & 0.64 & 0.58 & 0.57 & 0.20 & [0.16, 0.94] & 1.00 & 1035\\
Place & CS Govt health professional & 0.61 & 0.60 & 0.59 & 0.20 & [0.18, 0.96] & 1.00 & 1036\\
Place & Other public sector & 0.73 & 0.73 & 0.73 & 0.24 & [0.24, 1.19] & 1.00 & 1264\\
Place & Private hospital/clinic & 0.33 & 0.37 & 0.36 & 0.20 & [-0.05, 0.74] & 1.00 & 1062\\
Place & CS private health facility & 0.84 & 0.90 & 0.90 & 0.29 & [0.33, 1.49] & 1.00 & 1789\\
Size & Very large & 0.45 & 0.45 & 0.45 & 0.07 & [0.30, 0.59] & 1.00 & 1698\\
Size & Larger than average & 0.58 & 0.58 & 0.58 & 0.07 & [0.43, 0.72] & 1.00 & 1601\\
Size & Average & 0.63 & 0.64 & 0.64 & 0.06 & [0.52, 0.76] & 1.00 & 1400\\
Size & Smaller than average & 0.51 & 0.50 & 0.50 & 0.07 & [0.36, 0.64] & 1.00 & 1621\\
Size & Very small & 0.00 & 0.01 & 0.01 & 0.08 & [-0.15, 0.15] & 1.00 & 1751\\
\multicolumn{2}{|c|}{Order} & 0.00 & 0.00 & 0.00 & 0.01 & [-0.02, 0.01] & 1.00 & 5308\\
\multicolumn{2}{|c|}{$k$} & 1.87 & 1.86 & 1.86 & 0.04 & [1.79, 1.93] & 1.00 & 3258\\
\multicolumn{2}{|c|}{$\phi$} & 0.08 & 0.10 & 0.11 & 0.05 & [0.05, 0.22] & 1.00 & 1256\\
\multicolumn{2}{|c|}{RMST$_{12-19 group}$} & 40.13 & 40.58 & 40.44 & 2.57 & [35.09, 45.01] &  & \\
\multicolumn{2}{|c|}{RMST$_{31+ group}$} & 36.17 & 37.56 & 37.43 & 3.13 & [31.18, 43.19] &  & \\
\multicolumn{2}{|c|}{RMST$_{diff}$} & -2.64 & -2.96 & -3.01 & 1.10 & [-5.25, -0.97] &  & \\\hline
\end{tabular}
  \footnotesize{95\%CI: 95\% credible interval, ESS: Effective sample size}
  \end{center}
\end{table}

\begin{table}[H]
  \begin{center}
\caption{Results of log-logistic frailty model (12--19 group vs 31+ group)\label{t:DHS31_LL_f}}
\begin{tabular}{|c|c|c|c|c|c|c|c|c|}\hline
\multicolumn{2}{|c|}{Parameter} & Mode & Median & Mean & SE & 95\%CI & Rhat & ESS\\\hline
\multicolumn{2}{|c|}{Intercept} & 3.55 & 3.60 & 3.60 & 0.26 & [3.10, 4.10] & 1.00 & 1260\\
\multicolumn{2}{|c|}{31+ group} & -0.25 & -0.22 & -0.22 & 0.07 & [-0.35, -0.08] & 1.00 & 4729\\
\multicolumn{2}{|c|}{Sex} & 0.05 & 0.05 & 0.05 & 0.03 & [0.00, 0.10] & 1.00 & 4304\\
Place & Respondents home & 0.51 & 0.52 & 0.51 & 0.19 & [0.11, 0.85] & 1.00 & 1273\\
Place & Other home & 0.45 & 0.37 & 0.38 & 0.36 & [-0.28, 1.11] & 1.00 & 2561\\
Place & Public sector & 0.80 & 0.81 & 0.80 & 0.22 & [0.36, 1.19] & 1.00 & 1467\\
Place & Government hospital & 0.54 & 0.56 & 0.55 & 0.19 & [0.15, 0.89] & 1.00 & 1282\\
Place & CS Govt health professional & 0.62 & 0.58 & 0.57 & 0.19 & [0.17, 0.91] & 1.00 & 1267\\
Place & Other public sector & 0.70 & 0.71 & 0.71 & 0.23 & [0.24, 1.14] & 1.00 & 1558\\
Place & Private hospital/clinic & 0.31 & 0.36 & 0.35 & 0.19 & [-0.06, 0.69] & 1.00 & 1258\\
Place & CS private health facility & 0.81 & 0.90 & 0.90 & 0.29 & [0.35, 1.46] & 1.00 & 1746\\
Size & Very large & 0.42 & 0.44 & 0.44 & 0.07 & [0.31, 0.58] & 1.00 & 1802\\
Size & Larger than average & 0.60 & 0.57 & 0.57 & 0.07 & [0.43, 0.71] & 1.00 & 1761\\
Size & Average & 0.63 & 0.64 & 0.64 & 0.06 & [0.52, 0.75] & 1.00 & 1537\\
Size & Smaller than average & 0.48 & 0.50 & 0.50 & 0.07 & [0.36, 0.63] & 1.00 & 1847\\
Size & Very small & 0.03 & 0.02 & 0.02 & 0.07 & [-0.13, 0.16] & 1.00 & 1914\\
\multicolumn{2}{|c|}{Order} & 0.00 & 0.00 & 0.00 & 0.01 & [-0.02, 0.01] & 1.00 & 5181\\
\multicolumn{2}{|c|}{$k$} & 1.82 & 1.83 & 1.83 & 0.04 & [1.76, 1.90] & 1.00 & 3725\\
\multicolumn{2}{|c|}{$\phi$} & 1.16 & 2.15 & 2.90 & 2.74 & [0.20, 10.01] & 1.00 & 1302\\
\multicolumn{2}{|c|}{RMST$_{12-19 group}$} & 33.30 & 34.02 & 33.86 & 3.98 & [25.70, 41.10] &  & \\
\multicolumn{2}{|c|}{RMST$_{31+ group}$} & 31.44 & 30.50 & 30.40 & 4.27 & [21.91, 38.46] &  & \\
\multicolumn{2}{|c|}{RMST$_{diff}$} & -3.79 & -3.46 & -3.46 & 1.13 & [-5.65, -1.16] &  & \\\hline
\end{tabular}
  \footnotesize{95\%CI: 95\% credible interval, ESS: Effective sample size}
  \end{center}
\end{table}

\subsubsection{Values of widely applicable information criterion}\label{App:DHS_WAIC}
We presented the calculated WAICs in Tables \ref{t:DHS2030_WAIC} and \ref{t:DHS31_WAIC}. For the comparison between the "12-19 group" and the "20-30 group," the mixed effects Weibull model yielded the minimum WAIC. For the comparison between the "12-19 group" and the "31+ group," the Weibull frailty model had the minimum WAIC. The WAIC for the log-logistic frailty model was larger than that for the other models. We further investigated the WAIC using different data in Appendix \ref{App:LeukSurv}.
\begin{table}[H]
  \begin{center}
\caption{Results of WAIC (12--19 group vs 20-30 group)\label{t:DHS2030_WAIC}}
\begin{tabular}{|c|c|}\hline
 & WAIC\\\hline
exponential model & 70797.8912255\\
mixed effects exponential model & 70729.22940\\
exponential frailty model & 70728.64805\\
Weibull model & 69233.90573\\
{\bf mixed effects Weibull model} & {\bf 69160.98410}\\
Weibull frailty model& 69162.08765\\
log-normal model & 70373.81205\\
mixed effects log-normal model & 70302.03889\\
log-normal frailty model & 69379.64799\\
log-logistic model & 69349.66898\\
mixed effects log-logistic model & 69277.20932\\
log-logistic frailty model & 1006049.864\\
\hline
\end{tabular}
  \end{center}
\end{table}

\begin{table}[H]
  \begin{center}
\caption{Results of WAIC (12--19 group vs 31+ group)\label{t:DHS31_WAIC}}
\begin{tabular}{|c|c|}\hline
 & WAIC\\\hline
exponential model & 29204.82671\\
mixed effects exponential model & 29169.61918\\
exponential frailty model & 29169.12216\\
Weibull model & 28398.84546\\
mixed effects Weibull model & 28363.75397\\
{\bf Weibull frailty model} & {\bf 28363.7075}\\
log-normal model & 28902.55726\\
mixed effects log-normal model & 28862.84461\\
log-normal frailty model & 28512.54361\\
log-logistic model & 28448.37898\\
mixed effects log-logistic model & 28412.54262\\
log-logistic frailty model & 64523.48299\\
\hline
\end{tabular}
  \end{center}
\end{table}

\subsection{Sensitivity analysis for log-normal frailty model}\label{App:LeukSurv}
We conducted a sensitivity analysis of the log-normal frailty model using the LeukSurv dataset from the spBayesSurv package in R. We compared the 'Greater than or equal to 65 years old' group with the 'Less than 65 years old' group. The covariates included sex, white blood cell count at diagnosis (Wbc), and the Townsend score, where higher values indicate less affluent areas (Tpi). The clusters were the twenty-four administrative districts of residence. The sample size was 1043 (540 in the 'Greater than or equal to 65 years old' group and 503 in the 'Less than 65 years old' group). The distribution of the number of subjects in each group is as follows:
\begin{table}[H]
  \begin{center}
\caption{Disposition of the number of subjects in each distinct\label{App:Num_distinct}}
\begin{tabular}{|c|c|c|c|c|c|c|c|c|c|c|c|c|}\hline
Distinct ID &  1 & 2 & 3 & 4 & 5 & 6 & 7 & 8 & 9 & 10 & 11 & 12  \\\hline
\# subjects & 35 & 69 & 44 & 15 & 46 & 12 & 71 & 25 & 34 & 12 & 17 & 27 \\\hline\hline
Distinct ID &  13 & 14 & 15 & 16 & 17 & 18 & 19 & 20 & 21 & 22 & 23 & 24 \\\hline
\# subjects & 18 & 58 & 37 & 52 & 84 & 45 & 61 & 53 & 63 & 27 & 36 & 102  \\\hline
\end{tabular}
  \end{center}
\end{table}

The Kaplan-Meier plot comparing the 'Greater than or equal to 65 years old' group with the 'Less than 65 years old' group is illustrated below.
\begin{figure}[H]
  \begin{center}
  \includegraphics[width=15cm]{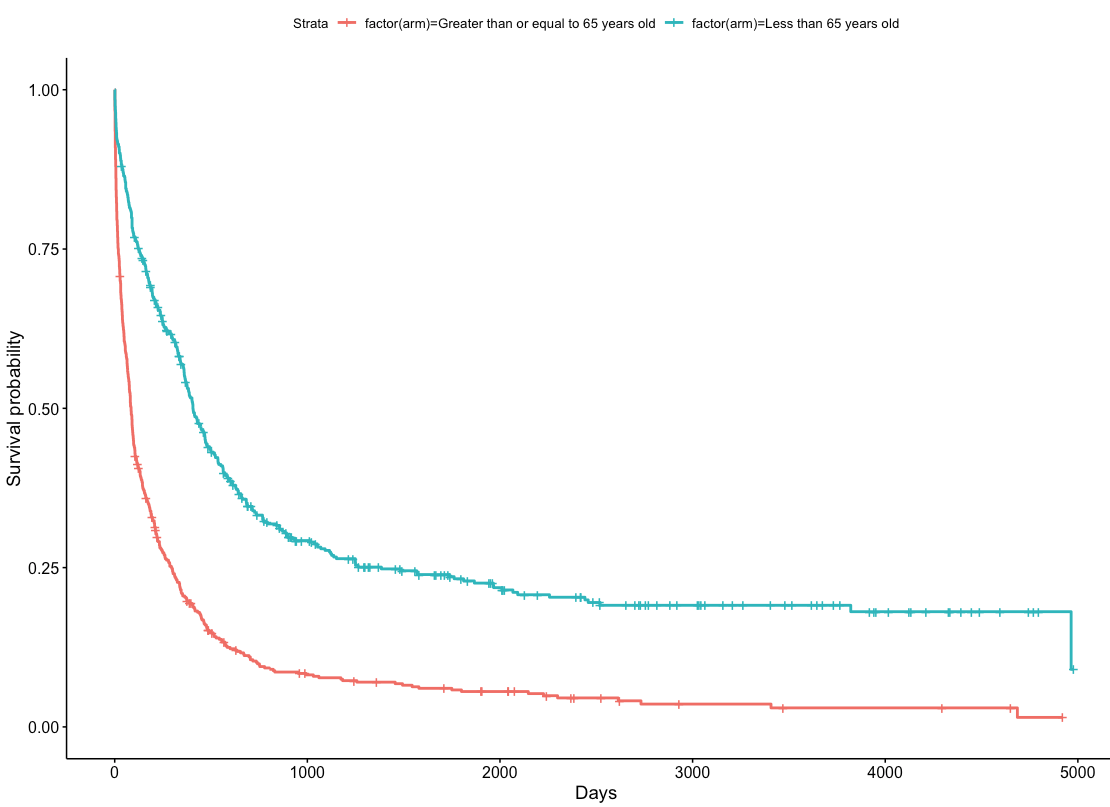}
  \caption{Kaplan Meier plot of LeukSurv ('Greater than or equal to 65 years old' group vs 'Less than 65 years old' group)}
  \label{App:KM_LeukSurv}
        \footnotesize{}
  \end{center}
\end{figure}

The estimation results are shown in Tables \ref{App:LeukSurv_LN},\ref{App:LeukSurv_LN_n}, and \ref{App:LeukSurv_LN_f}. The log-normal frailty model is consistent with the other estimated results.
\begin{table}[H]
  \begin{center}
\caption{Results of log-normal model\label{App:LeukSurv_LN}}
\begin{tabular}{|c|c|c|c|c|c|}\hline
Parameter & Mode & Median & Mean & SE & 95\%CI\\\hline
Intercept & 4.58 & 4.60 & 4.60 & 0.12 & [4.37, 4.84]\\
Less than 65 years old & 1.69 & 1.69 & 1.69 & 0.13 & [1.45, 1.93]\\
Sex & -0.01 & -0.03 & -0.04 & 0.13 & [-0.30, 0.21]\\
Wbc & -0.01 & -0.01 & -0.01 & 0.00 & [-0.01, -0.01]\\
Tpi & -0.06 & -0.06 & -0.06 & 0.02 & [-0.10, -0.03]\\
$\si$ & 2.05 & 2.05 & 2.05 & 0.05 & [1.95, 2.15]\\\hline
\end{tabular}
  \footnotesize{95\%CI: 95\% credible interval}
  \end{center}
\end{table}

\begin{table}[H]
  \begin{center}
\caption{Results of mixed effects log-normal model\label{App:LeukSurv_LN_n}}
\begin{tabular}{|c|c|c|c|c|c|}\hline
Parameter & Mode & Median & Mean & SE & 95\%CI\\\hline
Intercept & 4.61 & 4.60 & 4.60 & 0.12 & [4.36, 4.84]\\
Less than 65 years old & 1.69 & 1.70 & 1.70 & 0.13 & [1.44, 1.96]\\
Sex & -0.05 & -0.04 & -0.03 & 0.13 & [-0.28, 0.22]\\
Wbc & -0.01 & -0.01 & -0.01 & 0.00 & [-0.01, -0.01]\\
Tpi & -0.07 & -0.06 & -0.06 & 0.02 & [-0.10, -0.03]\\
$\si$ & 2.05 & 2.04 & 2.04 & 0.05 & [1.94, 2.14]\\
$\phi$ & 0.14 & 0.16 & 0.16 & 0.10 & [0.01, 0.37]\\\hline
\end{tabular}
  \footnotesize{95\%CI: 95\% credible interval}
  \end{center}
\end{table}

\begin{table}[H]
  \begin{center}
\caption{Results of log-normal frailty model\label{App:LeukSurv_LN_f}}
\begin{tabular}{|c|c|c|c|c|c|}\hline
Parameter & Mode & Median & Mean & SE & 95\%CI\\\hline
Intercept & 4.56 & 4.60 & 4.60 & 0.14 & [4.33, 4.87]\\
Less than 65 years old & 1.71 & 1.71 & 1.71 & 0.13 & [1.46, 1.96]\\
Sex & 0.02 & -0.03 & -0.03 & 0.13 & [-0.28, 0.21]\\
Wbc & -0.01 & -0.01 & -0.01 & 0.00 & [-0.01, -0.01]\\
Tpi & -0.06 & -0.07 & -0.07 & 0.02 & [-0.10, -0.03]\\
$\si$ & 2.02 & 2.03 & 2.03 & 0.06 & [1.92, 2.14]\\
$\phi$ & 0.02 & 0.03 & 0.03 & 0.02 & [0.01, 0.09]\\\hline
\end{tabular}
  \footnotesize{95\%CI: 95\% credible interval}
  \end{center}
\end{table}

We displayed the WAICs in Table \ref{App:LeukSurv_WAIC} and confirmed that the WAIC of the log-logistic frailty model is almost the same as the other WAICs.
\begin{table}[H]
  \begin{center}
\caption{Results of WAIC\label{App:LeukSurv_WAIC}}
\begin{tabular}{|c|c|}\hline
& WAIC\\\hline
log-normal model & 11983.72\\
mixed effects log-normal model & 11983.72\\
log-normal frailty model& 11977.70\\
log-logistic frailty model & 12068.60\\
\hline
\end{tabular}
  \end{center}
\end{table}